\documentclass[aps,prc,twocolumn,showpacs,floatfix,nofootinbib,preprintnumbers,superscriptaddress,amsmath,amssymb]{revtex4-1}

\usepackage{graphicx}
\usepackage{epsfig}
\usepackage{bm}
\usepackage{color}
\usepackage{float}
\usepackage{dcolumn}
\usepackage{multirow}

\newcommand{\ket}[1]{\left| #1 \right\rangle}
\newcommand{\braket}[2]{\left\langle #1 \vert #2 \right\rangle}
\newcommand{\vare}{\varepsilon} 
\newcommand{\ma}[1]{\mathbf{#1}}

\usepackage{amsmath}
\usepackage{tabularx}

\begin{document}

\title{Coupled-cluster studies of infinite  nuclear matter}

\date{\today}

\author{G.~Baardsen} \affiliation{Department of Physics and Center of
  Mathematics for Applications, University of Oslo, N-0316 Oslo,
  Norway}

\author{A.~Ekstr{\"o}m} \affiliation{Department of Physics and Center of
  Mathematics for Applications, University of Oslo, N-0316 Oslo,
  Norway} \affiliation{National Superconducting Cyclotron Laboratory,
  Michigan State University, East Lansing, MI 48824, USA}

\author{G.~Hagen}
\affiliation{Physics Division, Oak Ridge National Laboratory,
Oak Ridge, TN 37831, USA}
\affiliation{Department of Physics and Astronomy, University of
Tennessee, Knoxville, TN 37996, USA}

\author{M.~Hjorth-Jensen} \affiliation{Department of Physics
  and Center of Mathematics for Applications, University of Oslo,
  N-0316 Oslo, Norway} \affiliation{National Superconducting Cyclotron
  Laboratory, Michigan State University, East Lansing, MI 48824,
  USA} \affiliation{Department of Physics and Astronomy, Michigan
  State University, East Lansing, MI 48824, USA}

\begin{abstract}
\noindent
{\bf Background:} Coupled-cluster (CC) theory is a widely used
many-body method for studying strongly correlated many-fermion
systems.  It allows for systematic inclusions of complicated many-body
correlations beyond a mean-field. Recent applications to finite nuclei
have shown that first principle approaches like CC theory can be
extended to studies of medium-heavy nuclei, with excellent agreement
with experiment. However, CC calculations of properties of infinite
nuclear matter are rather few and date back more than 30 years in
time. \\
{\bf Purpose:} The aim of this work is thus to develop the
relevant formalism for performing coupled-cluster calculations in
nuclear matter and neutron star matter, including thereby important
correlations to infinite order in the interaction and testing modern
nuclear forces based on chiral effective field theory.  Our formalism
includes also the exact treatment of the so-called Pauli operator in a
partial wave expansion of the equation of state.  \\
{\bf Methods:} Nuclear and neutron matter calculations are done using a coupled
particle-particle and hole-hole ladder approximation. The coupled
ladder equations are derived as an approximation of CC theory, leaving
out particle-hole and non-linear diagrams from the CC doubles
amplitude equation. This study is a first step toward CC calculations
for nuclear and neutron matter.  \\
{\bf Results:} We present results for
both symmetric nuclear matter and pure neutron matter employing
state-of-the-art nucleon-nucleon interactions based on chiral
effective field theory. We employ also the newly optimized chiral
interaction [A. Ekstr{\"o}m \emph{et al.},
  Phys. Rev. Lett. $\mathbf{110}$, 192502 (2013)] to study infinite
nuclear matter. The ladder approximation method and corresponding
results are compared with conventional Brueckner-Hartree-Fock
theory. The ladder approximation is derived and studied using both
exact and angular-averaged Pauli exclusion operators, with
angular-averaged input momenta for the single-particle potentials in
all caluclations.  The inclusion of an exact treatment of the Pauli
operators in a partial wave expansion yields corrections of the order
of $1.7-2\%$ of the total energy in symmetric nuclear
matter. Similarly, the inclusion of both hole-hole and
particle-particle ladders result in corrections of the order $0.7-2\%$
compared to the approximation with only particle-particle
ladders. Including these effects, we get at most almost a $6\%$
difference between our CC calculation and the standard
Brueckner-Hartree-Fock approach. \\  
{\bf Conclusions:} We have performed
CC calculations of symmetric nuclear matter and pure neutron matter
including particle-particle and hole-hole diagrams to infinite order
using an exact Pauli operator and angular-averaged single-particle
energies. The contributions from hole-hole diagrams and exact Pauli
operators add important changes to the final energies per particle.

\end{abstract}


\maketitle

\section{Introduction}

Nuclear matter is defined as an isotropic system consisting of
infinitely many nucleons which interact only by nuclear forces.  
The Schr{\"o}dinger equation of this system has been solved
approximately using many different \emph{ab initio} many-body methods 
\cite{day1978,day_cc,akmal1997,dickhoff2004,gandolfi2007,li2006,lovato2011,carbone2013}. 
As an example, diagrammatic partial summations have been derived from
many-body perturbation theory to calculate the binding energy. One
approach belonging to this family of methods is the
Brueckner-Goldstone (BG) expansion \cite{brueckner}, which is a
Goldstone expansion where the interaction has been replaced by a
well-behaved reaction matrix \cite{day1967}. The
Brueckner-Hartree-Fock (BHF) scheme \cite{brueckner1955b, brueckner}, which 
is one of the standard methods of nuclear matter theory 
\cite{haftel_tabakin,song,schiller,suzuki,li2006,baldo2012}, is a first-order approximation in BG theory.

Unfortunately, the BG expansion does not converge very well when using
the number of reaction matrices as the order parameter \cite{raja}.  A
more appropriate way to include higher-order correlations is the
hole-line expansion \cite{day1978}, where the perturbation truncation
is determined by the number of independent hole lines in the the
Brueckner-Goldstone diagrams. The two-hole-line approximation is then
equivalent to the BHF method. Calculations including up to
three-hole-line diagrams indicate that the hole-line expansion
converges \cite{day1981,song}. Despite encouraging results for the
hole-line expansion, it would still be desirable to get a deeper
understanding of the accuracy of the many-body methods applied to
nuclear matter. Better knowledge about the convergence of the
many-body methods in nuclear matter is also necessary to validate the
quality of calculations including three- and many-body interactions
\cite{soma2008,vidana2009,hebeler2011,gandolfi2012,hebeler2013,tews,kruger2013}.
 
An approach that is related to the above-mentioned perturbative
techniques is the coupled-cluster (CC) method
\cite{coester1958,coester1960,cizek1966,
  cizek1969,cizek1971,bartlett_review}. As perturbation theory,
coupled-cluster theory gives a non-variational and size-extensive
method.  However, in contrast to many-body perturbation theory, CC
theory sums to infinite order, depending on the level of truncation,
selected many-body contributions. It is thus a non-perturbative
method.  The coupled-cluster truncation is physical in the sense that
a so-called $T_{n}$ truncation includes all possible correlations
arising from sets of $n$-particle-$n$-hole clusters
\cite{bartlett_book, crawford, harris}. During the last one and a half
decade, CC theory has been successfully applied to structure
calculations of finite nuclei
\cite{heisenberg1999,heisenberg2000,dean2004,gour2006,gour2008,hagen2010,
  jansen2011,jensen2010,roth2012}.  Inspired by the success of the CC
approach in finite nuclei, we hope that CC theory could also provide
accurate results for infinite nuclear matter.

Already in the
early 1980's, Day and Zabolitzky performed CC calculations for nuclear
matter \cite{day_cc} using the Bochum truncation
\cite{kummel1978, bishop_cc}. The theory of nuclear interactions has evolved a
lot since that time, see for example Ref.~\cite{machleidt2001}, with   
the construction of high-precision potentials exhibiting a $\chi^2$ per datum close to one  with respect to nucleon-nucleon scattering data \cite{machleidt2001}. Later, two- and three-body interactions have
been developed based on effective field theory (EFT), which is derived
using symmetries arising from quantum chromodynamics (QCD)
\cite{machleidt2011}. In addition, soft low-momentum interactions
based on renormalization group theory have been introduced
\cite{bogner2005}. It seems therefore necessary to perform new CC
calculations for nuclear matter with modern two- and three-body
interactions.

In the Bochum truncation scheme \cite{kummel1978}, the exact CC amplitude 
equations are approximated by setting all so-called reduced $n$-particle 
subsystem amplitudes $\chi_{n}$, with $n$ larger than a chosen truncation
level, to zero. The justification for using this truncation is that 
all subsystem amplitudes $\chi_{n}$ become small inside a radius
where the interaction may be very strong \cite{kummel1978}. This
truncation scheme therefore ensures that the contribution from
particles interacting strongly at short distances becomes small. 
In their CC calculations for nuclear matter, Day and Zabolitzky 
included the two- and three-body subsystem amplitudes $\chi_{2}$ and $\chi_{3}$, 
and incorporated also parts of the four-body amplitude $\chi_{4}$ 
\cite{day_cc,day1981}. Given the computing capacity in the early 
1980s, it was necessary to do further approximations in the 
CC equations. As is explained in Refs.~\cite{day_cc,day1981}, 
the authors used angular-averaged Pauli exclusion operators, 
other averages over angles, and first-order Taylor expansions 
to approximate the dependence of the single-particle potential 
on the $G$-matrix. In addition, they replaced some diagrams 
by estimates, which were simpler to calculate
than the original diagrams \cite{day_cc}. Before this study,
Manzke \cite{manzke1974} did CC calculations for 
nuclear matter in the two-body subsystem approximation.  

Coupled cluster theory is widely used within the quantum 
chemistry community \cite{bartlett_review}. A commonly used 
CC truncation scheme in quantum chemistry is to set all cluster
amplitudes in the exponential CC wave function ansatz beyond a 
given excitation level to zero (see Sec.~\ref{sec:cclad} and 
Refs.~\cite{crawford,bartlett_book}).  
These approximations are called for example CC doubles (CCD), CC
singles-doubles (CCSD), etc.~\cite{bartlett_review}, or alternatively 
SUB$n$ approximations \cite{bishop1}. 
This truncation scheme has been applied in studies of finite nuclei
\cite{dean2004,gour2006,hagen2010,jansen2011,jensen2010,roth2012},
mainly using soft interactions. Our aim is to apply the same CC method
to studies of the equation of state of symmetric nuclear matter and
neutron matter.

The SUB$n$ approximation includes long-range contributions, 
such as the forward and other ring diagrams, already at 
the CCD or SUB2 level \cite{freeman1977,bishop1}.
In contrast, these correlations are included in the Bochum
scheme in the subsystem amplitudes $\chi_{3}$ and $\chi_{4}$ 
\cite{kummel1978}.   
Another difference between the two CC truncation schemes is the
treatment of the single-particle state potentials in the energy denominator. 
In the Bochum CC method, the energy denominator contains 
single-hole state potentials with summations of particle-particle ladder diagrams
to infinite order, whereas the single-particle state potentials are zero in the
energy denominator. Instead, the single-particle state 
potential terms are part of the $\chi_{3}$ and $\chi_{4}$ 
subsystem amplitudes \cite{kummel1978}. In contrast, in the 
SUB$n$ approximation, which we employ in this work, the energy denominator
contains single-particle potentials at the Hartree-Fock level
for both particles and holes.


Previously, Freeman \cite{freeman_pplad} has studied the
two-dimensional electron gas including particle-particle ladder 
diagrams from the CC doubles approximation. In a similar way, as 
a first step toward CC calculations for nuclear matter, we will here include 
particle-particle and hole-hole correlations at the 2-particle-2-hole, or $T_{2}$, level. 
In this scheme, only the linear ladder diagrams have been included 
in the CC $T_{2}$ amplitude equation, whereas the linear particle-hole 
diagrams and all non-linear diagrams have been neglected. When leaving 
out certain diagrams of the $T_{2}$ amplitude equation, our scheme may
be considered as strictly not a coupled-cluster approximation. However,
the method shows a proof of principle of an iterative coupled-cluster
numerical scheme, where particle-particle and hole-hole ladder
diagrams are coupled and summed to infinite order. The coupled ladder approximation
is similar to the Bochum CC method including only the two-body subsystem
amplitude $\chi_{2}$, but the single-particle potentials are different. 
It ought  also to be emphasized that the calculations of Day and 
Zabolitzky \cite{day_cc} included a larger set of diagrams, and thereby correlations,  than
the approximation used in this work.

According to the hole-line expansion 
calculations by Song \emph{et al.}, the contribution of particle-particle 
diagrams is considerably larger than that of particle-hole diagrams 
\cite{song}. Still, the contribution of particle-hole diagrams is 
clearly non-negligible in the hole-line expansion \cite{song}. 
The results of Ref.~\cite{dickhoff1982} show that ring 
(particle-hole) diagrams are significant for the binding energy
of nuclear matter. The aim is therefore 
to include all $T_{2}$ diagrams in a future CC calculation to get a 
proper CC approximation at the $T_{2}$ level. 

Summation of particle-particle and hole-hole ladder diagrams is also a
common approximation in self-consistent Green's function (SCGF) theory
\cite{dickhoff2004}. The SCGF ladder approximation has been
extensively applied to studies of nuclear and neutron matter
\cite{ramos_polls,bozek2002,dewulf2003,frick2003,rios2009}, lately
including either three-body interactions or density dependent two-body
operators arising from three-body interactions
\cite{soma2008,carbone2013}.  In the SCGF ladder approximation, the
energy denominator contains self-consistently solved complete 
off-shell self-energies including both particle-particle and hole-hole ladder
diagrams. As the SCGF method, the CC ladder approximation also treats
particle and hole interactions symmetrically, but from the definition
of the CC equations it follows that the single-particle potentials
occur in the energy denominator only up to the Hartree-Fock level.

Another similar method is the particle-particle and hole-hole ring
diagram approximation \cite{song1987,jiang1988,engvik1997a,siu2009},
where the particle-particle and hole-hole diagrams are derived from
Green's function theory, and a momentum model space is used to avoid
poles in the energy denominator. The binding energy obtained in this
approximation depends however on the model space momentum cutoff
\cite{engvik1997a}. In the particle-particle and hole-hole ring
diagram method the authors employed the standard angular-average
approximations in order to decrease the computational complexity of
the calculations \cite{song1987}.

In the present work, we will analyze the partial wave expansion of 
the equation of state using an exact treatment of the intermediate 
states, avoiding thereby the standard angle-average approximation 
of Pauli exclusion operators \cite{haftel_tabakin}. Finally, we 
perform calculations of the above
systems using modern nucleon-nucleon interactions based on  
chiral perturbation theory to next-to-next-to-leading order  (NNLO)
\cite{ekstrom2013} and next-to-next-to-next-to-leading order 
(N$^3$LO) \cite{n3lo}.

After these introductory remarks, we present our formalism in the next 
three sections, followed subsequently by our results and discussion 
thereof in Sec.~\ref{sec:results}. Concluding remarks and perspectives 
are presented in Sec.~\ref{sec:conclusions}.

\section{Formalism: Many-body methods}

The general form of the Hamiltonian operator of infinite nuclear
matter is
\begin{align}
  \hat{H} &= \hat{K} + \hat{V}_{NN} + \hat{V}_{NNN} + \dots \nonumber
  \\ &= -\frac{\hbar^{2}}{2M}\sum_{i=1}^{A}\nabla_{i}^{2} +
  \sum_{i<j}^{A}\hat{v}_{NN}(\mathbf{r}_{i},\mathbf{r}_{j}) \nonumber
  \\ &+
  \sum_{i<j<k}^{A}\hat{v}_{NNN}(\mathbf{r}_{i},\mathbf{r}_{j},\mathbf{r}_{k})
  + \dots,
\end{align}
where $A$ is the number of nucleons, $\hat{K}$ is the kinetic energy
operator, $\hat{V}_{NN}$ is a two-particle interaction operator,
$\hat{V}_{NNN}$ is a three-particle interaction operator, $M$ is the
nucleon mass, and $\mathbf{r}_{l}$ is the position vector of particle
$l$. In this paper, we neglect $n$-body interactions for $n$ larger
than two and define the Hamiltonian operator as
\[
  \hat{H} = \hat{K} + \hat{V} =-\frac{\hbar^{2}}{2M}\sum_{i=1}^{A}\nabla_{i}^{2} +
  \sum_{i<j}^{A}\hat{v}(r_{ij}),
\]
where $\hat{v}$ is a two-body interaction and $r_{ij} =
|\mathbf{r}_{i}-\mathbf{r}_{j}|$.

In our calculations, we use the nucleon-nucleon interaction of
Ref.~\cite{n3lo}. This interaction model is given by an N$^3$LO approximation of
chiral perturbation theory. Nuclear interactions based on effective
field theory have the advantage that two- and many-body interactions
can be derived in a mutually consistent way \cite{machleidt2011}.
Furthermore, we present also results obtained with a recent 
nucleon-nucleon interaction at order NNLO in chiral perturbation
theory. This interaction results from an optimization-based re-parameterization
to the available body of experimental data using the model-based, derivative-free 
optimization algorithm POUNDerS developed at Argonne National Laboratory
\cite{taoman}.  The resulting new chiral interaction, labeled
NNLO$_{\mathrm{opt}}$ hereafter, exhibits a $\chi^2$ per datum close to one for
laboratory scattering-energies below approximately 125 MeV in the two-body
proton-proton and neutron-proton channels, see Ref.~\cite{ekstrom2013}
for further details.  In the $A=3$ and $A=4$ nucleon systems, this interaction
gives binding energies that differ by 20 keV and 45 keV from the experimental values, respectively. Thus, the  contributions of three-nucleon forces appear smaller than for previous parametrizations of chiral interactions.

We model infinite nuclear matter as a system of $A$ interacting
nucleons confined by a cubic box potential.  The cubic box
boundary condition together with the free nucleon Hamiltonian equation
\[
  -\frac{\hbar^{2}}{2M}\nabla^{2}\varphi(\mathbf{r}) = \vare
  \varphi(\mathbf{r}),
\]
gives the plane wave eigenfunctions $\varphi_{\mathbf{k}}(\mathbf{r})
= e^{i\mathbf{k}\cdot \mathbf{r}}/\sqrt{\Omega }$ and eigenenergies
$\vare_{\mathbf{k}} = \hbar^{2} k^{2}/(2M)$. Here $\hbar \mathbf{k}$
is the momentum, $\mathbf{r}$ is the position coordinate, and $\Omega
$ is the volume of the box. We therefore use plane waves as our
single-particle basis, from which the Slater determinants are
constructed.

\subsection{Coupled ladder approximation} \label{sec:cclad}
In this subsection, the coupled ladder equations will be derived as an
approximation of the coupled-cluster method. The coupled-cluster
formalism is presented in a momentum basis. In the general
expressions, we omit spin and isospin degrees of freedom. 

 In coupled-cluster theory, the $A$-fermion state vector is expressed
 using the exponential ansatz
\[
  \ket{\Psi } \equiv e^{\hat{T}}\ket{\Phi_{0} },
\]
where  $\ket{\Phi_{0} }$ is the uncorrelated free Fermi vacuum,
and the cluster operator $\hat{T}$ is defined as the sum
\[
  \hat{T} = \sum_{m=1}^{A}\hat{T}_{m},
\]
of $m$-particle $m$-hole excitation operators
\begin{align}
  \hat{T}_{m} &= \left( \frac{1}{m!}\right)^{2}
  \sum_{\mathbf{k}_{i_{1}}, \dots ,\mathbf{k}_{i_{m}}\atop
    \mathbf{k}_{a_{1}}, \dots ,
    \mathbf{k}_{a_{m}}}t_{\mathbf{k}_{i_{1}}\dots
    \mathbf{k}_{i_{m}}}^{\mathbf{k}_{a_{1}}\dots \mathbf{k}_{a_{m}}}
  \nonumber \\ & \times c_{\mathbf{k}_{a_{1}}}^{\dagger }\dots
  c_{\mathbf{k}_{a_{m}}}^{\dagger }c_{\mathbf{k}_{i_{m}}}\dots
  c_{\mathbf{k}_{i_{1}}}.
\end{align}
We label single-particle states occupied in the Fermi vacuum determinant
$\Phi_{0}$ (holes) by $i, j, k, \dots $ and excited states of the same
single-particle basis (particles) by $a, b, c, $ and so on. Indices $p, q, r, \dots $ are used to label single-particle states that may be either holes or particles.
The operators $c^{\dagger }$ and $c$ are fermion creation and annihilation
operators, respectively.

Given that the single-particle basis is complete, the $A$-particle
Schr{\"o}dinger equation can be written equivalently as the CC energy
equation
\begin{equation} \label{eq:ccene}
  \langle \Phi_{0}|e^{-\hat{T}}\hat{H}e^{\hat{T}}|\Phi_{0}\rangle = E,
\end{equation}  
where the cluster operator $\hat{T}$ is obtained from the
corresponding set of CC amplitude equations
\begin{equation} \label{eq:ccAmpl}
  \langle \Phi_{\mathbf{k}_{i_{i}}\mathbf{k}_{i_{2}}\dots
    \mathbf{k}_{i_{k}}}^{\mathbf{k}_{a_{1}}\mathbf{k}_{a_{2}}\dots
    \mathbf{k}_{a_{k}}}|e^{-\hat{T}}\hat{H}e^{\hat{T}}|\Phi_{0}\rangle
  = 0
\end{equation}
 for $k = 1, 2, 3, \dots , A$. Here we have used the notation
\begin{align}
  |\Phi_{\mathbf{k}_{i_{1}}\mathbf{k}_{i_{2}}\dots
    \mathbf{k}_{i_{k}}}^{\mathbf{k}_{a_{1}}\mathbf{k}_{a_{2}}\dots
    \mathbf{k}_{a_{k}}}\rangle &\equiv c_{\mathbf{k}_{a_{1}}}^{\dagger
  }c_{\mathbf{k}_{a_{2}}}^{\dagger }\dots
  c_{\mathbf{k}_{a_{k}}}^{\dagger } \nonumber \\ & \times
  c_{\mathbf{k}_{i_{k}}}\dots
  c_{\mathbf{k}_{i_{2}}}c_{\mathbf{k}_{i_{1}}}|\Phi_{0}\rangle ,
\end{align}
which means that the bra vector in Eq.~(\ref{eq:ccAmpl}) is a
$k$-particle $k$-hole excitation of the Fermi vacuum state.

In almost all practical calculations, except for some very simple
model systems, it is necessary to do a truncation both in the cluster
operator $\hat{T}$ and in the single-particle basis. To derive the
ladder expansion, we need only the approximation $\hat{T} \approx
\hat{T}_{2}$, which is commonly called the CC doubles approximation
(CCD). In fact, the $\hat{T}_{1}$ operator is found to vanish for
infinite nuclear and neutron matter \cite{kummel1978}.
By symmetry considerations, the total momentum of the system of nucleons 
is zero. Both the kinetic energy operator $\hat{K}$ and the total 
Hamiltonian $\hat{H}$ are assumed to
be diagonal in total momentum $\mathbf{K}$. Hence, both the reference state
$\Phi_{0}$ and the correlated ground state $\Psi $
must be eigenfunctions of the operator $\mathbf{\hat{K}}$ with the
corresponding eigenvalue $\mathbf{K} = \mathbf{0}$ \cite{day1967}.

Using abstract vectors and a momentum single-particle basis, the CC
ansatz can be written as
  \begin{align} \label{eq:expccmom}
    |\Psi_{CC} \rangle &= |\Phi_{0}\rangle +
    \sum_{\mathbf{k}_{i}\mathbf{k}_{a}}t_{\mathbf{k}_{i}}^{\mathbf{k}_{a}}|\Phi_{\mathbf{k}_{i}}^{\mathbf{k}_{a}}\rangle
    +
    \frac{1}{4}\sum_{\mathbf{k}_{i}\mathbf{k}_{j}\mathbf{k}_{a}\mathbf{k}_{b}}t_{\mathbf{k}_{i}\mathbf{k}_{j}}^{\mathbf{k}_{a}\mathbf{k}_{b}}|\Phi_{\mathbf{k}_{i}\mathbf{k}_{j}}^{\mathbf{k}_{a}\mathbf{k}_{b}}\rangle
    \nonumber \\ & +
    \frac{1}{2}\sum_{\mathbf{k}_{i}\mathbf{k}_{a}}\sum_{\mathbf{k}_{j}\mathbf{k}_{b}}t_{\mathbf{k}_{i}}^{\mathbf{k_{a}}}t_{\mathbf{k}_{j}}^{\mathbf{k}_{b}}|\Phi_{\mathbf{k}_{i}\mathbf{k}_{j}}^{\mathbf{k}_{a}\mathbf{k}_{b}}\rangle
    + \ldots ,
  \end{align}
where $\mathbf{k}_{m}$ is the momentum of the single-particle state
$m$. From Eq.~(\ref{eq:expccmom}) and the conditions
\[
  \hat{\mathbf{K}}|\Phi_{0}\rangle = \mathbf{0}|\Phi_{0}\rangle,
\]
and
\[
  \hat{\mathbf{K}}|\Psi_{CC}\rangle = \mathbf{0}|\Psi_{CC}\rangle,
\]
it follows that
 \[
   \hat{\mathbf{K}}t_{\mathbf{k}_{i}}^{\mathbf{k}_{a}}|\Phi_{\mathbf{k}_{i}}^{\mathbf{k}_{a}}\rangle
   = \left( \mathbf{k}_{a} - \mathbf{k}_{i}\right)
   t_{\mathbf{k}_{i}}^{\mathbf{k}_{a}}|\Phi_{\mathbf{k}_{i}}^{\mathbf{k}_{a}}\rangle,
 \]
and
\[
    \hat{\mathbf{K}}t_{\mathbf{k}_{i}\mathbf{k}_{j}}^{\mathbf{k}_{a}\mathbf{k}_{b}}|\Phi_{\mathbf{k}_{i}\mathbf{k}_{j}}^{\mathbf{k}_{a}\mathbf{k}_{b}}\rangle
    = \left( \mathbf{k}_{a} + \mathbf{k}_{b} - \mathbf{k}_{i} -
    \mathbf{k}_{j}\right)t_{\mathbf{k}_{i}\mathbf{k}_{j}}^{\mathbf{k}_{a}\mathbf{k}_{b}}|\Phi_{\mathbf{k}_{i}\mathbf{k}_{j}}^{\mathbf{k}_{a}\mathbf{k}_{b}}\rangle.
\]
Since by definition $|\mathbf{k}_{a}| > |\mathbf{k}_{i}|$, the CC
exponential ansatz becomes an eigenfunction of the operator
$\mathbf{\hat{K}}$ with the eigenvalue $\mathbf{0}$ only if all
coefficients $t_{\mathbf{k}_{i}}^{\mathbf{k}_{a}}$ are zero and the
restriction
\[
  \mathbf{k}_{a} + \mathbf{k}_{b} = \mathbf{k}_{i} + \mathbf{k}_{j},
\]
is fulfilled. This implies that the operator $\hat{T}_{1}$ is zero.
In the same way, the contribution of $\hat{T}_{1}$ vanishes in
coupled-cluster calculations for the three-dimensional electron gas
\cite{bishop1}.

\begin{figure} 
  \centering
  \includegraphics[scale=1.0]{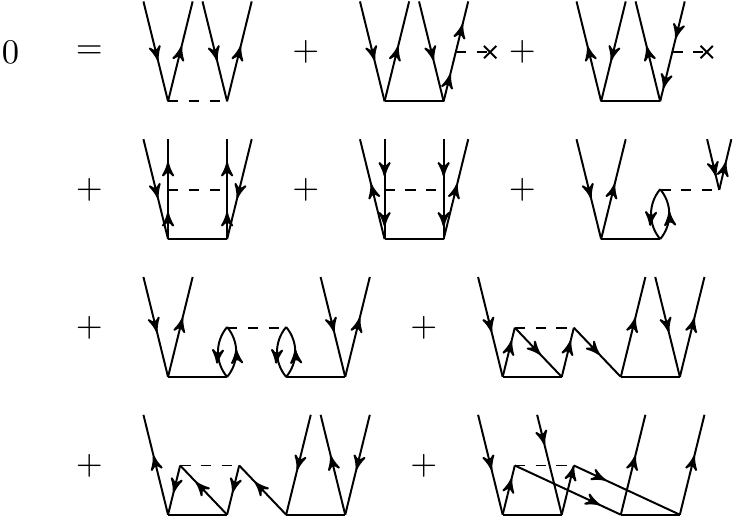}
  \caption{Diagrammatic representation of the $\hat{T}_{2}$ amplitude
    equation in the coupled-cluster doubles approximation. The coupled
    particle-particle and hole-hole ladder equations (PPHH-LAD) are
    obtained by leaving out the sixth diagram, which has summation
    over one particle and one hole state, and all the non-linear
    diagrams, that is, the four last diagrams. The particle-particle
    ladder equations (PP-LAD) is otherwise equal to the PPHH-LAD
    equations, apart from that in the former case also the fifth
    diagram is neglected. The two-particle interaction is illustrated
    by a dashed line and the $t$-amplitude by a solid line. The dashed
    line with a cross at one vertex represents the Fock
    operator. Lines with arrows pointing upwards represent particles
    whereas lines with arrows pointing downwards represent holes. The
    interaction and $t$-amplitude are assumed to be
    antisymmetric.} \label{fig:ampl_cc}
\end{figure}

The CCD $\hat{T}_{2}$ amplitude equation can be written in laboratory
momentum coordinates as, see for example Ref.~\cite{crawford},
\begin{eqnarray} \label{eq:t2ampl}
  0 &=&
  \braket{\mathbf{k}_{a}\mathbf{k}_{b}}{v|\mathbf{k}_{i}\mathbf{k}_{j}}
  \nonumber \\ 
  & + & P(\mathbf{k}_{a}\mathbf{k}_{b})
  \sum_{\mathbf{k}_{c}}\braket{\mathbf{k}_{b}}{f|\mathbf{k}_{c}}\braket{\mathbf{k}_{a}\mathbf{k}_{c}}{t|\mathbf{k}_{i}\mathbf{k}_{j}}
  \nonumber \\ 
  & - & 
  P(\mathbf{k}_{i}\mathbf{k}_{j})\sum_{\mathbf{k}_{k}}\braket{\mathbf{k}_{a}\mathbf{k}_{b}}{t|\mathbf{k}_{i}\mathbf{k}_{k}}\braket{\mathbf{k}_{k}}{f|\mathbf{k}_{j}}
  \nonumber \\ 
  & + & 
  \frac{1}{2}\sum_{\mathbf{k}_{c}\mathbf{k}_{d}}\braket{\mathbf{k}_{a}\mathbf{k}_{b}}{v|\mathbf{k}_{c}\mathbf{k}_{d}}\braket{\mathbf{k}_{c}\mathbf{k}_{d}}{t|\mathbf{k}_{i}\mathbf{k}_{j}} 
  \nonumber \\ 
  & + & 
  \frac{1}{2}\sum_{\mathbf{k}_{k}\mathbf{k}_{l}}\braket{\mathbf{k}_{a}\mathbf{k}_{b}}{t|\mathbf{k}_{k}\mathbf{k}_{l}}\braket{\mathbf{k}_{k}\mathbf{k}_{l}}{v|\mathbf{k}_{i}\mathbf{k}_{j}}
  \nonumber \\ 
  & + & P(\mathbf{k}_{i}\mathbf{k}_{j})P(\mathbf{k}_{a}\mathbf{k}_{b})\sum_{\mathbf{k}_{k}\mathbf{k}_{c}}\braket{\mathbf{k}_{a}\mathbf{k}_{c}}{t|\mathbf{k}_{i}\mathbf{k}_{k}}
  \braket{\mathbf{k}_{k}\mathbf{k}_{b}}{v|\mathbf{k}_{c}\mathbf{k}_{j}} 
  \nonumber  \\ 
  & + & 
  \frac{1}{2}P(\mathbf{k}_{i}\mathbf{k}_{j})P(\mathbf{k}_{a}\mathbf{k}_{b})\sum_{\mathbf{k}_{k}\mathbf{k}_{l}\mathbf{k}_{c}\mathbf{k}_{d}}\braket{\mathbf{k}_{k}\mathbf{k}_{l}}{v|\mathbf{k}_{c}\mathbf{k}_{d}}
  \nonumber \\ 
  & \times & 
  \braket{\mathbf{k}_{a}\mathbf{k}_{c}}{t|\mathbf{k}_{i}\mathbf{k}_{k}}\braket{\mathbf{k}_{d}\mathbf{k}_{b}}{t|\mathbf{k}_{l}\mathbf{k}_{j}}
  \nonumber \\
 & - & P(\mathbf{k}_{i}\mathbf{k}_{j})\frac{1}{2}\sum_{\mathbf{k}_{k}\mathbf{k}_{l}\mathbf{k}_{c}\mathbf{k}_{d}}\braket{\mathbf{k}_{k}\mathbf{k}_{l}}{v|\mathbf{k}_{c}\mathbf{k}_{d}}
  \nonumber \\ 
  & \times & 
  \braket{\mathbf{k}_{a}\mathbf{k}_{b}}{t|\mathbf{k}_{i}\mathbf{k}_{k}}\braket{\mathbf{k}_{c}\mathbf{k}_{d}}{t|\mathbf{k}_{j}\mathbf{k}_{l}}
  \nonumber \\
  & - & 
  P(\mathbf{k}_{a}\mathbf{k}_{b})\frac{1}{2}\sum_{\mathbf{k}_{k}\mathbf{k}_{l}\mathbf{k}_{c}\mathbf{k}_{d}}\braket{\mathbf{k}_{k}\mathbf{k}_{l}}{v|\mathbf{k}_{c}\mathbf{k}_{d}}
  \nonumber \\ 
  & \times & 
  \braket{\mathbf{k}_{a}\mathbf{k}_{c}}{t|\mathbf{k}_{i}\mathbf{k}_{j}}\braket{\mathbf{k}_{b}\mathbf{d}}{t|\mathbf{k}_{k}\mathbf{k}_{l}}
  \nonumber \\ 
  & + & 
  \frac{1}{4}\sum_{\mathbf{k}_{k}\mathbf{k}_{l}\mathbf{k}_{c}\mathbf{k}_{d}}\braket{\mathbf{k}_{k}\mathbf{k}_{l}}{v|\mathbf{k}_{c}\mathbf{k}_{d}}
  \nonumber \\ 
  & \times & 
  \braket{\mathbf{k}_{c}\mathbf{k}_{d}}{t|\mathbf{k}_{i}\mathbf{k}_{j}}\braket{\mathbf{k}_{a}\mathbf{k}_{b}}{t|\mathbf{k}_{k}\mathbf{k}_{l}},
\end{eqnarray}
  where all two-body matrix elements are antisymmetrized and $P(pq)$
  is a permutation operator that operates on a general function
  $y(p,q)$ according to
  \[
    P(pq)y(p,q) = y(p,q) - y(q,p).
  \]
The Fock operator is defined by
  \[
    \braket{\mathbf{k}_{p}}{f|\mathbf{k}_{q}} =
    \braket{\mathbf{k}_{p}}{h_{0}|\mathbf{k}_{q}} +
    \sum_{i}\braket{\mathbf{k}_{p}\mathbf{k}_{i}}{v|\mathbf{k}_{q}\mathbf{k}_{i}},
  \]
  where the sinlge-particle kinetic energy operator $h_{0}$ is
  $k^{2}/(2M)$ in momentum space. From the fact that the two-particle
  interaction conserves the total momentum, it follows that the Fock
  operator is diagonal in momentum basis. This also means that the
  plane wave basis is a Hartree-Fock basis for infinite nuclear matter
  and, as is well known, the Hartree-Fock energy for nuclear matter is
  simply the same as the ground state energy in first order many-body
  perturbation theory (MBPT(1)) \cite{mackenzie}. The CC $\hat{T}_{2}$
  amplitude equation from Eq.~(\ref{eq:t2ampl}) is given in diagrammatic
  representation in Fig.~\ref{fig:ampl_cc}. We use diagrammatic rules
  as defined in Ref.~\cite{crawford}.

\begin{figure}
  \centering
  \includegraphics[scale=1.0]{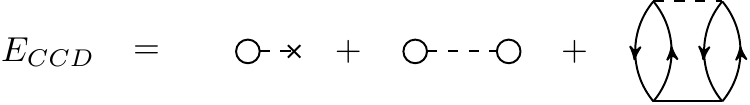}
  \caption{The CCD energy equation, given in terms of diagrams. The
    two-particle interaction is given by a dashed line and the
    $t$-amplitude by a solid line. The dashed line with a cross at one
    vertex represents the kinetic energy operator. Lines with arrows
    pointing upwards represent particles, lines with arrows pointing
    downwards represent holes, and circles are always hole
    lines.} \label{fig:ene_cc}
\end{figure}

The particle-particle and hole-hole ladder approximation (PPHH-LAD) is
obtained by leaving out from the CCD amplitude equation all nonlinear
terms, as well as the linear term with summation over one particle and
one hole index. The coupled ladder equations are
\begin{align} \label{eq:pphhlad}
  0 &=
  \braket{\mathbf{k}_{a}\mathbf{k}_{b}}{v|\mathbf{k}_{i}\mathbf{k}_{j}}
  \nonumber \\ & +
  P(\mathbf{k}_{a}\mathbf{k}_{b})\sum_{\mathbf{k}_{c}}\braket{\mathbf{k}_{b}}{f|\mathbf{k}_{c}}\braket{\mathbf{k}_{a}\mathbf{k}_{c}}{t|\mathbf{k}_{i}\mathbf{k}_{j}}
  \nonumber \\ & -
  P(\mathbf{k}_{i}\mathbf{k}_{j})\sum_{\mathbf{k}_{k}}\braket{\mathbf{k}_{a}\mathbf{k}_{b}}{t|\mathbf{k}_{i}\mathbf{k}_{k}}\braket{\mathbf{k}_{k}}{f|\mathbf{k}_{j}}
  \nonumber \\ & +
  \frac{1}{2}\sum_{\mathbf{k}_{c}\mathbf{k}_{d}}\braket{\mathbf{k}_{a}\mathbf{k}_{b}}{v|\mathbf{k}_{c}\mathbf{k}_{d}}\braket{\mathbf{k}_{c}\mathbf{k}_{d}}{t|\mathbf{k}_{i}\mathbf{k}_{j}}
  \nonumber \\ & +
  \frac{1}{2}\sum_{\mathbf{k}_{k}\mathbf{k}_{l}}\braket{\mathbf{k}_{a}\mathbf{k}_{b}}{t|\mathbf{k}_{k}\mathbf{k}_{l}}\braket{\mathbf{k}_{k}\mathbf{k}_{l}}{v|\mathbf{k}_{i}\mathbf{k}_{j}}.
\end{align}
We define the particle-particle ladder equations (PP-LAD) as
Eq.~(\ref{eq:pphhlad}) where the last line, the hole-hole ladder
diagram, is removed. In the PPHH-LAD approximation, the five first
diagrams of Fig.~\ref{fig:ampl_cc} are retained. The PP-LAD
approximation uses only the four first diagrams in the same figure.
  
  The coupled-cluster energy equation (\ref{eq:ccene}) becomes in the
  CCD approximation
  \[
    E_{CCD} = E_{REF} + \Delta E_{CCD},
  \]
  where the reference energy is written as
  \begin{equation} \label{eq:eneref}
    E_{REF} =
    \sum_{\mathbf{k}_{i}}\braket{\mathbf{k}_{i}}{h_{0}|\mathbf{k}_{i}}
    +
    \frac{1}{2}\sum_{\mathbf{k}_{i}\mathbf{k}_{j}}\braket{\mathbf{k}_{i}\mathbf{k}_{j}}{v|\mathbf{k}_{i}\mathbf{k}_{j}},
  \end{equation}
  and the correlation correction term is simply
  \begin{equation} \label{eq:eneccd}
    \Delta E_{CCD} =
    \frac{1}{4}\sum_{\mathbf{k}_{a}\mathbf{k}_{b}\mathbf{k}_{i}\mathbf{k}_{j}}\braket{\mathbf{k}_{i}\mathbf{k}_{j}}{v|\mathbf{k}_{a}\mathbf{k}_{b}}\braket{\mathbf{k}_{a}\mathbf{k}_{b}}{t|\mathbf{k}_{i}\mathbf{k}_{j}}.
  \end{equation}
  The general expressions for the CC energy are derived in for example
  Ref.~\cite{crawford}. A diagrammatic representation of the energy
  equation is given in Fig.~\ref{fig:ene_cc}. The correlation energy
  has more terms in a general case when the $\hat{T}_{1}$ amplitude
  does not vanish.

  \begin{table}
    \begin{tabular}{p{8cm}}

    ALGORITHM I. Fixed-point iteration scheme for solving the ladder
    equations, as explained in Ref.~\cite{crawford}. The amplitude matrix
    $T$ and the function $z$ are defined in the text. \\ \hline\hline
    \parbox[c]{8cm}{
      \begin{enumerate}
      \item Initialize $E_{\text{old}}$ to a large number.
      \item Initialize the amplitude matrix $T_{\text{old}}$ to zero.
      \item Loop until convergence:
        \begin{enumerate}
        \item \label{22} Calculate $T_{\text{new}} = z(T_{\text{old}})$. 
        \item Calculate a new binding energy \\ $E_{\text{new}} =
           \Delta E_{CCD}\left( T_{\text{new}}\right)$.
           \\ If
           $|E_{\text{new}}-E_{\text{old}}| $ is smaller than a given
           tolerance, stop. \\ Else, set $E_{\text{old}} =
           E_{\text{new}}$ and $T_{\text{old}} = T_{\text{new}}$
           \\ and return to \ref{22}.
        \end{enumerate}
      \end{enumerate}
    } \\ \hline\hline

    \end{tabular}
  \end{table}

  Let us define $T$ as the amplitude matrix, with the matrix elements  
  \[
  [T]_{\alpha , \beta } = \braket{\mathbf{k}_{p(\alpha )}\mathbf{k}_{q(\alpha )}}{t|\mathbf{k}_{r(\beta )}\mathbf{k}_{s(\beta )}},  
  \]
  where $p$, $q$, $r$, and $s$ are functions of the two-body 
  configurations $\alpha $ and $\beta $.
  As explained in Ref.~\cite{crawford}, the ladder equations can be
  written as the more convenient matrix equation 
  \begin{equation} \label{eq:fixedp}
    T = z(T),
  \end{equation} 
  where the left hand side consists of only an amplitude matrix and
  the rest of the ladder equations, here written as the function
  $z$ of the amplitude matrix $T$, is on the right hand side.
  Utilizing the representation of Eq.~(\ref{eq:fixedp}), 
  the amplitude equation can be solved by a fixed-point iteration
  scheme. Algorithm I is commonly used in CC calculations
  \cite{crawford}, and this is the procedure we have employed.
  
  \subsection{Brueckner-Hartree-Fock approximation}
  In Brueckner-Hartree-Fock (BHF) theory
  \cite{brueckner_gammel,haftel_tabakin}, the total energy is
  approximated by
  \[
    E_{BHF} =
    \sum_{\mathbf{k}_{i}}\braket{\mathbf{k}_{i}}{h_{0}|\mathbf{k}_{i}}
    +
    \frac{1}{2}\sum_{\mathbf{k}_{i}\mathbf{k}_{j}}\braket{\mathbf{k}_{i}\mathbf{k}_{j}}{g|\mathbf{k}_{i}\mathbf{k}_{j}},
  \]
  where the $G$-matrix is defined as
  \begin{align} \label{eq:gmat1}
    &\braket{\mathbf{k}_{p}\mathbf{k}_{q}}{g|\mathbf{k}_{r}\mathbf{k}_{s}}
    =
    \braket{\mathbf{k}_{p}\mathbf{k}_{q}}{v|\mathbf{k}_{r}\mathbf{k}_{s}}
    \nonumber \\ &+
    \frac{1}{2}\sum_{\mathbf{k}_{c}\mathbf{k}_{d}}\frac{\braket{\mathbf{k}_{p}\mathbf{k}_{q}}{v|\mathbf{k}_{c}\mathbf{k}_{d}}\braket{\mathbf{k}_{c}\mathbf{k}_{d}}{g|\mathbf{k}_{r}\mathbf{k}_{s}}}{\vare_{\mathbf{k}_{p}}+\vare_{\mathbf{k}_{q}}-\vare_{\mathbf{k}_{c}}-\vare_{\mathbf{k}_{d}}}.
  \end{align}
  and the single-particle energy is
  \begin{equation} \label{eq:bhf_sp}
    \vare_{\mathbf{k}_{p}} =
    \braket{\mathbf{k}_{p}}{h_{0}|\mathbf{k}_{p}} +
    \sum_{\mathbf{k}_{i}}\braket{\mathbf{k}_{p}\mathbf{k}_{i}}{g|\mathbf{k}_{p}\mathbf{k}_{i}}.
  \end{equation}
  In the so-called continuous option \cite{mahaux1985, mahaux1989},
  which we use, the single-particle energy has the form given in
  Eq.~(\ref{eq:bhf_sp}) for both particle and hole states.
\begin{figure}[!t]
  \centering
  \includegraphics[scale=1.0]{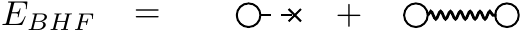}
  \caption{The BHF energy equation in terms of diagrams. The waved
    line represents the $G$-matrix, and a circle means summation over
    hole states.} \label{fig:ene_bhf}
\end{figure}


  \begin{figure}
    \centering
    \includegraphics[scale=1.0]{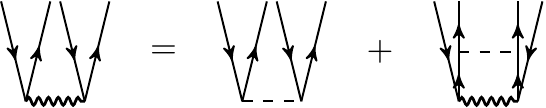}
    \caption{Definition of the $G$-matrix, expressed using
      diagrams. The wavy line represents the $G$-matrix
      interaction. All other parts are defined as in
      Fig.~\ref{fig:ampl_cc}.} \label{fig:gmatrix}
  \end{figure}

  Diagrams of the BHF equations are given in Figs.~\ref{fig:ene_bhf}
  and \ref{fig:gmatrix}. From the diagrammatic expressions one can see
  that the PP-LAD and BHF equations are similar. In fact, one can show
  that the BHF equations become equal to the particle-particle ladder
  equations (PP-LAD), provided that the $G$-matrix in the
  single-particle energy expression (\ref{eq:bhf_sp}) is replaced by
  the interaction matrix. We used this similarity to verify our
  implementation of the PP-LAD equations as well. In the PPHH-LAD
  approximation we have in addition to particle-particle ladders 
  hole-hole ladder contributions, and the two types of ladder diagrams 
  couple to each other.

  \section{Further approximations}\label{sec:approx}

  Explicit expressions of the nuclear interaction are commonly given
  in relative momentum coordinates, whereas the coupled-cluster
  equations are defined in laboratory coordinates. Either the
  interaction may be transformed to laboratory coordinates, or the
  many-body equations must be rewritten to relative coordinates. We
  chose to transform the ladder equations from laboratory to
  relative and center-of-mass (RCM) coordinates. Details of the 
  transformation are shown in App.~\ref{sec:relmom}. 

  In addition to transforming the equations from laboratory to RCM
  coordinates, 
  we write the equations in a basis of coupled angular
  momenta. The basis we use is
  $|k(lS)\mathcal{J}m_{\mathcal{J}}M_{T}\rangle$, where $k$ is the
  radial component of the relative momentum, $l$ is the relative
  orbital angular momentum, $S$ is the total two-particle spin,
  $\mathcal{J}$ is the angular momentum sum $l + S$, $m_{\mathcal{J}}$
  is the $z$ projection of $\mathcal{J}$, and $M_{T}$ is the
  projection of the two-particle isospin. The same representation is
  commonly used in nuclear structure calculations, such as the
  Brueckner-Hartree-Fock method \cite{haftel_tabakin}, but due to
  angular-average approximations the matrix elements are normally
  diagonal in $\mathcal{J}$ and $m_{\mathcal{J}}$.

  As can be seen from Eq.~(\ref{eq:ladrcm}), 
  the ladder equations in RCM coordinates depend on
  the particle-particle Pauli exclusion operator
  \[
    Q_{pp}^{(e)}(\mathbf{k}, \mathbf{K}, k_{F}) = \theta(|\mathbf{k}+\mathbf{K}/2|-k_{F})\theta(|-\mathbf{k}+\mathbf{K}/2|-k_{F})
  \]
  and the hole-hole Pauli exclusion operator
  \[
    Q_{hh}^{(e)}(\mathbf{k}, \mathbf{K}, k_{F})=\theta(k_{F}-|\mathbf{k}+\mathbf{K}/2|)\theta(k_{F}-|-\mathbf{k}+\mathbf{K}/2|),
  \]
  where $k_{F}$ is the Fermi momentum, 
  $\mathbf{k}$ and $\mathbf{K}$ are relative and CM momenta, respectively, 
  defined in Eq.~(\ref{eq:rcm_lab}), and the superscript $(e)$ emphasizes that these 
  are the exact Pauli operators. A common approximation for nuclear 
  matter calculations 
  is to replace the exact Pauli operators by an operator averaged over the
  angle between the relative and CM momentum vectors 
  \cite{brueckner_gammel,haftel_tabakin,ramos_polls}.

  In this paper, we will use a technique introduced by Suzuki
  \emph{et al.}~\cite{suzuki} to expand the exact Pauli 
  operator in partial waves. Using the exact Pauli operator in
  a partial wave expansion, we derive the CC ladder equations. 
  We will also derive the ladder equations using angular-average
  approximations of the Pauli operator. In both cases we will use
  an angular-average approximation of the single-particle energies.
  
  \subsection{Exact Pauli operator} \label{sec:exactpauli}

\begin{figure}
  \centering
  \includegraphics[scale=0.55]{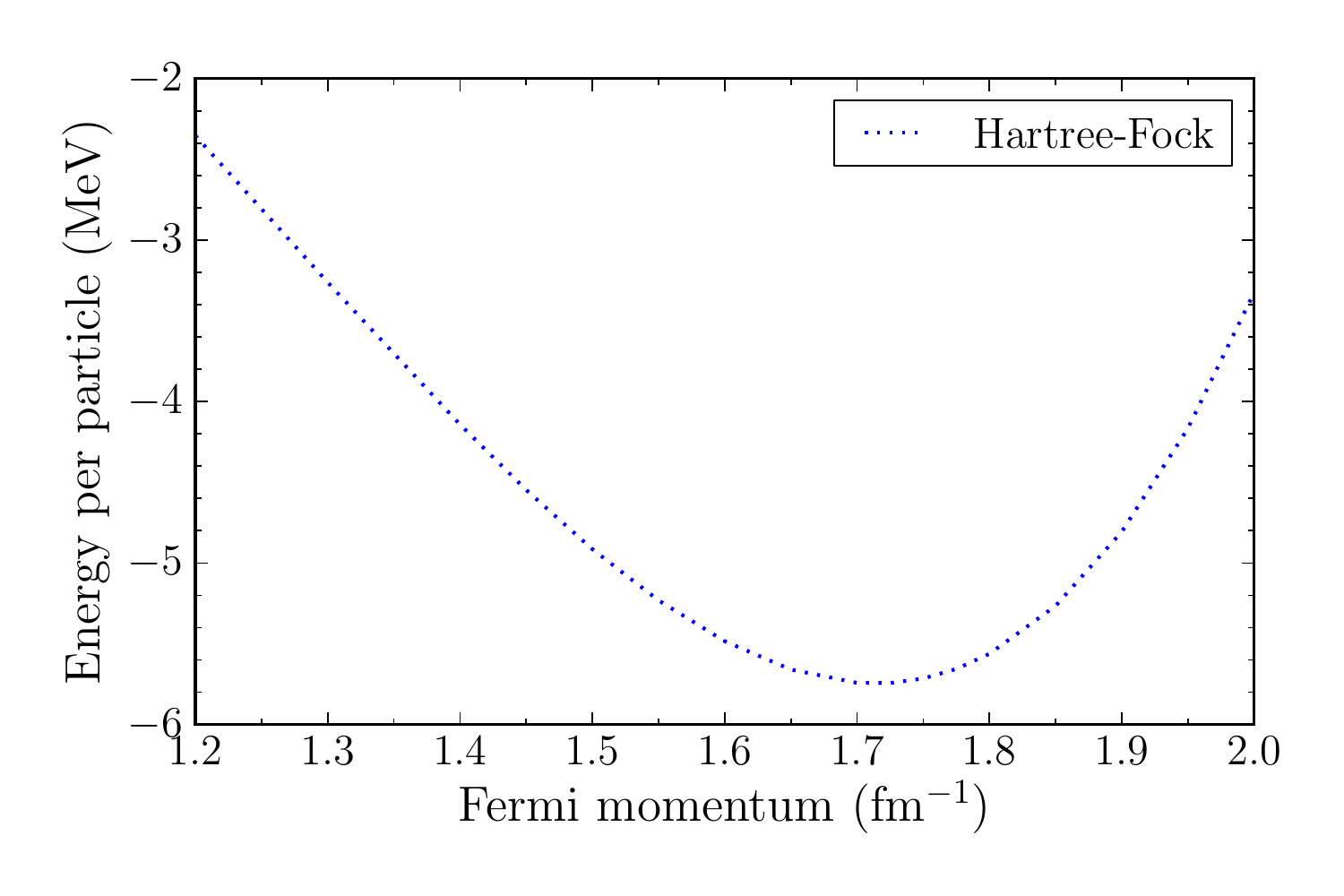}
  \caption{(Color online) Total energy per nucleon of symmetric nuclear 
    matter in the Hartree-Fock approximation, given as a function of 
    Fermi momentum $k_{F}$. The calculation was done with a bare N$^3$LO 
    two-body interaction, and the total angular momentum truncation was 
    set to $\mathcal{J} \leq $ 24.}~\label{fig:hf_ene}
\end{figure}

  Expressed in the coupled partial wave basis, the reference energy per
  particle is
  \begin{align}
    E_{REF}/A = & \frac{3\hbar^{2}k_{F}^{2}}{10m} +
    \frac{3C}{4k_{F}^{3}}\sum_{\mathcal{J}Sl}\sum_{M_{T}}(2\mathcal{J}+1)
    \nonumber \\ & \times \int_{0}^{2k_{F}}dK
    K^{2}\int_{0}^{\sqrt{k_{F}^{2}-K^{2}/4}}dk k^{2} \nonumber \\ &
    \times \braket{k(lS)\mathcal{J}M_{T}}{v|k(lS)\mathcal{J}M_{T}}
    \nonumber \\ & \times x_{hh}(k,K,k_{F}),
    \label{eq:hfenePw}
  \end{align}
  where $A$ is the number of particles, $k_{F}$ is the Fermi 
  momentum, $k$ and $K$ are the radial coordinates of the relative 
  and CM momentum, respectively, and $C$ is 1 for symmetric 
  nuclear matter and 2 for pure neutron
  matter. The function $x_{hh}$ is defined as
\begin{align} \label{eq:xhh}
  x_{hh} = \left\{ \begin{array}{ll} 0, & \text{ if } k >
    \sqrt{k_{F}^{2}-K^{2}/4}, \\ -\frac{k^{2}-k_{F}^{2}+K^{2}/4}{kK}, &
    \text{ if } k_{F}-K/2<k<\sqrt{k_{F}^{2}-K^{2}/4}, \\ 1, & \text{
      otherwise, }
  \end{array} \right.
\end{align}
and similarly we define a function 
\begin{align} \label{eq:xpp}
  x_{pp} = \left\{ \begin{array}{ll} 0, & \text{ if } k <
    \sqrt{k_{F}^{2}-K^{2}/4}, \\ \frac{k^{2}-k_{F}^{2}+K^{2}/4}{kK}, &
    \text{ if } \sqrt{k_{F}^{2}-K^{2}/4}<k<k_{F}+K/2, \\ 1, & \text{
      otherwise. }
  \end{array} \right.
\end{align}
 In Sec.~\ref{sec:approx}, all interaction and $t$-amplitude 
 matrix elements are 
 assumed to be multiplied by the antisymmetrization factor 
 $\mathcal{A}^{l'lSM_{T}}$ given in Eq.~(\ref{eq:antisymm_app}).
Since the Pauli exclusion operator is the only factor in 
the potential energy part of the reference energy 
that depends on the angle between $\mathbf{k}$ and 
$\mathbf{K}$, the expression for the reference energy is 
the same when using exact and angular-averaged Pauli operators.
 As mentioned above, the reference energy is also the Hartree-Fock energy when using the plane wave basis for this particular system. In the limit of an untruncated basis, the reference energy expressed in a partial wave basis, given in Eq. (\ref{eq:hfenePw}), equals the Hartree-Fock energy. Since we calculate the reference energy with a high cutoff in angular momentum, we will refer to the reference energy (\ref{eq:hfenePw}) as the Hartree-Fock energy. However, one should notice that the partial wave basis is not a Hartree-Fock basis for infinite nuclear matter. The reference energy is plotted in Fig.~\ref{fig:hf_ene}.

In the algorithm used to solve the PP-LAD and PPHH-LAD equations, it
is necessary to store all $t$-amplitude matrix elements to be able to
calculate the binding energy at each iteration step. When using a
sufficient number of integration grid points and angular momentum
partial waves to reach necessary accuracy, the size of the CC amplitude
matrix would typically exceed standard memory 
capacities at most high-performance computing facilities.
When the ladder equations are written in the coupled partial wave
basis we used, it is possible to calculate the $t$-amplitude matrix
for only one angular direction of the CM momentum $\mathbf{K}$, and
then obtain the other matrix elements by performing a rotation
\cite{suzuki}. This is a major advantage, since the memory
requirements for storing the $t$-amplitude decreases
significantly. 

The size of the $t$-amplitude matrix can be further decreased by
removing the dependency on the angular parts of the 
relative momenta, that is, $\hat{\mathbf{k}}$ and $\hat{\mathbf{k}}'$. 
As can be seen from Eq.~(\ref{eq:ladrcm}), the only dependency on these
vectors in the ladder amplitude equations that cannot be 
separated out occurs in the single-particle
energies. In RCM coordinates, the single-particle energies are
functions of $|\pm \mathbf{k}+\mathbf{K}/2|$ or 
$|\pm \mathbf{k}'+\mathbf{K}/2|$, as shown in Eq.~(\ref{eq:enedenom}).
To remove the dependency of the $t$-amplitude
matrix on $\hat{\mathbf{k}}$ and $\hat{\mathbf{k}}'$, we use an
angular-average approximation of the arguments $|\pm
\mathbf{k}+\mathbf{K}/2|$ in the single-particle energy
\cite{brueckner_gammel,ramos_phd,ramos_polls}. Since nuclear matter is
an isotropic medium, the single-particle energy must be a symmetric
function, and the single-particle energy can be approximated as a
finite polynomial with only even powers. Following Ramos
\cite{ramos_phd,ramos_polls}, we replace the input momentum  
\[
  k_{p} = |\pm \mathbf{k}+\mathbf{K}/2|,
\]
by the angular-average approximation
\begin{align}
  \overline{k_{p}} = \sqrt{ k^{2} + K^{2}/4 \pm kK\sqrt{\langle
    \cos^{2}\theta_{\mathbf{k}\mathbf{K}}\rangle }},
  \label{eq:k_lab_ave}
\end{align}
where
\[
  \langle \cos^{2}\theta_{\mathbf{k}\mathbf{K}}\rangle =
  \left\{ \begin{array}{ll} x_{hh}^{3}(k,K)/3, & \text{ if } k_{p}
    \leq k_{F}, \\ x_{pp}^{3}(k,K)/3, & \text{ if } k_{p} > k_{F},
  \end{array} \right.
\]
and $\theta_{\mathbf{k}\mathbf{K}}$ is the angle between the 
relative and CM momentum vectors.

If we use the above mentioned rotation of the $t$-amplitude matrix,
given explicitly in Eq.~(\ref{eq:rotation}), and apply the angular-average approximation 
in Eq.~(\ref{eq:k_lab_ave}), we can write the correlation energy per particle as
  \begin{align} \label{eq:ene_exact}
    \Delta E_{CCD}/A &=
    \frac{3C}{32k_{F}^{3}}\sum_{\mathcal{J}m_{\mathcal{J}}}\sum_{\mathcal{J}''m_{\mathcal{J}''}}\sum_{\mathcal{J}'''m_{\mathcal{J}'''}}\sum_{m_{\mathcal{J}'}}\sum_{SM_{T}}\sum_{l
      l'l''l'''} \nonumber \\ &\times
    \int_{0}^{2k_{F}}K^{2}dK\int_{-1}^{1}d\cos \theta_{K} \nonumber
    \\ &\times \int_{0}^{\sqrt{k_{F}^{2}-K^{2}/4}}k^{2}dk
    \int_{\sqrt{k_{F}^{2}-K^{2}/4}}^{\infty }k'^{2}dk' \nonumber
    \\ &\times
    d_{m_{\mathcal{J}''}m_{\mathcal{J}'}}^{\mathcal{J}''}(\theta_{K})d_{m_{\mathcal{J}'''}m_{\mathcal{J}'}}^{\mathcal{J}'''}(\theta_{K})
    \nonumber \\ &\times
    \braket{k(lS)\mathcal{J}M_{T}}{v|k'(l'S)\mathcal{J}M_{T}}
    \nonumber \\ &\times
    \braket{k'(l''S)\mathcal{J}''m_{\mathcal{J}'}M_{T}}{t(K)|k(l'''S)\mathcal{J}'''m_{\mathcal{J}'}M_{T}}
    \nonumber \\ &\times
    Q_{hh}(l'''\mathcal{J}'''m_{\mathcal{J}'''},l\mathcal{J}m_{\mathcal{J}};SM_{T}kK\theta_{K})
    \nonumber \\ &\times
    Q_{pp}(l'\mathcal{J}m_{\mathcal{J}},l''\mathcal{J}''m_{\mathcal{J}''};SM_{T}k'K\theta_{K}),
  \end{align}
  where the Wigner $D$-function has been defined through
  \begin{equation}
    D_{MM'}^{J}(\alpha ,\beta ,\gamma ) = e^{-iM\alpha
    }d_{MM'}^{J}(\beta )e^{-iM'\gamma },
  \end{equation}
  and the function $d_{MM'}^{J}(\beta )$ is given in
  Ref.~\cite{varshalovich}. In Eq.~(\ref{eq:ene_exact}), the
  $t$-amplitude is independent on the angles $\hat{\mathbf{k}}$ and
  $\hat{\mathbf{k}}'$, and we have used the definitions
\begin{align} \label{eq:qhh}
    &
    Q_{hh}(l''\mathcal{J}''m_{\mathcal{J}''},l'''\mathcal{J}m_{\mathcal{J}};SM_{T}kK\theta_{K}\phi_{K})
    = \nonumber \\ & \mathcal{A}^{l''l'''SM_{T}}\frac{1}{2}\sum_{m_{l'''}m_{l''}}\sum_{M_{S}}\int
    d\hat{\mathbf{k}}Y_{l''m_{l''}}^{*}(\hat{\mathbf{k}})Y_{l'''
      m_{l'''}}(\hat{\mathbf{k}}) \nonumber \\ & \times \braket{l'''
      m_{l'''}SM_{S}}{\mathcal{J}m_{\mathcal{J}}}\braket{l''m_{l''}SM_{S}}{\mathcal{J}''m_{\mathcal{J''}}}
    \nonumber \\ & \times
    \theta(k_{F}-|\mathbf{k}+\mathbf{K}/2|)\theta(k_{F}-|-\mathbf{k}+\mathbf{K}/2|)
  \end{align}
  and
  \begin{align} \label{eq:qpp}
   &
    Q_{pp}(l''\mathcal{J}'m_{\mathcal{J}'},l'''\mathcal{J}''m_{\mathcal{J}''};SM_{T}k'K\theta_{K}\phi_{K})
    = \nonumber \\ & \mathcal{A}^{l''l'''SM_{T}}\frac{1}{2}\sum_{m_{l''}m_{l'''}}\sum_{M_{S}'}\int
    d\hat{\mathbf{k}}'
    Y_{l''m_{l''}}^{*}(\hat{\mathbf{k}}')Y_{l'''m_{l'''}}(\hat{\mathbf{k}}')
    \nonumber \\ & \times
    \braket{l''m_{l''}SM_{S}'}{\mathcal{J}'m_{\mathcal{J}'}}\braket{l'''m_{l'''}SM_{S}'}{\mathcal{J}''m_{\mathcal{J}''}}
    \nonumber \\ & \times
    \theta(|\mathbf{k}'+\mathbf{K}/2|-k_{F})\theta(|-\mathbf{k}'+\mathbf{K}/2|-k_{F})
  \end{align}
  in a similar way as Suzuki \emph{et al.}~did in their BHF study
  \cite{suzuki}. In Eqs.~(\ref{eq:qhh}) and (\ref{eq:qpp}), the brackets
  denote Clebsch-Gordan coefficients and the functions 
  $Y_{lm_{l}}(\hat{\mathbf{k}})$ are spherical harmonics \cite{varshalovich}.

In Ref.~\cite{suzuki}, the authors have derived an expression for $Q_{pp}$ which
avoids the complicated integration limits over the space angle of
$\mathbf{k}$ \cite{suzuki}, namely
\begin{align} 
  &
  Q_{pp}(l,\mathcal{J},m_{\mathcal{J}},l'\mathcal{J}'m_{\mathcal{J}'};SM_{T}kK\theta_{K}\phi_{K})
  = \nonumber \\ &
  \mathcal{A}^{ll'SM_{T}}\frac{1}{2}\bigg[ x_{pp}\delta_{ll'}\delta_{\mathcal{J}\mathcal{J}'}\delta_{m_{\mathcal{J}}m_{\mathcal{J}'}} \nonumber \\ &
  + \sum_{L>0; L = \text{even}}(-1)^{S+m_{\mathcal{J}}} \nonumber \\ &
  \times \frac{(4\pi
    )^{1/2}\hat{l}\hat{l'}\hat{\mathcal{J}}\hat{\mathcal{J}'}}{\hat{L}}\braket{l0l'0}{L0}\braket{\mathcal{J}-m_{\mathcal{J}}\mathcal{J}'m_{\mathcal{J}'}}{LM_{L}}
  \nonumber \\ & \times
  Y_{LM_{L}}(\theta_{K},\phi_{K})W(l\mathcal{J}l'\mathcal{J}';SL)
  \nonumber \\ & \times \left[ P_{L+1}(x_{pp})-P_{L-1}(x_{pp})\right] \bigg]
  ,
  \label{eq:qpp_simple}
\end{align}
where $\hat{x} \equiv \sqrt{2x+1}$, $P_{L}(x)$ is the Legendre polynomial,
$W(l\mathcal{J}l'\mathcal{J}';SL)$ denotes the Racah coefficient
\cite{varshalovich}, and $x_{pp}$ is as defined in Eq.~(\ref{eq:xpp}). 
We have used Eq.~(\ref{eq:qpp_simple}) to evaluate $Q_{pp}$ and 
a similar expression for evaluating $Q_{hh}$. The expression for
$Q_{hh}$ was obtained by replacing $x_{pp}$ by the 
equivalent for hole-hole states, 
i.e.~$x_{hh}$ given in Eq.~(\ref{eq:xhh}). Observe that
the simplified but exact Pauli operator expression 
(\ref{eq:qpp_simple}) and the corresponding expression
for the hole-hole Pauli operator could not have been used
if the $t$-amplitude was not independent on the angular
vectors $\hat{\mathbf{k}}$ and $\hat{\mathbf{k}}'$. 
The reason for this restriction is that the Pauli 
operator expressions, when expanded in partial waves, as defined in
Eqs.~(\ref{eq:qhh}) and (\ref{eq:qpp}), are integral
operators and not just real functions. The effect of
the integral operators $Q_{hh}$ and $Q_{pp}$ can be
seen clearly from the correlation energy expression in 
Eq.~(\ref{eq:enecc_exact1}).
The angular-average approximation of the single-particle
energies therefore simplifies the correlation energy
expression significantly.

Since the exact Pauli operator is not
diagonal in $\mathcal{J}$ and $m_{\mathcal{J}}$, both the $G$-matrix
\cite{suzuki} and the $t$-amplitude matrix are not diagonal in the
total angular momentum $\mathcal{J}$ and its projection
$m_{\mathcal{J}}$. The fact that the amplitude matrix is not diagonal
in total angular momentum makes the ladder equations more
complicated. Writing the PPHH-LAD equations in the basis
$|k(lS)\mathcal{J}m_{\mathcal{J}}M_{T}\rangle$, we get
  \begin{align} 
    &\Delta \tilde{\vare}(k,k',K)\langle
    k'(l'S)\mathcal{J}'m_{\mathcal{J}'}M_{T}|t(K)|k(lS)\mathcal{J}m_{\mathcal{J}}M_{T}\rangle
    \nonumber
    \\ &=\braket{k'(l'S)\mathcal{J}'m_{\mathcal{J}'}M_{T}}{v|k(lS)\mathcal{J}m_{\mathcal{J}}M_{T}}\delta_{\mathcal{J}\mathcal{J}'}\delta_{m_{\mathcal{J}}m_{\mathcal{J}'}}
    \nonumber \\ &+
    \frac{1}{2}\sum_{\mathcal{J}''m_{\mathcal{J}''}}\sum_{l''l'''}\int_{0}^{k_{F}}h^{2}dh
    \nonumber \\ &\times \langle
    k'(l'S)\mathcal{J}'m_{\mathcal{J}'}M_{T}|t(K)|h(l''S)\mathcal{J}''m_{\mathcal{J}''}M_{T}\rangle
    \nonumber \\ &\times
    \braket{h(l'''S)\mathcal{J}m_{\mathcal{J}}M_{T}}{v|k(lS)\mathcal{J}m_{\mathcal{J}}M_{T}}
    \nonumber \\ &\times
    Q_{hh}(l''\mathcal{J}''m_{\mathcal{J}''},l'''\mathcal{J}m_{\mathcal{J}};SM_{T}hK\theta_{K}\phi_{K})
    \nonumber \\ &+
    \frac{1}{2}\sum_{\mathcal{J}''m_{\mathcal{J}''}}\sum_{l''l'''}\int_{0}^{\infty
    }p^{2}dp \nonumber \\ &\times
    \braket{k'(l'S)\mathcal{J}'m_{\mathcal{J}'}M_{T}}{v|p(l''S)\mathcal{J}'m_{\mathcal{J}'}M_{T}}
    \nonumber \\ &\times \langle
    p(l'''S)\mathcal{J}''m_{\mathcal{J}''}M_{T}|t(K)|k(lS)\mathcal{J}m_{\mathcal{J}}M_{T}\rangle
    \nonumber \\ &\times
    Q_{pp}(l''\mathcal{J}'m_{\mathcal{J}'},l'''\mathcal{J}''m_{\mathcal{J}''};SM_{T}pK\theta_{K}\phi_{K}),
    \label{eq:pphhlad_exact}
  \end{align}
  where $\ket{(l S)\mathcal{J} m_{\mathcal{J}}}$ denotes a vector
  where $l$ and $S$ are coupled to $\mathcal{J}$. In Eq.~(\ref{eq:pphhlad_exact}) we have 
  used the angular-averaged energy denominator $\Delta \tilde{\vare}(k,k',K)$, which 
  is defined in Eq.~(\ref{eq:angavdenom}). When using the rotation of the ladder amplitude matrix,
  given in Eq.~(\ref{eq:rotation}), the amplitude matrix needs to be evaluated
  only at a single angular coordinate of the CM momentum. 
  The amplitude equation (\ref{eq:pphhlad_exact}) is therefore given as a function
  of only the radial part of the CM momentum. 

In Eq.~(\ref{eq:qpp_simple}), the restriction that $L$ must be even
ensures that parity is conserved \cite{suzuki}. This follows from the
properties of the first Clebsch-Gordan coefficient in
Eq.~(\ref{eq:qpp_simple}), which vanishes when $(-1)^{l} \neq
(-1)^{l'}$, provided that $L$ is even. Since the operators $Q_{hh}$
and $Q_{pp}$ conserve parity, one can see from
Eq.~(\ref{eq:pphhlad_exact}) that the $t$-amplitude of the ladder
equation also conserves parity. On the contrary, the exact Pauli
operators (both particle-particle and hole-hole ones) do not conserve the total angular momentum
$\mathcal{J}$. Consequently, the $t$-amplitude also becomes
non-diagonal in the total angular momentum. The exact Pauli
operators become diagonal in the projection $m_{\mathcal{J}}$
in the special case when the angular part of the CM momentum 
is zero. The $t$-amplitude matrix elements are therefore also diagonal
in $m_{\mathcal{J}}$ when $\mathbf{K}$ is parallel with the
$z$-axis.

  We will later refer to the approximation with exact Pauli operators
  (both particle-particle and hole-hole ones) and angular-averaged
  single-particle energies as 'exact'. The angular-average
  approximation of the single-particle energies, given in
  Eq.~(\ref{eq:k_lab_ave}), is used in all calculations presented in
  this work, including both the coupled ladder approximations and the
  Brueckner-Hartree-Fock method.

\subsection{Angular-averaged Pauli operators}

The ladder equations can be simplified substantially by doing an
angular-average approximation of the Pauli exclusion operators. The
hole-hole and particle-particle exclusion operators become
\cite{suzuki}, respectively,
\begin{align} \label{eq:q_exact_ave}
  &
  Q_{hh}(l\mathcal{J}m_{\mathcal{J}},l'\mathcal{J}'m_{\mathcal{J}'};SM_{T}hK\theta_{K}\phi_{K})
  \nonumber \\ &\rightarrow \overline{Q}_{hh} \equiv \frac{1}{4\pi
  }\int d\hat{\mathbf{K}}
  Q_{hh}(l\mathcal{J}m_{\mathcal{J}},l'\mathcal{J}'m_{\mathcal{J}'};SM_{T}hK\theta_{K}\phi_{K})
  \nonumber \\ &=
  \mathcal{A}^{llSM_{T}}\frac{1}{2}x_{hh}\delta_{ll'}\delta_{\mathcal{J}\mathcal{J}'}\delta_{m_{\mathcal{J}}m_{\mathcal{J}'}},
\end{align}
and
\begin{align}
  &
  Q_{pp}(l\mathcal{J}m_{\mathcal{J}},l'\mathcal{J}'m_{\mathcal{J}'};SM_{T}pK\theta_{K}\phi_{K})
  \nonumber \\ &\rightarrow \overline{Q}_{pp} \equiv \frac{1}{4\pi
  }\int d\hat{\mathbf{K}}
  Q_{pp}(l\mathcal{J}m_{\mathcal{J}},l'\mathcal{J}'m_{\mathcal{J}'};SM_{T}pK\theta_{K}\phi_{K})
  \nonumber \\ &=
  \mathcal{A}^{llSM_{T}}\frac{1}{2}x_{pp}\delta_{ll'}\delta_{\mathcal{J}\mathcal{J}'}\delta_{m_{\mathcal{J}}m_{\mathcal{J}'}},
\end{align} 
where $x_{hh}$ and $x_{pp}$ are as defined in Eqs.~(\ref{eq:xhh}) and (\ref{eq:xpp}). We note that the angular-averaged Pauli
exclusion operators are diagonal in the total angular momentum
$\mathcal{J}$, in contrast to the exact operators.

  When using the angular-average approximation, the PPHH-LAD equations
  simplify to
  \begin{align}
    &\Delta \tilde{\vare
    }(k,k',K)\braket{k'(l'S)\mathcal{J}M_{T}}{t(K)|k(lS)\mathcal{J}M_{T}}
    \nonumber \\ &=
    \braket{k'(l'S)\mathcal{J}M_{T}}{v|k(lS)\mathcal{J}M_{T}}
    \nonumber \\ &+ \frac{1}{2}\sum_{l''}\int_{0}^{k_{F}}h^{2}dh
    \braket{k'(l'S)\mathcal{J}M_{T}}{t(K)|h(l''S)\mathcal{J}M_{T}}
    \nonumber \\ & \times
    \braket{h(l''S)\mathcal{J}M_{T}}{v|k(lS)\mathcal{J}M_{T}}\overline{Q}_{hh}(h,K)
    \nonumber \\ &+ \frac{1}{2}\sum_{l''}\int_{0}^{\infty }p^{2}dp
    \braket{k'(l'S)\mathcal{J}M_{T}}{v|p(l''S)\mathcal{J}M_{T}}
    \nonumber \\ & \times
    \braket{p(l''S)\mathcal{J}M_{T}}{t(K)|k(lS)\mathcal{J}M_{T}}\overline{Q}_{pp}(p,K),
  \end{align}
  where $\Delta \tilde{\vare }(k,k',K)$ is the energy denominator with
  angular-averaged arguments. From the properties of the
  angular-averaged Pauli operators, it follows that the $t$-amplitude
  is diagonal in $\mathcal{J}$, and independent on $m_{\mathcal{J}}$
  and the CM momentum angles $\theta_{K}$ and $\phi_{K}$. Because of
  these symmetries, the CC amplitude matrix is orders of magnitude
  smaller in the angular-averaged approximation than when using exact
  Pauli exclusion operators. 

  The CCD correlation energy per particle becomes in the angular-averaged
  approximation
  \begin{align}
    \Delta E_{CCD}^{ave}/A &= \frac{3C}{16k_{F}^{3}}
    \sum_{\mathcal{J}}\sum_{SM_{T}}\sum_{l l'}(2\mathcal{J}+1)
    \nonumber \\ &\times
    \int_{0}^{\sqrt{k_{F}^{2}-K^{2}/4}}k^{2}dk\int_{\sqrt{k_{F}^{2}-K^{2}/4}}^{\infty
    }k'^{2}dk' \nonumber \\ &\times \int_{0}^{2k_{F}}K^{2}dK
    \braket{k(lS)\mathcal{J}M_{T}}{v|k'(l'S)\mathcal{J}M_{T}}
    \nonumber \\ & \times
    \braket{k'(l'S)\mathcal{J}M_{T}}{t(K)|k(lS)\mathcal{J}M_{T}}
    \nonumber \\ &\times \overline{Q}_{hh}(k,K)\overline{Q}_{pp}(k',K).
  \end{align}
  The approximation using both
  angular-averaged Pauli operators and angular-averaged arguments in
  the single-particle energies will in the following be referred to as
  'average'.

\section{Results and discussion}\label{sec:results}
In the following we will present results of numerical calculations
using the above mentioned ladder approximations. These approximations are
compared with conventional Brueckner-Hartree-Fock theory. We investigate also
the  role of  angular-averaged Pauli exclusion
operators and compare this with the exact treatment discussed above. 
In addition, we compare results obtained using the
optimized NNLO$_{\text{opt}}$ two-body interaction \cite{ekstrom2013}
with calculations done with the N$^{3}$LO interaction
\cite{n3lo}. The different interaction models and many-body methods
are applied to both symmetric nuclear matter and neutron matter
systems. 

In all our calculations, we have taken into account charge symmetry 
breaking and charge independence breaking of the chiral interactions. 
The BHF calculations were done using continuous single-particle
energies \cite{baldo}, which here means that single-particle energies
for both particles and holes were calculated using
Eq.~(\ref{eq:bhf_sp}). The singularities in the $G$-matrix equation
due to the continuous single-particle energies were avoided by
calculating the principal value of the integral in the $G$-matrix
equation. The $G$-matrix equation (\ref{eq:gmat1}) was solved in a
coupled angular momentum basis
$|k(lS)\mathcal{J}m_{\mathcal{J}}M_{T}\rangle$ using the matrix
inversion method of Haftel and Tabakin \cite{haftel_tabakin}. Unless
stated explicitly otherwise, the BHF calculations have been calculated
with a truncation of the total angular momentum at $\mathcal{J} \leq $
24 in the Born approximation and $\mathcal{J} \leq $ 9 for the full
$G$-matrix. The BHF calculations were done with angular-averaged Pauli
operators, as described in Ref.~\cite{haftel_tabakin}. 

\begin{figure} 
  \centering
  \includegraphics[scale=0.55]{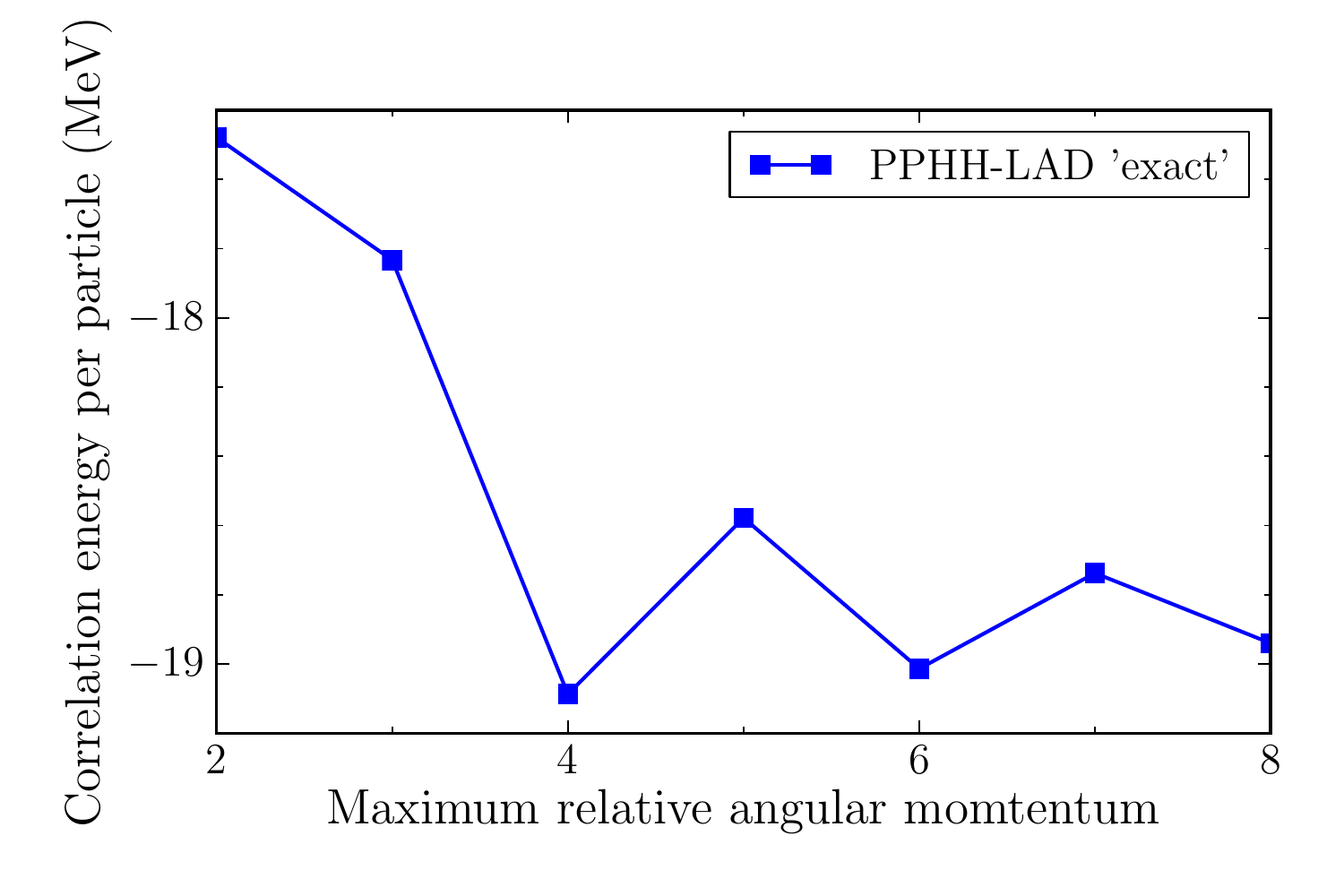}
  \caption{(Color online) Convergence of the correlation energy of 
    symmetric nuclear matter, given as a function of maximum total 
    angular momentum $\mathcal{J}_{max}$. The correlation energy was 
    calculated for the Fermi momentum $k_{F} = $ 1.8 fm$^{-1}$, using 
    the PPHH-LAD approximation and exact Pauli exclusion
    operators.}~\label{fig:conv_exact}

\end{figure}

The coupled ladder equations were solved both with exact and
angular-averaged Pauli exclusion operators. We refer to these two
approximations as 'exact' and 'average', respectively. All
calculations were done with an angular-average approximation of the
single-particle energies, as was explained in
Sec.~\ref{sec:exactpauli}. The Hartree-Fock energy was calculated with a
cutoff in total angular momentum at $\mathcal{J} \leq $ 24. The
correlation energy was calculated with a trunction at $\mathcal{J}
\leq $ 16 in the 'average' approximation and $\mathcal{J} \leq $ 8 in
the 'exact' approximation. Fig.~\ref{fig:conv_exact} shows the
convergence of the correlation energy as a function of the total
angular momentum cutoff $\mathcal{J}_{max}$. Because of the high
density, the angular momentum barrier cannot keep particles far apart
from each other, and therefore the convergence as a function of total
angular momentum is slow in infinite nuclear matter. At high angular
momenta, the interaction is dominated by the one-pion exchange part,
and the convergence behavior as a function of total angular momentum
is therefore similar for different interaction models. Due to
restrictions in computer memory, we were not able to calculate with
$\mathcal{J}_{max}$ higher than 8 in the 'exact' approximation.

\begin{figure}
  \centering
  \includegraphics[scale=0.55]{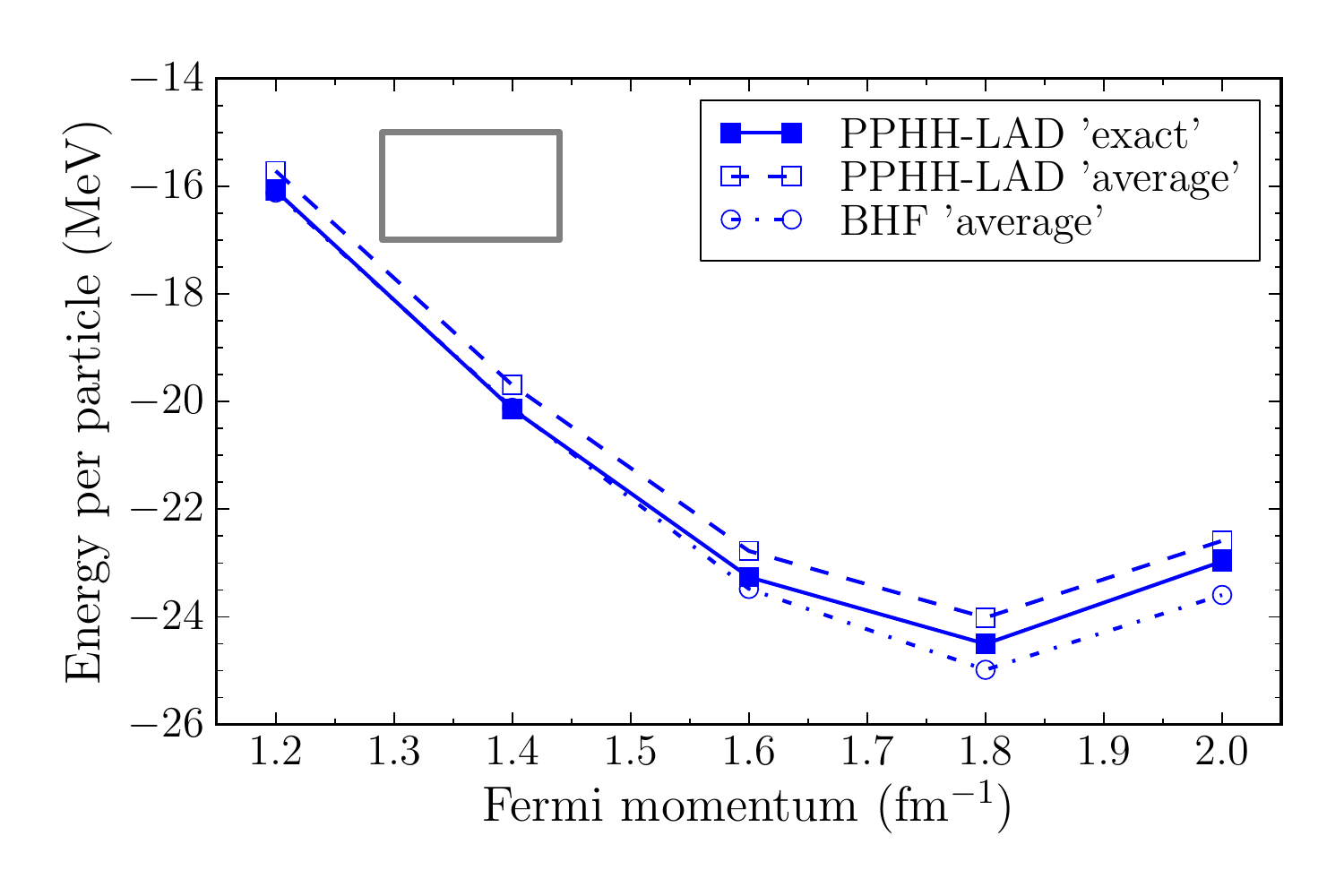}
  \caption{(Color online) Total energy per particle as a function of the Fermi
    momentum, calculated for symmetric nuclear matter. The two
    approximations of the PPHH-LAD equations, 'average' and 'exact',
    are compared with a Brueckner-Hartree-Fock calculation with
    angular-averaged Pauli exclusion operator. These calculations were
    done with the N$^{3}$LO \cite{n3lo} two-body interaction. The box
    denotes the uncertainty region for the experimental saturation
    point of symmetric nuclear matter, as obtained by extrapolating
    from observables of finite nuclei.}~\label{fig:ave_exact}
\end{figure}

  \begin{table}
    \caption{Total energy per nucleon at selected Fermi momenta
      $k_{F}$, as obtained with different approximations. For easier
      comparison, all these energies were calculated with the same
      cutoff in total angular momentum, i.e. $\mathcal{J}_{max} = $ 24
      for the Hartree-Fock / Born approximation and $\mathcal{J}_{max}
      = $ 8 for the correlation contribution. All results were obtained with the N$^{3}$LO interaction
\cite{n3lo}. Energies in units of MeV.}
    \begin{ruledtabular}
    \begin{tabular}{l c c c c c}

    $k_{F}$ (fm$^{-1}$) & 1.2 & 1.4 & 1.6 & 1.8 & 2.0
    \\ \hline PPHH 'exact' & -16.08 & -20.14 & -23.26 & -24.50 & -22.97
    \\ PPHH 'average' & -15.74 & -19.74 & -22.84 & -24.09 & -22.56 \\ PP
    'exact' & -15.76 & -19.83 & -22.98 & -24.27 & -22.82 \\ PP 'average' &
    -15.45 & -19.45 & -22.57 & -23.86 & -22.40 \\ PT2 'average' & -15.11 &
    -19.81 & -23.35 & -24.80 & -23.25 \\ BHF 'average' & -16.18 & -20.25 &
    -23.74 & -25.47 & -24.42 \\ 
    \end{tabular}
    \label{tab:table1}
    \end{ruledtabular}
  \end{table}

\begin{figure}
  \centering
  \includegraphics[scale=0.55]{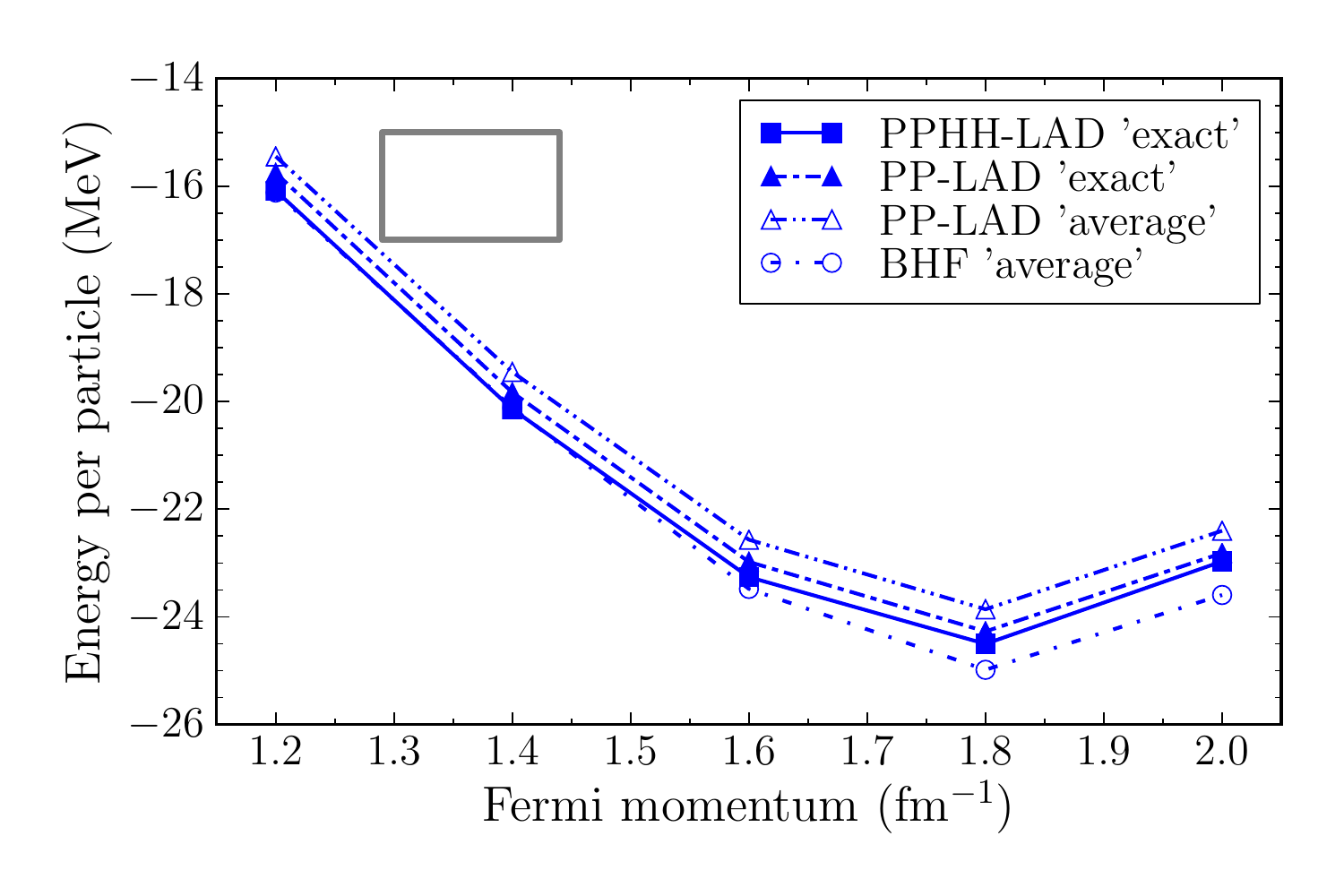}
  \caption{(Color online) The 'exact' PPHH-LAD and PP-LAD calculations 
    of energy per nucleon as a function of Fermi momentum. The equation of state of a Brueckner-Hartree-Fock calculation is also given. The box denotes
    the uncertainty region for the experimental saturation point of
    symmetric nuclear matter. All results were obtained with the N$^{3}$LO interaction
\cite{n3lo}.}~\label{fig:pphh_pp_bhf}
\end{figure}

\begin{figure}
  \centering \includegraphics[scale=0.55]{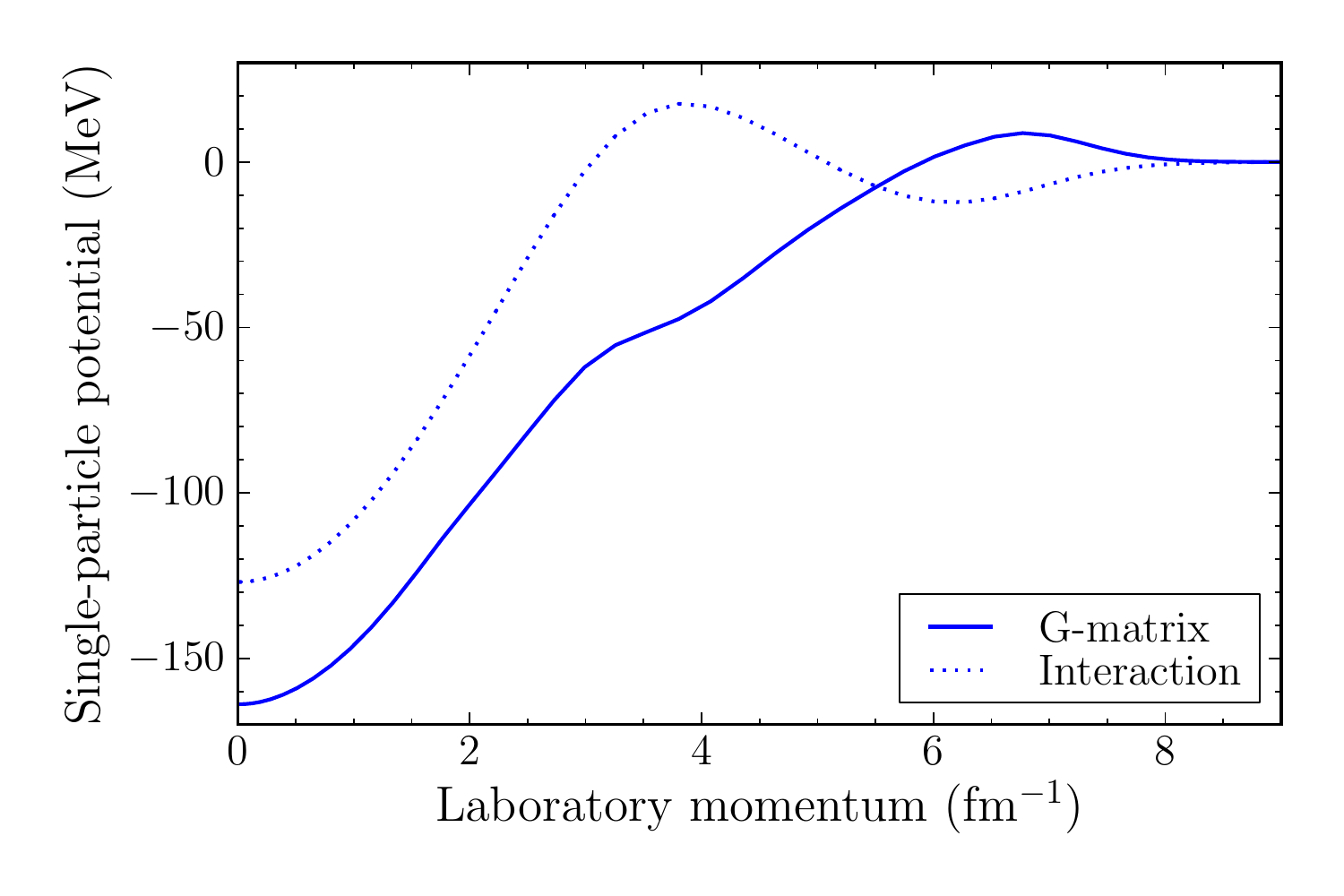}
  \caption{(Color online) Single-particle potential as a function of laboratory frame
    momentum, calculated at Fermi momentum 1.8 fm$^{-1}$ with a
    $G$-matrix and with a bare interaction. In both cases, we used the
    N$^{3}$LO two-body interaction and the total angular momentum cutoff 
    $\mathcal{J}_{\text{max}} = 9$. The only difference between the
    Brueckner-Hartree-Fock and PP-LAD equations is the single-particle
    energy: In BHF theory the single-particle potential is calculated
    using a self-consistent $G$-matrix, whereas in the PP-LAD
    approximation the single-particle energy is obtained by replacing
    the $G$-matrix with a bare interaction.}~\label{fig:sppot}
\end{figure}

Let us first consider symmetric nuclear matter using the N$^{3}$LO 
interaction. In Fig.~\ref{fig:ave_exact}, we compare the energy per 
nucleon as a function of the Fermi momentum for different approximations. 
The Fermi momentum at saturation is equal for all the 
three methods, whereas there are differences in the 
binding energies. The general form of the equation of state is very 
similar for the coupled ladder and BHF methods. As can be seen from 
the figure, the PPHH-LAD approximation 
gives less binding than the BHF approximation both when the PPHH-LAD 
calculation is done with angular-averaged and exact Pauli operators. 
The binding energy at saturation obtained with the BHF method is
approximately 0.5 MeV lower than the corresponding result of Li
\emph{et al.}~\cite{li2006}. There are several factors that may have
contributed to the difference between our BHF results and those of Li
\emph{et al}~\cite{li2006}. For example, in Ref.~\cite{li2006} they
used a complex $G$-matrix, whereas we have used a real $G$-matrix and
treated the singularities by using a principal value integral
\cite{haftel_tabakin}. It is also possible that we have used different
angular-average approximations in the single-particle energies. As
seen from Fig.~\ref{fig:ave_exact}, an exact treatment of the Pauli
operators gives more binding than when using an angular-average
approximation. This is in agreement with the results of Suzuki
\emph{et al.}~\cite{suzuki} and Schiller \emph{et
  al.}~\cite{schiller}, where an exact treatment of the Pauli operator
gave approximately $0.2-0.5$ MeV more binding energy in the BHF
approximation.

Carbone \emph{et al.}~\cite{carbone2013} have compared correlation energies
obtained with the the self-consistent Green's function (SCGF) method 
at finite temperature with BHF calculations, using the 
same N$^{3}$LO two-body interaction as we have used. Similarly as in 
Fig.~\ref{fig:ave_exact}, they got slightly higher energies with 
the SCGF method, which contains both particle-particle and hole-hole 
ladders, compared to the BHF results. In previous studies where the
SCGF method has been compared with the BHF approximation 
\cite{bozek2002,dewulf2003,baldo2012} using other two-body interactions, 
the saturation energies obtained using the SCGF method have been 
located several MeVs higher than the corresponding BHF result, 
and the saturation densities have been shifted towards lower values.
As will be shown systematically in a future publication \cite{polls2013}, 
we find a similar difference between the PPHH-LAD and BHF methods
when using the hard-core Argonne $v_{18}$ interaction \cite{av18}. 
In fact, when using the Argonne $v_{18}$ potential, the saturation 
energy of the PPHH-LAD approximation is found to be only about 
1 MeV below the SCGF saturation energy shown in Fig.~3 of 
Ref.~\cite{baldo2012}, and the saturation density is almost the 
same for both ladder approximations. 
It is interesting to note that we observed a larger difference between
our CC and BHF results in systems with a hard interaction compared to
systems with a soft interaction. If we relate these findings with a
hard-core interaction to those obtained with for example the Bochum CC
truncation scheme \cite{kummel1978} mentioned in the introduction, it
may be possible that the Bochum scheme will give a faster convergence
than the SUB$n$ truncation scheme when using hard-core interaction
models.

Finally, we ought to mention that the way the CC equations are solved
here, and in most other CC applications as well, no self-consistent
solution of the pairing gap equations is performed. In practical terms
this means that we never face instabilities in the denominators of the
CC expansions due to poles arising in the two-hole sector. In CC
theory, in contrast to various SCGF approaches, there is never an
explicit energy dependence in the denominators of the different
amplitudes. The energy differences in the two-particle-two-hole
energies that enter the computation of various denominators are never
zero, by construction.  There are no terms in our present formalism
which thus could account for pairing instabilities in the
denominators, as discussed in for example
Refs.~\cite{vonderfecht1993,dean2003}. The effects of pairing
instabilities and the self-consistent solution of the gap equation
together with the full CCD equations (including the particle-hole
terms as well), await therefore further investigations.


In Table \ref{tab:table1}, we list the total energies for symmetric 
nuclear matter calculated with the N$^{3}$LO two-body interaction. For 
easier comparison, all results are computed with the same cutoffs 
in total angular momentum. We find that the difference between the PPHH-LAD
energies with angular-averaged and exact Pauli operators is approximately
0.4 MeV at the saturation Fermi momentum. This makes a difference of roughly 
$1.7\%$. At the same Fermi momentum, the PPHH-LAD method with exact Pauli 
operators gives approximately  1 MeV more binding than the BHF method with 
angular-averaged Pauli operators. The 1 MeV difference corresponds to 
$3.8\%$ of the total energy.
 
In Fig.~\ref{fig:pphh_pp_bhf} we compare the coupled particle-particle
and hole-hole ladder approximation, PPHH-LAD, with the particle-particle
ladder approximation, PP-LAD. From Fig.~\ref{fig:pphh_pp_bhf} one can see 
that the inclusion of hole-hole ladders gives slightly more binding 
compared to the pure particle-particle ladder approximation. From 
Table \ref{tab:table1} we find that the difference is approximately
0.2 MeV at saturation Fermi momentum, or about $1\%$ of the binding 
energy. At the saturation density, the contribution 
coming from including the hole-hole ladders is smaller than 
the error of an angular-average approximation of the Pauli exclusion 
operators.   

As mentioned earlier, the only difference between the
BHF and PP-LAD approximations is the single-particle energy, which in
the BHF method is calculated with a $G$-matrix and in the ladder 
approximation with a bare interaction. Single-particle potentials 
with a $G$-matrix and with a bare interaction are plotted in 
Fig.~\ref{fig:sppot}, as obtained using the N$^{3}$LO interaction.
\begin{figure}
  \centering
  \includegraphics[scale=0.55]{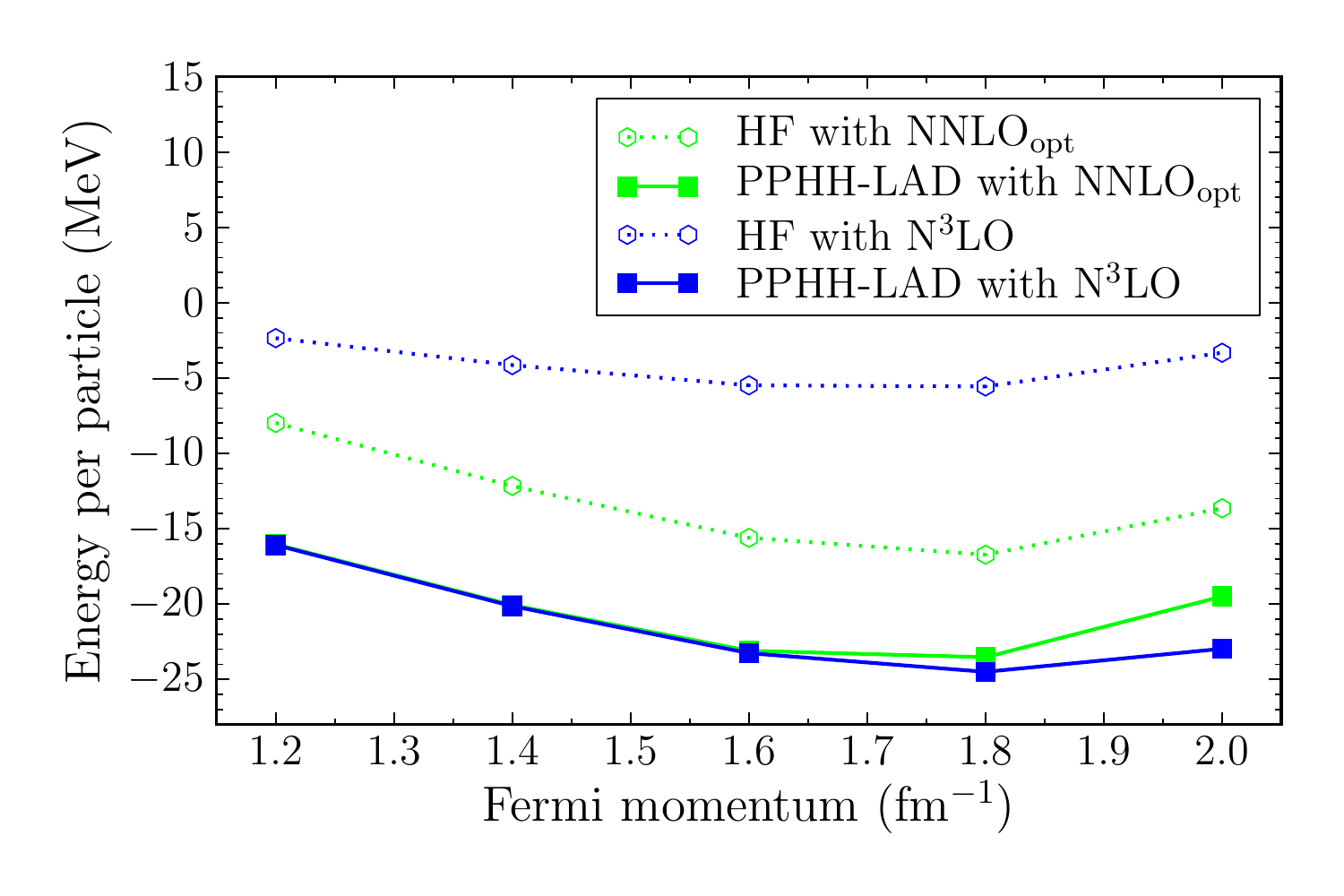}
  \caption{(Color online) Energy per particle for symmetric nuclear matter,
    as calculated in the HF and PPHH-LAD approximations using the
    N$^{3}$LO and NNLO$_{\text{opt}}$ two-body interactions. In the
    PPHH-LAD approximation, the angular momentum cutoff was set to
    $\mathcal{J} \leq 8$ and the calculations were done with exact
    Pauli exclusion operators.} \label{fig:n3lo_n2lo_snm}
\end{figure}
\begin{figure}
  \centering
  \includegraphics[scale=0.55]{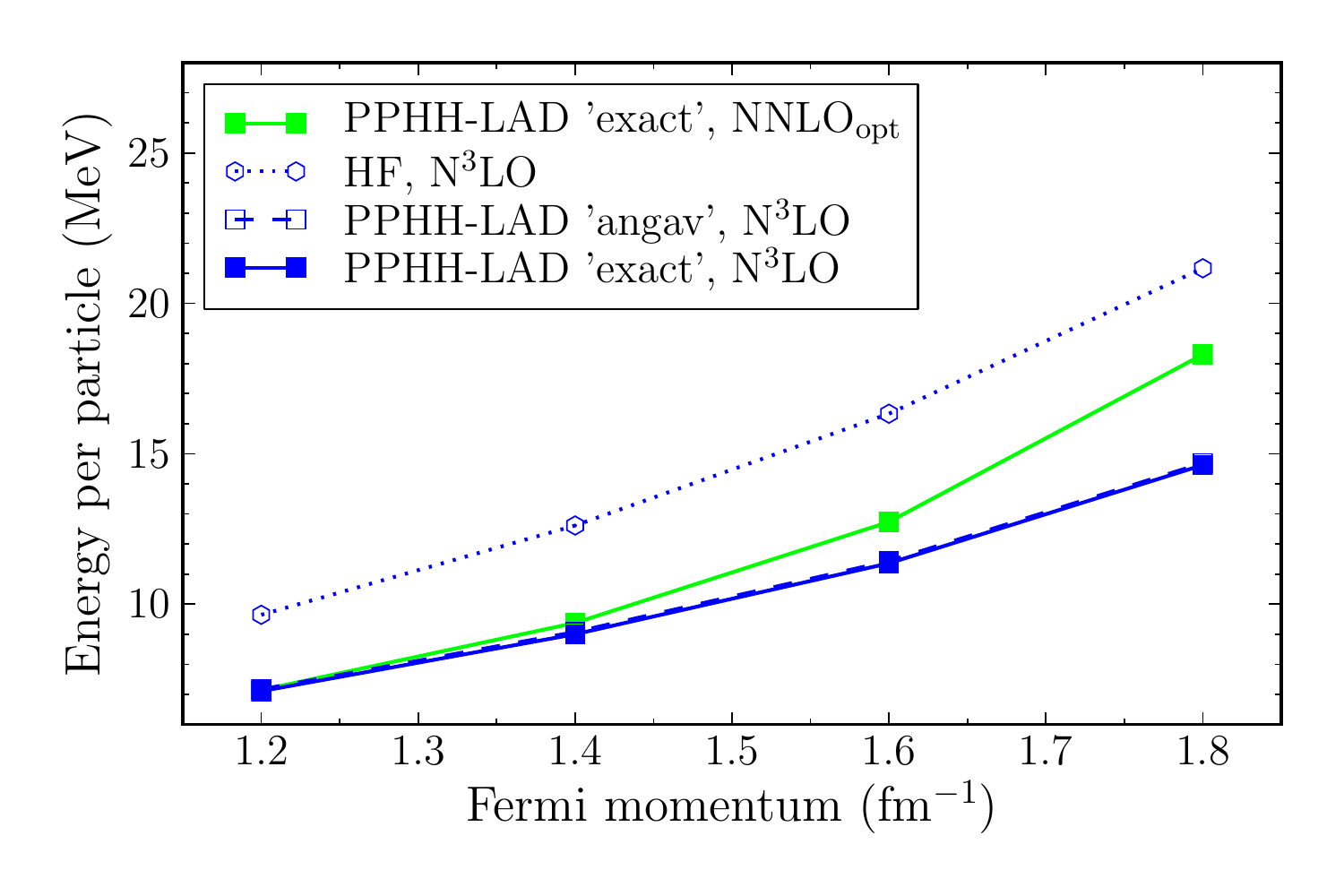}
  \caption{(Color online) Energy per particle for pure neutron matter given as
    a function of Fermi momentum $k_{F}$. The figure shows results of
    calculations in Hartree-Fock (HF) and coupled particle-particle
    and hole-hole ladder approximation
    (PPHH-LAD), using the N$^{3}$LO and NNLO$_{\text{opt}}$ two-body 
    interactions. The neutron matter results obtained with exact 
    Pauli operators have been published in Ref.~\cite{ekstrom2013}.
  } \label{fig:pnm_pphh_hf}
\end{figure}

Next we compare the two different chiral interactions N$^{3}$LO and 
NNLO$_{\text{opt}}$ when applied to infinite matter systems. In 
Fig.~\ref{fig:n3lo_n2lo_snm} we have plotted the equations of state for 
symmetric nuclear matter and in Fig.~\ref{fig:pnm_pphh_hf}  for
pure neutron matter, respectively, as obtained with the two different nuclear 
interaction models. When using the NNLO$_{\text{opt}}$ interaction for
symmetric nuclear matter, we find that the HF energy is much closer to the
PPHH-LAD approximation than is the case with the N$^{3}$LO
interaction. Thus, the optimized next-to-next-to-lowest-order
interaction provides a better starting point for the perturbation
series than the N$^{3}$LO interaction. As can be seen from
Fig.~\ref{fig:n3lo_n2lo_snm}, in the PPHH-LAD approximation the two
different interactions give almost the same equation of state for
Fermi momenta less than 1.6 fm$^{-1}$. At higher Fermi momenta, the
NNLO$_{\text{opt}}$ interaction gives less binding. 
Both the N$^{3}$LO and NNLO$_{\text{opt}}$ interactions overbinds
considerably and saturates at too high density in symmetric nuclear
matter. The similarity between the binding energies obtained with the
two different two-body interactions is in contrast to the results in
finite nuclei \cite{ekstrom2013}, where the NNLO$_{\text{opt}}$
interaction gave significantly better agreement with experiments than
the N$^{3}$LO interaction. However, even if this may indicate that
three-body forces could play a smaller role with the optimized
interaction, there is no clear indication that such correlations are
negligible.  The N$^{3}$LO and NNLO$_{\text{opt}}$ interactions have
also been compared in nuclear matter calculations using the SCGF
method at finite temperature \cite{carbone2013}. The SCGF method was
found to give slightly more binding when using the NNLO$_{\text{opt}}$
interaction compared to calculations with the N$^{3}$LO two-body
interaction.  The results obtained by Carbone {\em et al.}~\cite{carbone2013}, 
using the SCGF method, are rather close to those
obtained with our present CCD calculations.


Equations of state for neutron matter are given in Fig.~\ref{fig:pnm_pphh_hf}.
As can be seen from the figure, the differences between the
calculations with exact and angular-averaged Pauli operators are much
smaller for neutron matter than for symmetric nuclear
matter. According to these results, the angular-average approximation
of the Pauli operators is a fairly good approximation in neutron
matter systems.
In Ref.~\cite{ekstrom2013} we found that the
equation of state for neutron matter with the optimized NNLO
interaction was within the error estimates obtained with an N$^{3}$LO
interaction with three-body forces \cite{tews}, whereas a calculation
with a two-body N$^{3}$LO interaction gave an equation of state 
that was more attractive around the empirical saturation density. Below we
will show that the stronger repulsion seen when using the optimized
NNLO$_{\text{opt}}$ interaction stems from a poorer reproduction of the $^{3}P_0$ and $^{3}P_1$ partial wave phase shifts of the Nijmegen analysis.

\begin{figure}
  \centering
  \includegraphics[scale=0.55]{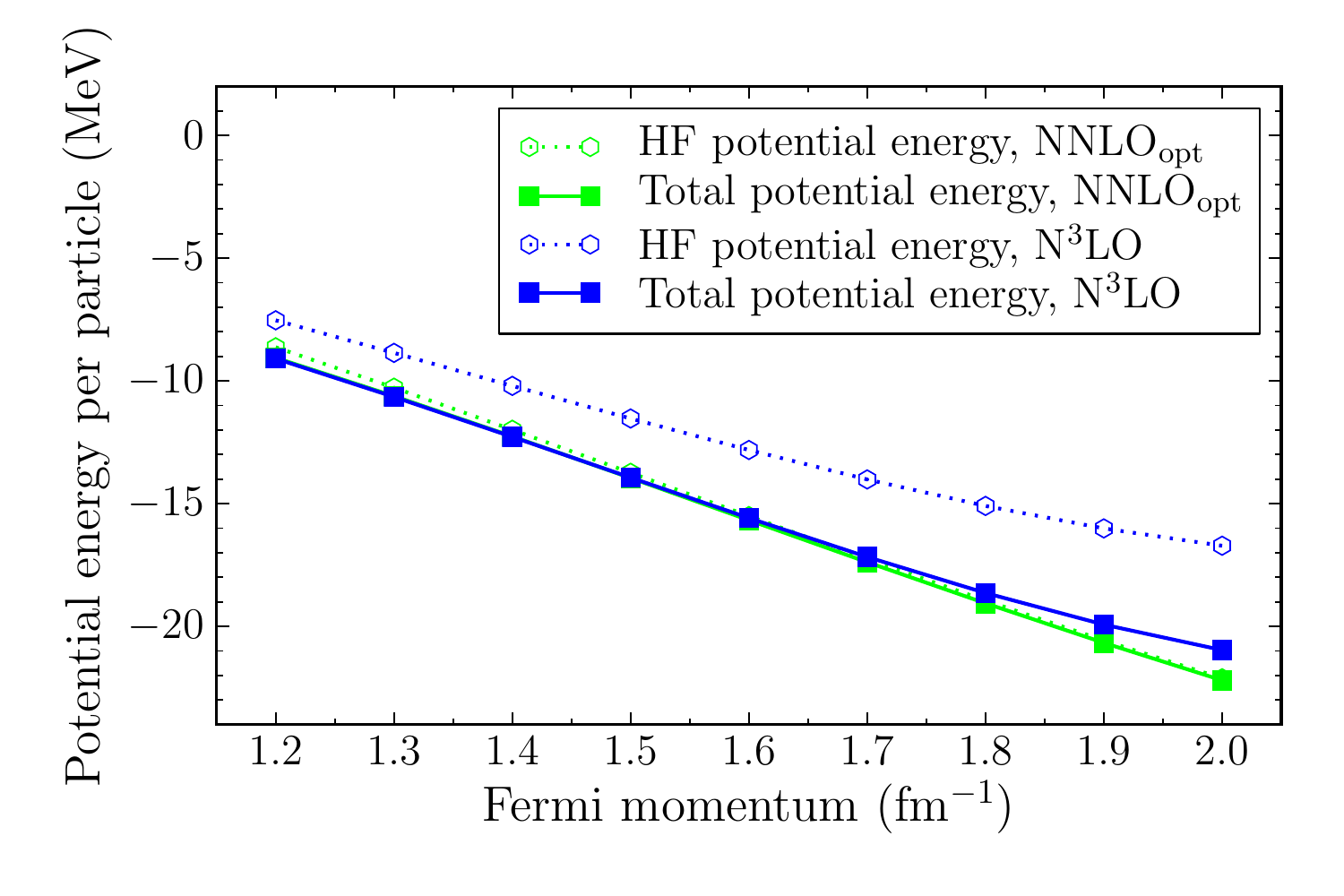}
  \caption{(Color online) Potential energy per particle for the $^{1}S_0$ partial wave 
as  function of the Fermi momentum  $k_F$ for pure neutron matter. 
    We plot the Hartree-Fock potential energy and the total potential energy obtained by adding
    the correlation energies obtained  
    from PPHH-LAD approximation with angular-averaged Pauli
    exclusion operators. Both potential models have been employed.} \label{fig:partialwavespnm1s0}
\end{figure}

\begin{figure}
  \centering
  \includegraphics[scale=0.55]{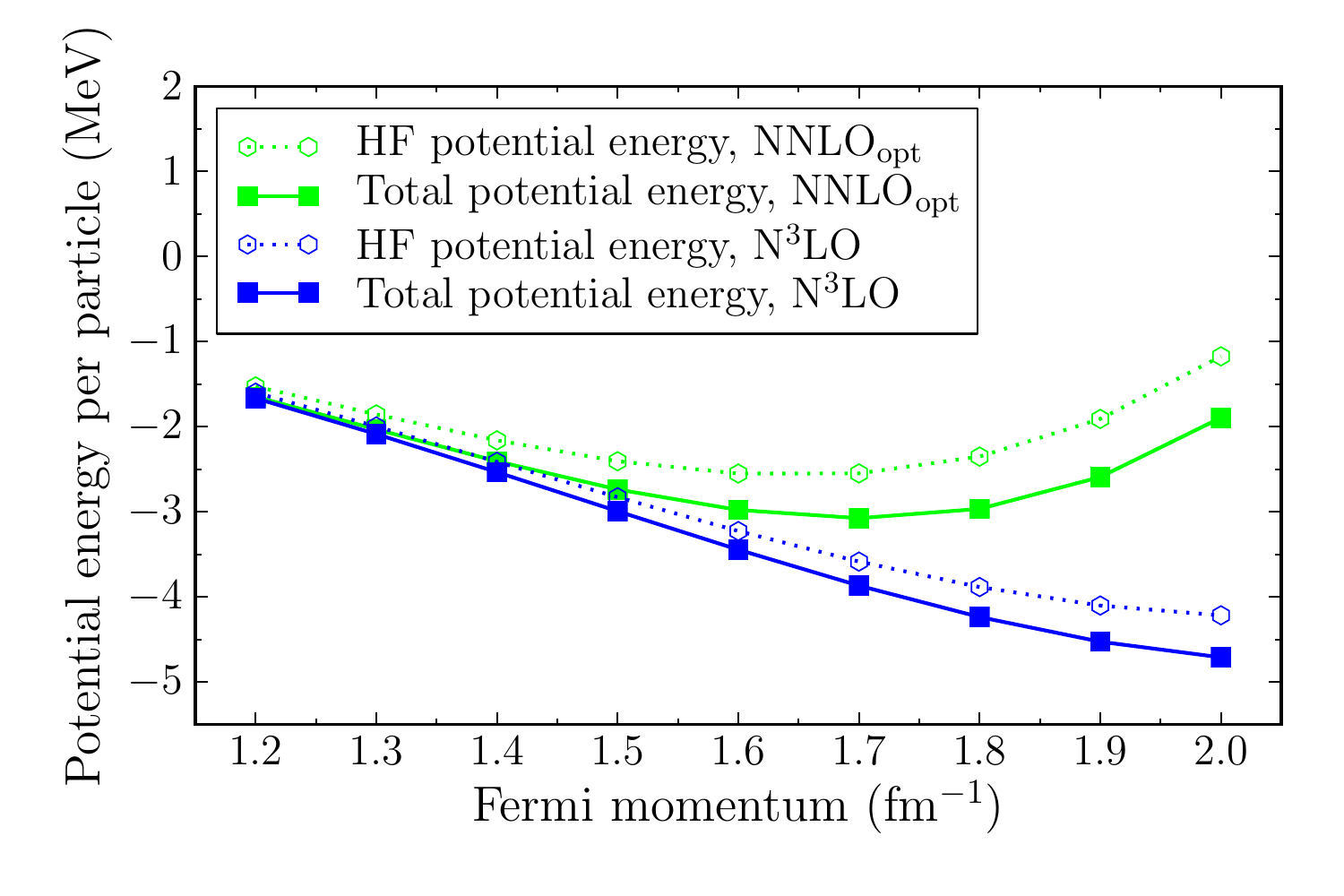}
  \caption{(Color online) Potential energy per particle for the $^{3}P_0$ partial wave 
as  function of the Fermi momentum  $k_F$ for pure neutron matter. 
    We plot the Hartree-Fock potential energy and the total potential energy obtained by adding
    the correlation energies obtained  
    from PPHH-LAD approximation with angular-averaged Pauli
    exclusion operators. Both potential models have been employed.} \label{fig:partialwavespnm3p0}
\end{figure}

The results for symmetric nuclear matter with the two potential models
result in energies that are very similar. This effect is largely due
to the excellent reproduction of various partial waves for the
proton-neutron channel, in particular the $^3S_1$ partial wave
\cite{ekstrom2013b}. However, for pure neutron matter we see a clear
deviation starting at Fermi momenta $k_F=1.4$ fm$^{-1}$. To better
understand this behavior, we have singled out two partial waves,
namely the $^{1}S_0$ and the $^{3}P_0$ partial waves. The results for
the potential energies per particle are shown in
Figs.~\ref{fig:partialwavespnm1s0} and \ref{fig:partialwavespnm3p0}
for the $^{1}S_0$ and the $^{3}P_0$ partial waves, respectively. We
show both the Hartree-Fock potential energy and the total potential
energy by adding the results from the PPHH-LAD correlations. At the
NNLO level of optimization, the $P$-waves show larger deviations from
the phase shifts deduced from the experimental cross sections
\cite{ekstrom2013b}, yielding a poorer agreement compared with the
N$^{3}$LO interaction at lab energies beyond 100 MeV in energy. This
applies in particular to the $^{3}P_0$ and the $^{3}P_1$ partial
waves, resulting in a $^{3}P_0$ wave which is less attractive for the
NNLO optimized interaction model. The contributions from the $^{3}P_1$
partial wave to the equation of state plays a smaller role compared
with the $^{3}P_0$ partial wave.  The differences for the $^{3}P_0$
partial wave is seen rather clearly in
Fig.~\ref{fig:partialwavespnm3p0}. The
N$^{3}$LO interaction results in more binding than the
NNLO$_{\text{opt}}$ interaction model for this particular partial
wave.  This applies both to the Hartree-Fock potential energy and to
the final potential energy that includes correlations.  The
discrepancy that arises from this partial wave is the main reason
behind the more repulsive equation of state obtained with the
NNLO$_{\text{opt}}$ interaction. It will thus be interesting to see
whether an optimization with respect to the experimental cross section
at both the NNLO and N$^{3}$LO levels will bring the results for pure
neutron matter closer to the results obtained with the N$^3$LO
interaction of Ref.~\cite{n3lo}.  These results will be presented in
Ref.~\cite{ekstrom2013b}. It is interesting to note also that the
Hartree-Fock potential energies for the $^{1}S_0$ channel are rather
similar to the fully correlated potential energy with the
NNLO$_{\text{opt}}$ interaction. This is also in line with our
analysis from nuclear structure of Ref.~\cite{ekstrom2013}, indicating
that this interaction is rather soft at the two-body level. Both
interaction models yield negligible differences for the full potential
energy for the $^{1}S_0$ partial wave.  In summary, the poorer
reproduction of the phase shifts for two selected $P$-waves, lead to a
more repulsive equation of state for pure neutron matter with the
newly optimized NNLO$_{\text{opt}}$ interaction. Whether three-body forces or more complicated 
correlations beyond the CCD approximation employed here, will improve the situation, remains however 
to be explored.

\begin{figure}
  \centering
  \includegraphics[scale=0.55]{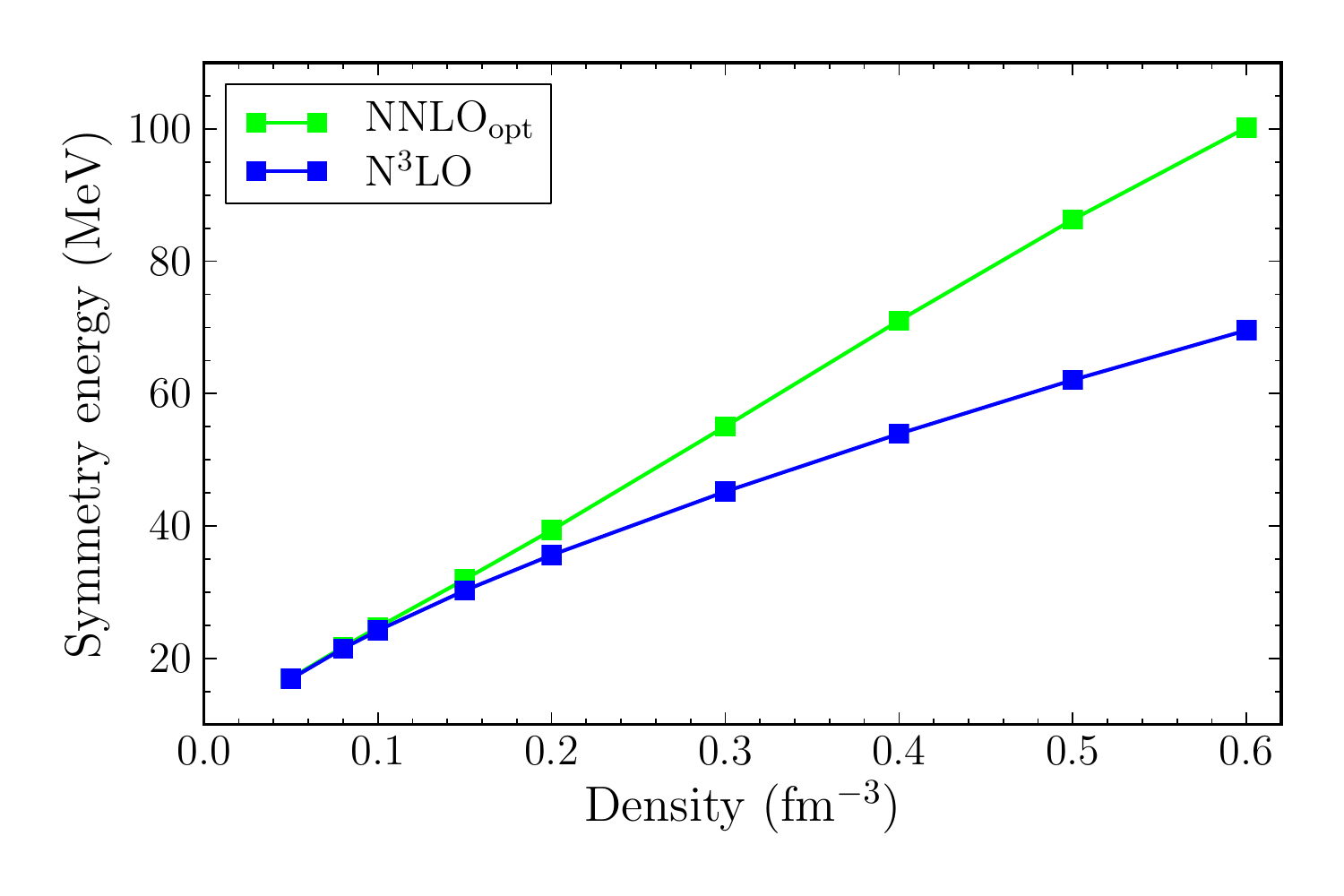}
  \caption{(Color online) Symmetry energy as function of density,
    calculated in the PPHH-LAD approximation with exact Pauli
    exclusion operators. The empirical saturation density of symmetric
    nuclear matter is approximately 0.17 fm$^{-3}$
    \cite{myers1990}.} \label{fig:pphh_symm_ene}
\end{figure}

Finally, we present our results for the symmetry energy in Fig.~\ref{fig:pphh_symm_ene}.
The symmetry energy $S$ is defined as the difference between the binding
energies of pure neutron matter and symmetric nuclear matter, that is
\begin{equation}
  S = (E_{\text{pnm}} - E_{\text{snm}})/A,
\end{equation}
where $E_{\text{pnm}}/A$ and $E_{\text{snm}}/A$ are the binding
energies per particle for pure neutron matter and symmetric nuclear
matter, respectively. The behavior of the symmetry energy at high
densities is important for the understanding of several physical
properties and processes of neutron stars (see \cite{lattimer2012,li2002, erler2013} and references
therein). In Fig.~\ref{fig:pphh_symm_ene}, the symmetry energy is
plotted as a function of nucleon density, as obtained from a PPHH-LAD
calculation with exact Pauli exclusion operators. The symmetry energies
are calculated with both the N$^{3}$LO and NNLO$_{\text{opt}}$
two-body interactions. The symmetry energy obtained with the N$^{3}$LO
interaction is slightly larger than what was reported in
Refs.~\cite{engvik1997,heiselberg}, where the calculations were done with BHF
theory using the CD-Bonn interaction. At densities lower than 0.1 fm$^{-1}$,
the two interaction models give almost the same symmetry energy. However,
above the saturation density, the difference between the two models
increases as a function of density.  As seen from Fig.~\ref{fig:pphh_symm_ene}, 
the NNLO$_{\text{opt}}$ interaction gives significantly larger symmetry
energies than the N$^{3}$LO interaction at high densities. Such a large
 deviation between the two different two-body interactions is possible
since the nuclear interactions are fitted to phase shifts for laboratory
energies only up to 290 MeV \cite{ekstrom2013,n3lo}.  \\

\section{Conclusions}\label{sec:conclusions}
We have studied infinite nuclear and neutron matter using a 
coupled-cluster ladder approximation, where the equations were derived
from the coupled-cluster doubles approximation. In the coupled 
ladder approximation, particle-hole and non-linear
diagrams were neglected from the coupled-cluster doubles amplitude
equations. Our approach can be seen as a first step in implementing 
CC theory for infinite matter. The coupled ladder equations
consist of particle-particle and hole-hole ladder diagrams which are
coupled together. As we have shown, this method is closely related to the
commonly used Brueckner-Hartree-Fock approximation.

We have derived coupled ladder equations both with exact and
angular-averaged Pauli exclusion operators, following the approach 
introduced by Suzuki \emph{et al.}~for the BHF approximation
\cite{suzuki}. In all calculations we have used angular-averaged 
input momenta for the single-particle energies. 
Our method was applied numerically to both
symmetric nuclear matter and pure neutron matter systems. The ladder
approximations for symmetric nuclear matter were found to give less
binding than the BHF approximation. In symmetric nuclear matter, 
the contribution from the hole-hole
ladder diagrams was found to be smaller than the error due to
angular-average approximations for the Pauli exclusion operators.
Generally, symmetric nuclear 
matter calculations with exact Pauli exclusion operators gave more binding 
than calculations with angular-averaged Pauli operators. This behavior 
is in agreement with observations made with  the BHF method 
\cite{suzuki,schiller}. The binding energy per particle of pure neutron 
matter was found to be less sensitive to the Pauli exclusion operator 
approximation than what was the case for symmetric nuclear matter.

The ladder approximations were applied to infinite neutron and
nuclear matter using two different chiral two-body interactions. An
optimized next-to-next-to-leading order (NNLO) interaction
\cite{ekstrom2013} was compared with the
next-to-next-to-next-to-leading order (N$^{3}$LO) interaction of Entem
and Machleidt \cite{n3lo}. In symmetric nuclear matter, we found that
the two interaction models gave similar binding energies. As was shown in
Ref.~\cite{ekstrom2013}, the optimized NNLO interaction gives more repulsion 
in neutron matter compared to the N$^{3}$LO two-body interaction. 
In the present work, we showed that the increased 
repulsion obtained with the NNLO interaction is due to differences 
in the $^{3}P_0$ and $^{3}P_1$ partial waves. We also calculated 
symmetry energies with the N$^{3}$LO and optimized NNLO 
two-body interactions. 

\begin{acknowledgments}
  We thank Scott Bogner, Boris Carlsson, Gustav Jansen, \O yvind Jensen, Simen Kvaal, and Thomas Papenbrock for several discussions. This work was supported
  by the Research Council of Norway under contract ISP-Fysikk/216699; 
  by the Office of Nuclear Physics, U.S. Department of Energy 
  (Oak Ridge National Laboratory), under DE-FG02-96ER40963 
  (University of Tennessee), and DE-SC0008499 (NUCLEI SciDAC collaboration).
  This research used computational resources of the Notur project in Norway.
\end{acknowledgments}


\begin{thebibliography}{10}%
\makeatletter
\providecommand \@ifxundefined [1]{%
 \ifx #1\undefined \expandafter \@firstoftwo
 \else \expandafter \@secondoftwo
\fi
}%
\providecommand \@ifnum [1]{%
 \ifnum #1\expandafter \@firstoftwo
 \else \expandafter \@secondoftwo
\fi
}%
\providecommand \enquote [1]{``#1''}%
\providecommand \bibnamefont  [1]{#1}%
\providecommand \bibfnamefont [1]{#1}%
\providecommand \citenamefont [1]{#1}%
\providecommand\href[0]{\@sanitize\@href}%
\providecommand\@href[1]{\endgroup\@@startlink{#1}\endgroup\@@href}%
\providecommand\@@href[1]{#1\@@endlink}%
\providecommand \@sanitize [0]{\begingroup\catcode`\&12\catcode`\#12\relax}%
\@ifxundefined \pdfoutput {\@firstoftwo}{%
 \@ifnum{\z@=\pdfoutput}{\@firstoftwo}{\@secondoftwo}%
}{%
 \providecommand\@@startlink[1]{\leavevmode}%
 \providecommand\@@endlink[0]{}%
}{%
 \providecommand\@@startlink[1]{%
  \leavevmode
  \pdfstartlink
   attr{/Border[0 0 1 ]/H/I/C[0 1 1]}%
   user{/Subtype/Link/A<</Type/Action/S/URI/URI(#1)>>}%
  \relax
 }%
 \providecommand\@@endlink[0]{\pdfendlink}%
}%
\providecommand \url  [0]{\begingroup\@sanitize \@url }%
\providecommand \@url [1]{\endgroup\@href {#1}{\urlprefix}}%
\providecommand \urlprefix [0]{URL }%
\providecommand \Eprint[0]{\href }%
\@ifxundefined \urlstyle {%
  \providecommand \doi [1]{doi:\discretionary{}{}{}#1}%
}{%
  \providecommand \doi [0]{doi:\discretionary{}{}{}\begingroup
  \urlstyle{rm}\Url }%
}%
\providecommand \doibase [0]{http://dx.doi.org/}%
\providecommand \Doi[1]{\href{\doibase#1}}%
\providecommand \bibAnnote [3]{%
  \BibitemShut{#1}%
  \begin{quotation}\noindent
    \textsc{Key:}\ #2\\\textsc{Annotation:}\ #3%
  \end{quotation}%
}%
\providecommand \bibAnnoteFile [2]{%
  \IfFileExists{#2}{\bibAnnote {#1} {#2} {\input{#2}}}{}%
}%
\providecommand \typeout [0]{\immediate \write \m@ne }%
\providecommand \selectlanguage [0]{\@gobble}%
\providecommand \bibinfo [0]{\@secondoftwo}%
\providecommand \bibfield [0]{\@secondoftwo}%
\providecommand \translation [1]{[#1]}%
\providecommand \BibitemOpen[0]{}%
\providecommand \bibitemStop [0]{}%
\providecommand \bibitemNoStop [0]{.\EOS\space}%
\providecommand \EOS [0]{\spacefactor3000\relax}%
\providecommand \BibitemShut [1]{\csname bibitem#1\endcsname}%
\bibitem{day1978}%
  \BibitemOpen
  \bibfield{author}{%
  \bibinfo {author} {\bibfnamefont{B.~D.}\ \bibnamefont{Day}},\ }%
  \bibfield{journal}{%
  \bibinfo {journal} {Rev. Mod. Phys.}\ }%
  \textbf{\bibinfo {volume} {50}},\ \bibinfo {pages} {495} (\bibinfo {year}
  {1978})%
  \bibAnnoteFile{NoStop}{day1978}%
\bibitem{day_cc}%
  \BibitemOpen
  \bibfield{author}{%
  \bibinfo {author} {\bibfnamefont{B.~D.}\ \bibnamefont{Day}}\ and\ \bibinfo
  {author} {\bibfnamefont{G.}~\bibnamefont{Zabolitzky}},\ }%
  \bibfield{journal}{%
  \bibinfo {journal} {Nucl. Phys. A}\ }%
  \textbf{\bibinfo {volume} {366}},\ \bibinfo {pages} {221} (\bibinfo {year}
  {1981})%
  \bibAnnoteFile{NoStop}{day_cc}%
\bibitem{akmal1997}%
  \BibitemOpen
  \bibfield{author}{%
  \bibinfo {author} {\bibfnamefont{A.}~\bibnamefont{Akmal}}\ and\ \bibinfo
  {author} {\bibfnamefont{V.~R.}\ \bibnamefont{Pandharipande}},\ }%
  \bibfield{journal}{%
  \bibinfo {journal} {Phys. Rev. C}\ }%
  \textbf{\bibinfo {volume} {56}},\ \bibinfo {pages} {2261} (\bibinfo {year}
  {1997})%
  \bibAnnoteFile{NoStop}{akmal1997}%
\bibitem{dickhoff2004}%
  \BibitemOpen
  \bibfield{author}{%
  \bibinfo {author} {\bibfnamefont{W.}~\bibnamefont{Dickhoff}}\ and\ \bibinfo
  {author} {\bibfnamefont{C.}~\bibnamefont{Barbieri}},\ }%
  \bibfield{journal}{%
  \bibinfo {journal} {Prog. Part. and Nucl. Phys.}\ }%
  \textbf{\bibinfo {volume} {52}},\ \bibinfo {pages} {377} (\bibinfo {year}
  {2004})%
  \bibAnnoteFile{NoStop}{dickhoff2004}%
\bibitem{gandolfi2007}%
  \BibitemOpen
  \bibfield{author}{%
  \bibinfo {author} {\bibfnamefont{S.}~\bibnamefont{Gandolfi}}, \bibinfo
  {author} {\bibfnamefont{F.}~\bibnamefont{Pederiva}}, \bibinfo {author}
  {\bibfnamefont{S.}~\bibnamefont{Fantoni}},\ and\ \bibinfo {author}
  {\bibfnamefont{K.~E.}\ \bibnamefont{Schmidt}},\ }%
  \bibfield{journal}{%
  \bibinfo {journal} {Phys. Rev. Lett.}\ }%
  \textbf{\bibinfo {volume} {98}},\ \bibinfo {pages} {102503} (\bibinfo {year}
  {2007})%
  \bibAnnoteFile{NoStop}{gandolfi2007}%
\bibitem{li2006}%
  \BibitemOpen
  \bibfield{author}{%
  \bibinfo {author} {\bibfnamefont{Z.~H.}\ \bibnamefont{Li}}, \bibinfo {author}
  {\bibfnamefont{U.}~\bibnamefont{Lombardo}}, \bibinfo {author}
  {\bibfnamefont{H.-J.}\ \bibnamefont{Schulze}}, \bibinfo {author}
  {\bibfnamefont{W.}~\bibnamefont{Zuo}}, \bibinfo {author}
  {\bibfnamefont{L.~W.}\ \bibnamefont{Chen}},\ and\ \bibinfo {author}
  {\bibfnamefont{H.~R.}\ \bibnamefont{Ma}},\ }%
  \bibfield{journal}{%
  \bibinfo {journal} {Phys. Rev. C}\ }%
  \textbf{\bibinfo {volume} {74}},\ \bibinfo {pages} {047304} (\bibinfo {year}
  {2006})%
  \bibAnnoteFile{NoStop}{li2006}%
\bibitem{lovato2011}%
  \BibitemOpen
  \bibfield{author}{%
  \bibinfo {author} {\bibfnamefont{A.}~\bibnamefont{Lovato}}, \bibinfo {author}
  {\bibfnamefont{O.}~\bibnamefont{Benhar}}, \bibinfo {author}
  {\bibfnamefont{S.}~\bibnamefont{Fantoni}}, \bibinfo {author}
  {\bibfnamefont{A.~Y.}\ \bibnamefont{Illarionov}},\ and\ \bibinfo {author}
  {\bibfnamefont{K.~E.}\ \bibnamefont{Schmidt}},\ }%
  \bibfield{journal}{%
  \bibinfo {journal} {Phys. Rev. C}\ }%
  \textbf{\bibinfo {volume} {83}},\ \bibinfo {pages} {054003} (\bibinfo {year}
  {2011})%
  \bibAnnoteFile{NoStop}{lovato2011}%
\bibitem{carbone2013}%
  \BibitemOpen
  \bibfield{author}{%
  \bibinfo {author} {\bibfnamefont{A.}~\bibnamefont{Carbone}}, \bibinfo
  {author} {\bibfnamefont{A.}~\bibnamefont{Polls}},\ and\ \bibinfo {author}
  {\bibfnamefont{A.}~\bibnamefont{Rios}},\ }%
  \bibfield{journal}{%
  \bibinfo {journal} {Phys. Rev. C}\ }%
  \textbf{\bibinfo {volume} {88}},\ \bibinfo {pages} {044302} (\bibinfo {year}
  {2013})%
  \bibAnnoteFile{NoStop}{carbone2013}%
\bibitem{brueckner}%
  \BibitemOpen
  \bibfield{author}{%
  \bibinfo {author} {\bibfnamefont{K.~A.}\ \bibnamefont{Brueckner}},\ }%
  \bibfield{journal}{%
  \bibinfo {journal} {Phys. Rev.}\ }%
  \textbf{\bibinfo {volume} {100}},\ \bibinfo {pages} {36} (\bibinfo {year}
  {1955})%
  \bibAnnoteFile{NoStop}{brueckner}%
\bibitem{day1967}%
  \BibitemOpen
  \bibfield{author}{%
  \bibinfo {author} {\bibfnamefont{B.~D.}\ \bibnamefont{Day}},\ }%
  \bibfield{journal}{%
  \bibinfo {journal} {Rev. Mod. Phys.}\ }%
  \textbf{\bibinfo {volume} {39}},\ \bibinfo {pages} {719} (\bibinfo {year}
  {1967})%
  \bibAnnoteFile{NoStop}{day1967}%
\bibitem{brueckner1955b}%
  \BibitemOpen
  \bibfield{author}{%
  \bibinfo {author} {\bibfnamefont{K.~A.}\ \bibnamefont{Brueckner}}\ and\
  \bibinfo {author} {\bibfnamefont{C.~A.}\ \bibnamefont{Levinson}},\ }%
  \bibfield{journal}{%
  \bibinfo {journal} {Phys. Rev.}\ }%
  \textbf{\bibinfo {volume} {97}},\ \bibinfo {pages} {1344} (\bibinfo {year}
  {1955})%
  \bibAnnoteFile{NoStop}{brueckner1955b}%
\bibitem{haftel_tabakin}%
  \BibitemOpen
  \bibfield{author}{%
  \bibinfo {author} {\bibfnamefont{M.~I.}\ \bibnamefont{Haftel}}\ and\ \bibinfo
  {author} {\bibfnamefont{F.}~\bibnamefont{Tabakin}},\ }%
  \bibfield{journal}{%
  \bibinfo {journal} {Nucl. Phys. A}\ }%
  \textbf{\bibinfo {volume} {158}},\ \bibinfo {pages} {1} (\bibinfo {year}
  {1970})%
  \bibAnnoteFile{NoStop}{haftel_tabakin}%
\bibitem{song}%
  \BibitemOpen
  \bibfield{author}{%
  \bibinfo {author} {\bibfnamefont{H.~Q.}\ \bibnamefont{Song}}, \bibinfo
  {author} {\bibfnamefont{M.}~\bibnamefont{Baldo}}, \bibinfo {author}
  {\bibfnamefont{G.}~\bibnamefont{Giansiracusa}},\ and\ \bibinfo {author}
  {\bibfnamefont{U.}~\bibnamefont{Lombardo}},\ }%
  \bibfield{journal}{%
  \bibinfo {journal} {Phys. Rev. Lett.}\ }%
  \textbf{\bibinfo {volume} {81}},\ \bibinfo {pages} {1584} (\bibinfo {year}
  {1998})%
  \bibAnnoteFile{NoStop}{song}%
\bibitem{schiller}%
  \BibitemOpen
  \bibfield{author}{%
  \bibinfo {author} {\bibfnamefont{E.}~\bibnamefont{Schiller}}, \bibinfo
  {author} {\bibfnamefont{H.}~\bibnamefont{M{\"u}ther}},\ and\ \bibinfo
  {author} {\bibfnamefont{P.}~\bibnamefont{Czerski}},\ }%
  \bibfield{journal}{%
  \bibinfo {journal} {Phys. Rev. C}\ }%
  \textbf{\bibinfo {volume} {60}},\ \bibinfo {pages} {059901(E)} (\bibinfo
  {year} {1999})%
  \bibAnnoteFile{NoStop}{schiller}%
\bibitem{suzuki}%
  \BibitemOpen
  \bibfield{author}{%
  \bibinfo {author} {\bibfnamefont{K.}~\bibnamefont{Suzuki}}, \bibinfo {author}
  {\bibfnamefont{R.}~\bibnamefont{Okamoto}}, \bibinfo {author}
  {\bibfnamefont{M.}~\bibnamefont{Kohno}},\ and\ \bibinfo {author}
  {\bibfnamefont{S.}~\bibnamefont{Nagata}},\ }%
  \bibfield{journal}{%
  \bibinfo {journal} {Nucl. Phys. A}\ }%
  \textbf{\bibinfo {volume} {665}},\ \bibinfo {pages} {92} (\bibinfo {year}
  {2000})%
  \bibAnnoteFile{NoStop}{suzuki}%
\bibitem{baldo2012}%
  \BibitemOpen
  \bibfield{author}{%
  \bibinfo {author} {\bibfnamefont{M.}~\bibnamefont{Baldo}}, \bibinfo {author}
  {\bibfnamefont{A.}~\bibnamefont{Polls}}, \bibinfo {author}
  {\bibfnamefont{A.}~\bibnamefont{Rios}}, \bibinfo {author}
  {\bibfnamefont{H.-J.}\ \bibnamefont{Schulze}},\ and\ \bibinfo {author}
  {\bibfnamefont{I.}~\bibnamefont{Vida\~na}},\ }%
  \bibfield{journal}{%
  \bibinfo {journal} {Phys. Rev. C}\ }%
  \textbf{\bibinfo {volume} {86}},\ \bibinfo {pages} {064001} (\bibinfo {year}
  {2012})%
  \bibAnnoteFile{NoStop}{baldo2012}%
\bibitem{raja}%
  \BibitemOpen
  \bibfield{author}{%
  \bibinfo {author} {\bibfnamefont{R.}~\bibnamefont{Rajaraman}}\ and\ \bibinfo
  {author} {\bibfnamefont{H.~A.}\ \bibnamefont{Bethe}},\ }%
  \bibfield{journal}{%
  \bibinfo {journal} {Rev. Mod. Phys.}\ }%
  \textbf{\bibinfo {volume} {39}},\ \bibinfo {pages} {745} (\bibinfo {year}
  {1967})%
  \bibAnnoteFile{NoStop}{raja}%
\bibitem{day1981}%
  \BibitemOpen
  \bibfield{author}{%
  \bibinfo {author} {\bibfnamefont{B.~D.}\ \bibnamefont{Day}},\ }%
  \bibfield{journal}{%
  \bibinfo {journal} {Phys. Rev. C}\ }%
  \textbf{\bibinfo {volume} {24}},\ \bibinfo {pages} {1203} (\bibinfo {year}
  {1981})%
  \bibAnnoteFile{NoStop}{day1981}%
\bibitem{soma2008}%
  \BibitemOpen
  \bibfield{author}{%
  \bibinfo {author} {\bibfnamefont{V.}~\bibnamefont{Som\`a}}\ and\ \bibinfo
  {author} {\bibfnamefont{P.}~\bibnamefont{Bozek}},\ }%
  \bibfield{journal}{%
  \bibinfo {journal} {Phys. Rev. C}\ }%
  \textbf{\bibinfo {volume} {78}},\ \bibinfo {pages} {054003} (\bibinfo {year}
  {2008})%
  \bibAnnoteFile{NoStop}{soma2008}%
\bibitem{vidana2009}%
  \BibitemOpen
  \bibfield{author}{%
  \bibinfo {author} {\bibfnamefont{I.}~\bibnamefont{Vida\~na}}, \bibinfo
  {author} {\bibfnamefont{C.}~\bibnamefont{Provid\^encia}}, \bibinfo {author}
  {\bibfnamefont{A.}~\bibnamefont{Polls}},\ and\ \bibinfo {author}
  {\bibfnamefont{A.}~\bibnamefont{Rios}},\ }%
  \bibfield{journal}{%
  \bibinfo {journal} {Phys. Rev. C}\ }%
  \textbf{\bibinfo {volume} {80}},\ \bibinfo {pages} {045806} (\bibinfo {year}
  {2009})%
  \bibAnnoteFile{NoStop}{vidana2009}%
\bibitem{hebeler2011}%
  \BibitemOpen
  \bibfield{author}{%
  \bibinfo {author} {\bibfnamefont{K.}~\bibnamefont{Hebeler}}, \bibinfo
  {author} {\bibfnamefont{S.~K.}\ \bibnamefont{Bogner}}, \bibinfo {author}
  {\bibfnamefont{R.~J.}\ \bibnamefont{Furnstahl}}, \bibinfo {author}
  {\bibfnamefont{A.}~\bibnamefont{Nogga}},\ and\ \bibinfo {author}
  {\bibfnamefont{A.}~\bibnamefont{Schwenk}},\ }%
  \bibfield{journal}{%
  \bibinfo {journal} {Phys. Rev. C}\ }%
  \textbf{\bibinfo {volume} {83}},\ \bibinfo {pages} {031301} (\bibinfo {year}
  {2011})%
  \bibAnnoteFile{NoStop}{hebeler2011}%
\bibitem{gandolfi2012}%
  \BibitemOpen
  \bibfield{author}{%
  \bibinfo {author} {\bibfnamefont{A.~W.}\ \bibnamefont{Steiner}}\ and\
  \bibinfo {author} {\bibfnamefont{S.}~\bibnamefont{Gandolfi}},\ }%
  \bibfield{journal}{%
  \bibinfo {journal} {Phys. Rev. Lett.}\ }%
  \textbf{\bibinfo {volume} {108}},\ \bibinfo {pages} {081102} (\bibinfo {year}
  {2012})%
  \bibAnnoteFile{NoStop}{gandolfi2012}%
\bibitem{hebeler2013}%
  \BibitemOpen
  \bibfield{author}{%
  \bibinfo {author} {\bibfnamefont{K.}~\bibnamefont{Hebeler}}\ and\ \bibinfo
  {author} {\bibfnamefont{R.~J.}\ \bibnamefont{Furnstahl}},\ }%
  \bibfield{journal}{%
  \bibinfo {journal} {Phys. Rev. C}\ }%
  \textbf{\bibinfo {volume} {87}},\ \bibinfo {pages} {031302} (\bibinfo {year}
  {2013})%
  \bibAnnoteFile{NoStop}{hebeler2013}%
\bibitem{tews}%
  \BibitemOpen
  \bibfield{author}{%
  \bibinfo {author} {\bibfnamefont{I.}~\bibnamefont{Tews}}, \bibinfo {author}
  {\bibfnamefont{T.}~\bibnamefont{Kr{\"u}ger}}, \bibinfo {author}
  {\bibfnamefont{K.}~\bibnamefont{Hebeler}},\ and\ \bibinfo {author}
  {\bibfnamefont{A.}~\bibnamefont{Schwenk}},\ }%
  \bibfield{journal}{%
  \bibinfo {journal} {Phys. Rev. Lett.}\ }%
  \textbf{\bibinfo {volume} {110}},\ \bibinfo {pages} {032504} (\bibinfo {year}
  {2013})%
  \bibAnnoteFile{NoStop}{tews}%
\bibitem{kruger2013}%
  \BibitemOpen
  \bibfield{author}{%
  \bibinfo {author} {\bibfnamefont{T.}~\bibnamefont{Kr\"uger}}, \bibinfo
  {author} {\bibfnamefont{I.}~\bibnamefont{Tews}}, \bibinfo {author}
  {\bibfnamefont{K.}~\bibnamefont{Hebeler}},\ and\ \bibinfo {author}
  {\bibfnamefont{A.}~\bibnamefont{Schwenk}},\ }%
  \bibfield{journal}{%
  \bibinfo {journal} {Phys. Rev. C}\ }%
  \textbf{\bibinfo {volume} {88}},\ \bibinfo {pages} {025802} (\bibinfo {year}
  {2013})%
  \bibAnnoteFile{NoStop}{kruger2013}%
\bibitem{coester1958}%
  \BibitemOpen
  \bibfield{author}{%
  \bibinfo {author} {\bibfnamefont{F.}~\bibnamefont{Coester}},\ }%
  \bibfield{journal}{%
  \bibinfo {journal} {Nucl. Phys.}\ }%
  \textbf{\bibinfo {volume} {7}},\ \bibinfo {pages} {421} (\bibinfo {year}
  {1958})%
  \bibAnnoteFile{NoStop}{coester1958}%
\bibitem{coester1960}%
  \BibitemOpen
  \bibfield{author}{%
  \bibinfo {author} {\bibfnamefont{F.}~\bibnamefont{Coester}}\ and\ \bibinfo
  {author} {\bibfnamefont{H.}~\bibnamefont{K{\"u}mmel}},\ }%
  \bibfield{journal}{%
  \bibinfo {journal} {Nucl. Phys.}\ }%
  \textbf{\bibinfo {volume} {17}},\ \bibinfo {pages} {477} (\bibinfo {year}
  {1960})%
  \bibAnnoteFile{NoStop}{coester1960}%
\bibitem{cizek1966}%
  \BibitemOpen
  \bibfield{author}{%
  \bibinfo {author} {\bibfnamefont{J.}~\bibnamefont{{\v {C}}{\'\i}{\v z}ek}},\
  }%
  \bibfield{journal}{%
  \bibinfo {journal} {J. Chem. Phys.}\ }%
  \textbf{\bibinfo {volume} {45}},\ \bibinfo {pages} {4256} (\bibinfo {year}
  {1966})%
  \bibAnnoteFile{NoStop}{cizek1966}%
\bibitem{cizek1969}%
  \BibitemOpen
  \bibfield{author}{%
  \bibinfo {author} {\bibfnamefont{J.}~\bibnamefont{{\v {C}}{\'\i}{\v z}ek}},\
  }%
  \enquote{\bibinfo {title} {{On the Use of the Cluster Expansion and the
  Technique of Diagrams in Calculations of Correlation Effects in Atoms and
  Molecules}},}\ in\ \emph{\bibinfo {booktitle} {Advances in Chemical
  Physics}}\ (\bibinfo {publisher} {John Wiley \& Sons, Inc.},\ \bibinfo {year}
  {2007})\ p.~\bibinfo {pages} {35}%
  \bibAnnoteFile{NoStop}{cizek1969}%
\bibitem{cizek1971}%
  \BibitemOpen
  \bibfield{author}{%
  \bibinfo {author} {\bibfnamefont{J.}~\bibnamefont{{\v {C}}{\'\i}{\v z}ek}}\
  and\ \bibinfo {author} {\bibfnamefont{J.}~\bibnamefont{Paldus}},\ }%
  \bibfield{journal}{%
  \bibinfo {journal} {Int. J. Quant. Chem.}\ }%
  \textbf{\bibinfo {volume} {5}},\ \bibinfo {pages} {359} (\bibinfo {year}
  {1971})%
  \bibAnnoteFile{NoStop}{cizek1971}%
\bibitem{bartlett_review}%
  \BibitemOpen
  \bibfield{author}{%
  \bibinfo {author} {\bibfnamefont{R.}~\bibnamefont{Bartlett}}\ and\ \bibinfo
  {author} {\bibfnamefont{M.}~\bibnamefont{Musia{\l}}},\ }%
  \bibfield{journal}{%
  \bibinfo {journal} {Rev. Mod. Phys.}\ }%
  \textbf{\bibinfo {volume} {79}},\ \bibinfo {pages} {291} (\bibinfo {year}
  {2007})%
  \bibAnnoteFile{NoStop}{bartlett_review}%
\bibitem{bartlett_book}%
  \BibitemOpen
  \bibfield{author}{%
  \bibinfo {author} {\bibfnamefont{I.}~\bibnamefont{Shavitt}}\ and\ \bibinfo
  {author} {\bibfnamefont{R.~J.}\ \bibnamefont{Bartlett}},\ }%
  \emph{\bibinfo {title} {Many-Body Methods in Chemistry and Physics}}\
  (\bibinfo {publisher} {Cambridge University Press},\ \bibinfo {year} {2009})%
  \bibAnnoteFile{NoStop}{bartlett_book}%
\bibitem{crawford}%
  \BibitemOpen
  \bibfield{author}{%
  \bibinfo {author} {\bibfnamefont{T.~D.}\ \bibnamefont{Crawford}}\ and\
  \bibinfo {author} {\bibfnamefont{H.~F.}\ \bibnamefont{Schaefer}},\ }%
  \bibfield{journal}{%
  \bibinfo {journal} {Rev. Comp. Chem.}\ }%
  \textbf{\bibinfo {volume} {14}},\ \bibinfo {pages} {33} (\bibinfo {year}
  {2000})%
  \bibAnnoteFile{NoStop}{crawford}%
\bibitem{harris}%
  \BibitemOpen
  \bibfield{author}{%
  \bibinfo {author} {\bibfnamefont{F.~E.}\ \bibnamefont{Harris}}, \bibinfo
  {author} {\bibfnamefont{H.~J.}\ \bibnamefont{Monkhorst}},\ and\ \bibinfo
  {author} {\bibfnamefont{D.~L.}\ \bibnamefont{Freeman}},\ }%
  \emph{\bibinfo {title} {Algebraic and diagrammatic methods in many-fermion
  theory}}\ (\bibinfo {publisher} {Oxford University Press},\ \bibinfo {year}
  {1992})%
  \bibAnnoteFile{NoStop}{harris}%
\bibitem{heisenberg1999}%
  \BibitemOpen
  \bibfield{author}{%
  \bibinfo {author} {\bibfnamefont{J.~H.}\ \bibnamefont{Heisenberg}}\ and\
  \bibinfo {author} {\bibfnamefont{B.}~\bibnamefont{Mihaila}},\ }%
  \bibfield{journal}{%
  \bibinfo {journal} {Phys. Rev. C}\ }%
  \textbf{\bibinfo {volume} {59}},\ \bibinfo {pages} {1440} (\bibinfo {year}
  {1999})%
  \bibAnnoteFile{NoStop}{heisenberg1999}%
\bibitem{heisenberg2000}%
  \BibitemOpen
  \bibfield{author}{%
  \bibinfo {author} {\bibfnamefont{B.}~\bibnamefont{Mihaila}}\ and\ \bibinfo
  {author} {\bibfnamefont{J.~H.}\ \bibnamefont{Heisenberg}},\ }%
  \bibfield{journal}{%
  \bibinfo {journal} {Phys. Rev. C}\ }%
  \textbf{\bibinfo {volume} {61}},\ \bibinfo {pages} {054309} (\bibinfo {year}
  {2000})%
  \bibAnnoteFile{NoStop}{heisenberg2000}%
\bibitem{dean2004}%
  \BibitemOpen
  \bibfield{author}{%
  \bibinfo {author} {\bibfnamefont{D.~J.}\ \bibnamefont{Dean}}\ and\ \bibinfo
  {author} {\bibfnamefont{M.}~\bibnamefont{Hjorth-Jensen}},\ }%
  \bibfield{journal}{%
  \bibinfo {journal} {Phys. Rev. C}\ }%
  \textbf{\bibinfo {volume} {69}},\ \bibinfo {pages} {054320} (\bibinfo {year}
  {2004})%
  \bibAnnoteFile{NoStop}{dean2004}%
\bibitem{gour2006}%
  \BibitemOpen
  \bibfield{author}{%
  \bibinfo {author} {\bibfnamefont{J.~R.}\ \bibnamefont{Gour}}, \bibinfo
  {author} {\bibfnamefont{P.}~\bibnamefont{Piecuch}}, \bibinfo {author}
  {\bibfnamefont{M.}~\bibnamefont{Hjorth-Jensen}}, \bibinfo {author}
  {\bibfnamefont{M.}~\bibnamefont{W\l{}och}},\ and\ \bibinfo {author}
  {\bibfnamefont{D.~J.}\ \bibnamefont{Dean}},\ }%
  \bibfield{journal}{%
  \bibinfo {journal} {Phys. Rev. C}\ }%
  \textbf{\bibinfo {volume} {74}},\ \bibinfo {pages} {024310} (\bibinfo {year}
  {2006})%
  \bibAnnoteFile{NoStop}{gour2006}%
\bibitem{gour2008}%
  \BibitemOpen
  \bibfield{author}{%
  \bibinfo {author} {\bibfnamefont{J.~R.}\ \bibnamefont{Gour}}, \bibinfo
  {author} {\bibfnamefont{M.}~\bibnamefont{Horoi}}, \bibinfo {author}
  {\bibfnamefont{P.}~\bibnamefont{Piecuch}},\ and\ \bibinfo {author}
  {\bibfnamefont{B.~A.}\ \bibnamefont{Brown}},\ }%
  \bibfield{journal}{%
  \bibinfo {journal} {Phys. Rev. Lett.}\ }%
  \textbf{\bibinfo {volume} {101}},\ \bibinfo {pages} {052501} (\bibinfo {year}
  {2008})%
  \bibAnnoteFile{NoStop}{gour2008}%
\bibitem{hagen2010}%
  \BibitemOpen
  \bibfield{author}{%
  \bibinfo {author} {\bibfnamefont{G.}~\bibnamefont{Hagen}}, \bibinfo {author}
  {\bibfnamefont{T.}~\bibnamefont{Papenbrock}}, \bibinfo {author}
  {\bibfnamefont{D.~J.}\ \bibnamefont{Dean}},\ and\ \bibinfo {author}
  {\bibfnamefont{M.}~\bibnamefont{Hjorth-Jensen}},\ }%
  \bibfield{journal}{%
  \bibinfo {journal} {Phys. Rev. C}\ }%
  \textbf{\bibinfo {volume} {82}},\ \bibinfo {pages} {034330} (\bibinfo {year}
  {2010})%
  \bibAnnoteFile{NoStop}{hagen2010}%
\bibitem{jansen2011}%
  \BibitemOpen
  \bibfield{author}{%
  \bibinfo {author} {\bibfnamefont{G.~R.}\ \bibnamefont{Jansen}}, \bibinfo
  {author} {\bibfnamefont{M.}~\bibnamefont{Hjorth-Jensen}}, \bibinfo {author}
  {\bibfnamefont{G.}~\bibnamefont{Hagen}},\ and\ \bibinfo {author}
  {\bibfnamefont{T.}~\bibnamefont{Papenbrock}},\ }%
  \bibfield{journal}{%
  \bibinfo {journal} {Phys. Rev. C}\ }%
  \textbf{\bibinfo {volume} {83}},\ \bibinfo {pages} {054306} (\bibinfo {year}
  {2011})%
  \bibAnnoteFile{NoStop}{jansen2011}%
\bibitem{jensen2010}%
  \BibitemOpen
  \bibfield{author}{%
  \bibinfo {author} {\bibfnamefont{O.}~\bibnamefont{Jensen}}, \bibinfo {author}
  {\bibfnamefont{G.}~\bibnamefont{Hagen}}, \bibinfo {author}
  {\bibfnamefont{T.}~\bibnamefont{Papenbrock}}, \bibinfo {author}
  {\bibfnamefont{D.~J.}\ \bibnamefont{Dean}},\ and\ \bibinfo {author}
  {\bibfnamefont{J.~S.}\ \bibnamefont{Vaagen}},\ }%
  \bibfield{journal}{%
  \bibinfo {journal} {Phys. Rev. C}\ }%
  \textbf{\bibinfo {volume} {82}},\ \bibinfo {pages} {014310} (\bibinfo {year}
  {2010})%
  \bibAnnoteFile{NoStop}{jensen2010}%
\bibitem{roth2012}%
  \BibitemOpen
  \bibfield{author}{%
  \bibinfo {author} {\bibfnamefont{R.}~\bibnamefont{Roth}}, \bibinfo {author}
  {\bibfnamefont{S.}~\bibnamefont{Binder}}, \bibinfo {author}
  {\bibfnamefont{K.}~\bibnamefont{Vobig}}, \bibinfo {author}
  {\bibfnamefont{A.}~\bibnamefont{Calci}}, \bibinfo {author}
  {\bibfnamefont{J.}~\bibnamefont{Langhammer}},\ and\ \bibinfo {author}
  {\bibfnamefont{P.}~\bibnamefont{Navr\'atil}},\ }%
  \bibfield{journal}{%
  \bibinfo {journal} {Phys. Rev. Lett.}\ }%
  \textbf{\bibinfo {volume} {109}},\ \bibinfo {pages} {052501} (\bibinfo {year}
  {2012})%
  \bibAnnoteFile{NoStop}{roth2012}%
\bibitem{kummel1978}%
  \BibitemOpen
  \bibfield{author}{%
  \bibinfo {author} {\bibfnamefont{H.}~\bibnamefont{K{\"u}mmel}}, \bibinfo
  {author} {\bibfnamefont{K.~H.}\ \bibnamefont{L{\"u}hrmann}},\ and\ \bibinfo
  {author} {\bibfnamefont{J.~G.}\ \bibnamefont{Zabolitzky}},\ }%
  \bibfield{journal}{%
  \bibinfo {journal} {Phys. Rep.}\ }%
  \textbf{\bibinfo {volume} {36}},\ \bibinfo {pages} {1} (\bibinfo {year}
  {1978})%
  \bibAnnoteFile{NoStop}{kummel1978}%
\bibitem{bishop_cc}%
  \BibitemOpen
  \bibfield{author}{%
  \bibinfo {author} {\bibfnamefont{R.~F.}\ \bibnamefont{Bishop}},\ }%
  \bibfield{journal}{%
  \bibinfo {journal} {Theor Chim Acta}\ }%
  \textbf{\bibinfo {volume} {80}},\ \bibinfo {pages} {95} (\bibinfo {year}
  {1991})%
  \bibAnnoteFile{NoStop}{bishop_cc}%
\bibitem{machleidt2001}%
  \BibitemOpen
  \bibfield{author}{%
  \bibinfo {author} {\bibfnamefont{R.}~\bibnamefont{Machleidt}}\ and\ \bibinfo
  {author} {\bibfnamefont{I.}~\bibnamefont{Slaus}},\ }%
  \bibfield{journal}{%
  \bibinfo {journal} {J. Phys. G: Nucl. Part. Phys.}\ }%
  \textbf{\bibinfo {volume} {27}},\ \bibinfo {pages} {R69} (\bibinfo {year}
  {2001})%
  \bibAnnoteFile{NoStop}{machleidt2001}%
\bibitem{machleidt2011}%
  \BibitemOpen
  \bibfield{author}{%
  \bibinfo {author} {\bibfnamefont{R.}~\bibnamefont{Machleidt}}\ and\ \bibinfo
  {author} {\bibfnamefont{D.~R.}\ \bibnamefont{Entem}},\ }%
  \bibfield{journal}{%
  \bibinfo {journal} {Phys. Rep.}\ }%
  \textbf{\bibinfo {volume} {503}},\ \bibinfo {pages} {1} (\bibinfo {year}
  {2011})%
  \bibAnnoteFile{NoStop}{machleidt2011}%
\bibitem{bogner2005}%
  \BibitemOpen
  \bibfield{author}{%
  \bibinfo {author} {\bibfnamefont{S.~K.}\ \bibnamefont{Bogner}}, \bibinfo
  {author} {\bibfnamefont{A.}~\bibnamefont{Schwenk}}, \bibinfo {author}
  {\bibfnamefont{R.~J.}\ \bibnamefont{Furnstahl}},\ and\ \bibinfo {author}
  {\bibfnamefont{A.}~\bibnamefont{Nogga}},\ }%
  \bibfield{journal}{%
  \bibinfo {journal} {Nucl. Phys. A}\ }%
  \textbf{\bibinfo {volume} {763}},\ \bibinfo {pages} {59} (\bibinfo {year}
  {2005})%
  \bibAnnoteFile{NoStop}{bogner2005}%
\bibitem{manzke1974}%
  \BibitemOpen
  \bibfield{author}{%
  \bibinfo {author} {\bibfnamefont{W.}~\bibnamefont{Manzke}},\ }%
  \bibinfo {type} {Diplomarbeit},\ \bibinfo {school} {Bochum University}
  (\bibinfo {year} {1974})%
  \bibAnnoteFile{NoStop}{manzke1974}%
\bibitem{bishop1}%
  \BibitemOpen
  \bibfield{author}{%
  \bibinfo {author} {\bibfnamefont{R.~F.}\ \bibnamefont{Bishop}}\ and\ \bibinfo
  {author} {\bibfnamefont{K.~H.}\ \bibnamefont{L{\"u}hrmann}},\ }%
  \bibfield{journal}{%
  \bibinfo {journal} {Phys. Rev.}\ }%
  \textbf{\bibinfo {volume} {17}},\ \bibinfo {pages} {3757} (\bibinfo {year}
  {1978})%
  \bibAnnoteFile{NoStop}{bishop1}%
\bibitem{freeman1977}%
  \BibitemOpen
  \bibfield{author}{%
  \bibinfo {author} {\bibfnamefont{D.~L.}\ \bibnamefont{Freeman}},\ }%
  \bibfield{journal}{%
  \bibinfo {journal} {Phys. Rev. B}\ }%
  \textbf{\bibinfo {volume} {15}},\ \bibinfo {pages} {5512} (\bibinfo {year}
  {1977})%
  \bibAnnoteFile{NoStop}{freeman1977}%
\bibitem{freeman_pplad}%
  \BibitemOpen
  \bibfield{author}{%
  \bibinfo {author} {\bibfnamefont{D.~L.}\ \bibnamefont{Freeman}},\ }%
  \bibfield{journal}{%
  \bibinfo {journal} {J. Phys. C: Solid State Phys.}\ }%
  \textbf{\bibinfo {volume} {16}},\ \bibinfo {pages} {711} (\bibinfo {year}
  {1983})%
  \bibAnnoteFile{NoStop}{freeman_pplad}%
\bibitem{dickhoff1982}%
  \BibitemOpen
  \bibfield{author}{%
  \bibinfo {author} {\bibfnamefont{W.}~\bibnamefont{Dickhoff}}, \bibinfo
  {author} {\bibfnamefont{A.}~\bibnamefont{Faessler}},\ and\ \bibinfo {author}
  {\bibfnamefont{H.}~\bibnamefont{M{\"u}ther}},\ }%
  \bibfield{journal}{%
  \bibinfo {journal} {Nucl. Phys. A}\ }%
  \textbf{\bibinfo {volume} {389}},\ \bibinfo {pages} {492} (\bibinfo {year}
  {1982})%
  \bibAnnoteFile{NoStop}{dickhoff1982}%
\bibitem{ramos_polls}%
  \BibitemOpen
  \bibfield{author}{%
  \bibinfo {author} {\bibfnamefont{A.}~\bibnamefont{Ramos}}, \bibinfo {author}
  {\bibfnamefont{A.}~\bibnamefont{Polls}},\ and\ \bibinfo {author}
  {\bibfnamefont{W.~H.}\ \bibnamefont{Dickhoff}},\ }%
  \bibfield{journal}{%
  \bibinfo {journal} {Nucl. Phys. A}\ }%
  \textbf{\bibinfo {volume} {503}},\ \bibinfo {pages} {1} (\bibinfo {year}
  {1989})%
  \bibAnnoteFile{NoStop}{ramos_polls}%
\bibitem{bozek2002}%
  \BibitemOpen
  \bibfield{author}{%
  \bibinfo {author} {\bibfnamefont{P.}~\bibnamefont{Bozek}},\ }%
  \bibfield{journal}{%
  \bibinfo {journal} {Eur. Phys. J. A}\ }%
  \textbf{\bibinfo {volume} {15}},\ \bibinfo {pages} {325} (\bibinfo {year}
  {2002}),\ ISSN \bibinfo {issn} {1434-6001}%
  \bibAnnoteFile{NoStop}{bozek2002}%
\bibitem{dewulf2003}%
  \BibitemOpen
  \bibfield{author}{%
  \bibinfo {author} {\bibfnamefont{Y.}~\bibnamefont{Dewulf}}, \bibinfo {author}
  {\bibfnamefont{W.~H.}\ \bibnamefont{Dickhoff}}, \bibinfo {author}
  {\bibfnamefont{D.}~\bibnamefont{Van~Neck}}, \bibinfo {author}
  {\bibfnamefont{E.~R.}\ \bibnamefont{Stoddard}},\ and\ \bibinfo {author}
  {\bibfnamefont{M.}~\bibnamefont{Waroquier}},\ }%
  \bibfield{journal}{%
  \bibinfo {journal} {Phys. Rev. Lett.}\ }%
  \textbf{\bibinfo {volume} {90}},\ \bibinfo {pages} {152501} (\bibinfo {year}
  {2003})%
  \bibAnnoteFile{NoStop}{dewulf2003}%
\bibitem{frick2003}%
  \BibitemOpen
  \bibfield{author}{%
  \bibinfo {author} {\bibfnamefont{T.}~\bibnamefont{Frick}}\ and\ \bibinfo
  {author} {\bibfnamefont{H.}~\bibnamefont{M\"uther}},\ }%
  \bibfield{journal}{%
  \bibinfo {journal} {Phys. Rev. C}\ }%
  \textbf{\bibinfo {volume} {68}},\ \bibinfo {pages} {034310} (\bibinfo {year}
  {2003})%
  \bibAnnoteFile{NoStop}{frick2003}%
\bibitem{rios2009}%
  \BibitemOpen
  \bibfield{author}{%
  \bibinfo {author} {\bibfnamefont{A.}~\bibnamefont{Rios}}, \bibinfo {author}
  {\bibfnamefont{A.}~\bibnamefont{Polls}},\ and\ \bibinfo {author}
  {\bibfnamefont{I.}~\bibnamefont{Vida\~na}},\ }%
  \bibfield{journal}{%
  \bibinfo {journal} {Phys. Rev. C}\ }%
  \textbf{\bibinfo {volume} {79}},\ \bibinfo {pages} {025802} (\bibinfo {year}
  {2009})%
  \bibAnnoteFile{NoStop}{rios2009}%
\bibitem{song1987}%
  \BibitemOpen
  \bibfield{author}{%
  \bibinfo {author} {\bibfnamefont{H.}~\bibnamefont{Song}}, \bibinfo {author}
  {\bibfnamefont{S.}~\bibnamefont{Yang}},\ and\ \bibinfo {author}
  {\bibfnamefont{T.}~\bibnamefont{Kuo}},\ }%
  \bibfield{journal}{%
  \bibinfo {journal} {Nucl. Phys. A}\ }%
  \textbf{\bibinfo {volume} {462}},\ \bibinfo {pages} {491 } (\bibinfo {year}
  {1987})%
  \bibAnnoteFile{NoStop}{song1987}%
\bibitem{jiang1988}%
  \BibitemOpen
  \bibfield{author}{%
  \bibinfo {author} {\bibfnamefont{M.~F.}\ \bibnamefont{Jiang}}, \bibinfo
  {author} {\bibfnamefont{T.~T.~S.}\ \bibnamefont{Kuo}},\ and\ \bibinfo
  {author} {\bibfnamefont{H.}~\bibnamefont{M\"uther}},\ }%
  \bibfield{journal}{%
  \bibinfo {journal} {Phys. Rev. C}\ }%
  \textbf{\bibinfo {volume} {38}},\ \bibinfo {pages} {2408} (\bibinfo {year}
  {1988})%
  \bibAnnoteFile{NoStop}{jiang1988}%
\bibitem{engvik1997a}%
  \BibitemOpen
  \bibfield{author}{%
  \bibinfo {author} {\bibfnamefont{L.}~\bibnamefont{Engvik}}, \bibinfo {author}
  {\bibfnamefont{E.}~\bibnamefont{Osnes}}, \bibinfo {author}
  {\bibfnamefont{M.}~\bibnamefont{Hjorth-Jensen}},\ and\ \bibinfo {author}
  {\bibfnamefont{T.}~\bibnamefont{Kuo}},\ }%
  \bibfield{journal}{%
  \bibinfo {journal} {Nucl. Phys. A}\ }%
  \textbf{\bibinfo {volume} {622}},\ \bibinfo {pages} {553} (\bibinfo {year}
  {1997})%
  \bibAnnoteFile{NoStop}{engvik1997a}%
\bibitem{siu2009}%
  \BibitemOpen
  \bibfield{author}{%
  \bibinfo {author} {\bibfnamefont{L.-W.}\ \bibnamefont{Siu}}, \bibinfo
  {author} {\bibfnamefont{J.~W.}\ \bibnamefont{Holt}}, \bibinfo {author}
  {\bibfnamefont{T.~T.~S.}\ \bibnamefont{Kuo}},\ and\ \bibinfo {author}
  {\bibfnamefont{G.~E.}\ \bibnamefont{Brown}},\ }%
  \bibfield{journal}{%
  \bibinfo {journal} {Phys. Rev. C}\ }%
  \textbf{\bibinfo {volume} {79}},\ \bibinfo {pages} {054004} (\bibinfo {year}
  {2009})%
  \bibAnnoteFile{NoStop}{siu2009}%
\bibitem{ekstrom2013}%
  \BibitemOpen
  \bibfield{author}{%
  \bibinfo {author} {\bibfnamefont{A.}~\bibnamefont{Ekstr\"om}}, \bibinfo
  {author} {\bibfnamefont{G.}~\bibnamefont{Baardsen}}, \bibinfo {author}
  {\bibfnamefont{C.}~\bibnamefont{Forss\'en}}, \bibinfo {author}
  {\bibfnamefont{G.}~\bibnamefont{Hagen}}, \bibinfo {author}
  {\bibfnamefont{M.}~\bibnamefont{Hjorth-Jensen}}, \bibinfo {author}
  {\bibfnamefont{G.~R.}\ \bibnamefont{Jansen}}, \bibinfo {author}
  {\bibfnamefont{R.}~\bibnamefont{Machleidt}}, \bibinfo {author}
  {\bibfnamefont{W.}~\bibnamefont{Nazarewicz}}, \bibinfo {author}
  {\bibfnamefont{T.}~\bibnamefont{Papenbrock}}, \bibinfo {author}
  {\bibfnamefont{J.}~\bibnamefont{Sarich}},\ and\ \bibinfo {author}
  {\bibfnamefont{S.~M.}\ \bibnamefont{Wild}},\ }%
  \bibfield{journal}{%
  \bibinfo {journal} {Phys. Rev. Lett.}\ }%
  \textbf{\bibinfo {volume} {110}},\ \bibinfo {pages} {192502} (\bibinfo {year}
  {2013})%
  \bibAnnoteFile{NoStop}{ekstrom2013}%
\bibitem{n3lo}%
  \BibitemOpen
  \bibfield{author}{%
  \bibinfo {author} {\bibfnamefont{D.~R.}\ \bibnamefont{Entem}}\ and\ \bibinfo
  {author} {\bibfnamefont{R.}~\bibnamefont{Machleidt}},\ }%
  \bibfield{journal}{%
  \bibinfo {journal} {Phys. Rev. C}\ }%
  \textbf{\bibinfo {volume} {68}},\ \bibinfo {pages} {041001(R)} (\bibinfo
  {year} {2003})%
  \bibAnnoteFile{NoStop}{n3lo}%
\bibitem{taoman}%
  \BibitemOpen
  \bibfield{author}{%
  \bibinfo {author} {\bibfnamefont{T.}~\bibnamefont{Munson}}, \bibinfo {author}
  {\bibfnamefont{J.}~\bibnamefont{Sarich}}, \bibinfo {author}
  {\bibfnamefont{S.~M.}\ \bibnamefont{Wild}}, \bibinfo {author}
  {\bibfnamefont{S.}~\bibnamefont{Benson}},\ and\ \bibinfo {author}
  {\bibfnamefont{L.}~\bibnamefont{{Curfman McInnes}}},\ }%
  \emph{\bibinfo {title} {{TAO} 2.0 Users Manual}},\ \bibinfo {type} {Technical
  Memorandum}\ \bibinfo {number} {ANL/MCS-TM-322}\ (\bibinfo {institution}
  {Argonne National Laboratory},\ \bibinfo {address} {Argonne, Illinois},\
  \bibinfo {year} {2012})\
  \url{http://www.mcs.anl.gov/uploads/cels/papers/TM-322.pdf}%
  \bibAnnoteFile{NoStop}{taoman}%
\bibitem{mackenzie}%
  \BibitemOpen
  \bibfield{author}{%
  \bibinfo {author} {\bibfnamefont{J.~J.}\ \bibnamefont{MacKenzie}},\ }%
  \bibfield{journal}{%
  \bibinfo {journal} {Phys. Rev.}\ }%
  \textbf{\bibinfo {volume} {179}},\ \bibinfo {pages} {1002} (\bibinfo {year}
  {1969})%
  \bibAnnoteFile{NoStop}{mackenzie}%
\bibitem{brueckner_gammel}%
  \BibitemOpen
  \bibfield{author}{%
  \bibinfo {author} {\bibfnamefont{K.~A.}\ \bibnamefont{Brueckner}}\ and\
  \bibinfo {author} {\bibfnamefont{J.~L.}\ \bibnamefont{Gammel}},\ }%
  \bibfield{journal}{%
  \bibinfo {journal} {Phys. Rev.}\ }%
  \textbf{\bibinfo {volume} {109}},\ \bibinfo {pages} {1023} (\bibinfo {year}
  {1958})%
  \bibAnnoteFile{NoStop}{brueckner_gammel}%
\bibitem{mahaux1985}%
  \BibitemOpen
  \bibfield{author}{%
  \bibinfo {author} {\bibfnamefont{C.}~\bibnamefont{Mahaux}}, \bibinfo {author}
  {\bibfnamefont{P.~F.}\ \bibnamefont{Bortignon}},\ and\ \bibinfo {author}
  {\bibfnamefont{R.~A.}\ \bibnamefont{Broglia}},\ }%
  \bibfield{journal}{%
  \bibinfo {journal} {Phys. Rep.}\ }%
  \textbf{\bibinfo {volume} {120}},\ \bibinfo {pages} {1} (\bibinfo {year}
  {1985})%
  \bibAnnoteFile{NoStop}{mahaux1985}%
\bibitem{mahaux1989}%
  \BibitemOpen
  \bibfield{author}{%
  \bibinfo {author} {\bibfnamefont{C.}~\bibnamefont{Mahaux}}\ and\ \bibinfo
  {author} {\bibfnamefont{R.}~\bibnamefont{Sartor}},\ }%
  \bibfield{journal}{%
  \bibinfo {journal} {Phys. Rev. C}\ }%
  \textbf{\bibinfo {volume} {40}},\ \bibinfo {pages} {1833} (\bibinfo {year}
  {1989})%
  \bibAnnoteFile{NoStop}{mahaux1989}%
\bibitem{ramos_phd}%
  \BibitemOpen
  \bibfield{author}{%
  \bibinfo {author} {\bibfnamefont{{\`A}.~G.}\ \bibnamefont{Ramos}},\ }%
  Ph.D. thesis,\ \bibinfo {school} {University of Barcelona} (\bibinfo {year}
  {1988})%
  \bibAnnoteFile{NoStop}{ramos_phd}%
\bibitem{varshalovich}%
  \BibitemOpen
  \bibfield{author}{%
  \bibinfo {author} {\bibfnamefont{D.~A.}\ \bibnamefont{Varshalovich}},
  \bibinfo {author} {\bibfnamefont{A.~N.}\ \bibnamefont{Moskalev}},\ and\
  \bibinfo {author} {\bibfnamefont{V.~K.}\ \bibnamefont{Khersonskii}},\ }%
  \emph{\bibinfo {title} {Quantum Theory of Angular Momentum}}\ (\bibinfo
  {publisher} {World Scientific},\ \bibinfo {year} {1988})%
  \bibAnnoteFile{NoStop}{varshalovich}%
\bibitem{baldo}%
  \BibitemOpen
  \bibfield{author}{%
  \bibinfo {author} {\bibfnamefont{M.}~\bibnamefont{Baldo}}, \bibinfo {author}
  {\bibfnamefont{I.}~\bibnamefont{Bombaci}}, \bibinfo {author}
  {\bibfnamefont{L.~S.}\ \bibnamefont{Ferreira}}, \bibinfo {author}
  {\bibfnamefont{G.}~\bibnamefont{Giansiracusa}},\ and\ \bibinfo {author}
  {\bibfnamefont{U.}~\bibnamefont{Lombardo}},\ }%
  \bibfield{journal}{%
  \bibinfo {journal} {Phys. Rev. C}\ }%
  \textbf{\bibinfo {volume} {43}},\ \bibinfo {pages} {2605} (\bibinfo {year}
  {1991})%
  \bibAnnoteFile{NoStop}{baldo}%
\bibitem{polls2013}%
  \BibitemOpen
  \bibfield{author}{%
  \bibinfo {author} {\bibfnamefont{A.}~\bibnamefont{Polls}} \emph{et~al.},\ }%
  \bibfield{journal}{%
  \bibinfo {journal} {in preparation}}%
   (\bibinfo {year} {2013})%
  \bibAnnoteFile{NoStop}{polls2013}%
\bibitem{av18}%
  \BibitemOpen
  \bibfield{author}{%
  \bibinfo {author} {\bibfnamefont{R.~B.}\ \bibnamefont{Wiringa}}, \bibinfo
  {author} {\bibfnamefont{V.~G.~J.}\ \bibnamefont{Stoks}},\ and\ \bibinfo
  {author} {\bibfnamefont{R.}~\bibnamefont{Schiavilla}},\ }%
  \bibfield{journal}{%
  \bibinfo {journal} {Phys. Rev. C}\ }%
  \textbf{\bibinfo {volume} {51}},\ \bibinfo {pages} {38} (\bibinfo {year}
  {1995})%
  \bibAnnoteFile{NoStop}{av18}%
\bibitem{vonderfecht1993}%
  \BibitemOpen
  \bibfield{author}{%
  \bibinfo {author} {\bibfnamefont{B.}~\bibnamefont{Vonderfecht}}, \bibinfo
  {author} {\bibfnamefont{W.}~\bibnamefont{Dickhoff}}, \bibinfo {author}
  {\bibfnamefont{A.}~\bibnamefont{Polls}},\ and\ \bibinfo {author}
  {\bibfnamefont{A.}~\bibnamefont{Ramos}},\ }%
  \bibfield{journal}{%
  \bibinfo {journal} {Nuclear Physics A}\ }%
  \textbf{\bibinfo {volume} {555}},\ \bibinfo {pages} {1} (\bibinfo {year}
  {1993})%
  \bibAnnoteFile{NoStop}{vonderfecht1993}%
\bibitem{dean2003}%
  \BibitemOpen
  \bibfield{author}{%
  \bibinfo {author} {\bibfnamefont{D.~J.}\ \bibnamefont{Dean}}\ and\ \bibinfo
  {author} {\bibfnamefont{M.}~\bibnamefont{Hjorth-Jensen}},\ }%
  \bibfield{journal}{%
  \bibinfo {journal} {Rev. Mod. Phys.}\ }%
  \textbf{\bibinfo {volume} {75}},\ \bibinfo {pages} {607} (\bibinfo {year}
  {2003})%
  \bibAnnoteFile{NoStop}{dean2003}%
\bibitem{ekstrom2013b}%
  \BibitemOpen
  \bibfield{author}{%
  \bibinfo {author} {\bibfnamefont{A.}~\bibnamefont{Ekstr{\"o}m}},\ }%
  \bibfield{journal}{%
  \bibinfo {journal} {in preparation}}%
   (\bibinfo {year} {2013})%
  \bibAnnoteFile{NoStop}{ekstrom2013b}%
\bibitem{lattimer2012}%
  \BibitemOpen
  \bibfield{author}{%
  \bibinfo {author} {\bibfnamefont{J.~M.}\ \bibnamefont{Lattimer}},\ }%
  \bibfield{journal}{%
  \bibinfo {journal} {Annu. Rev. Nucl. Part. Science}\ }%
  \textbf{\bibinfo {volume} {62}},\ \bibinfo {pages} {485} (\bibinfo {year}
  {2012})%
  \bibAnnoteFile{NoStop}{lattimer2012}%
\bibitem{li2002}%
  \BibitemOpen
  \bibfield{author}{%
  \bibinfo {author} {\bibfnamefont{B.-A.}\ \bibnamefont{Li}},\ }%
  \bibfield{journal}{%
  \bibinfo {journal} {Phys. Rev. Lett.}\ }%
  \textbf{\bibinfo {volume} {88}},\ \bibinfo {pages} {192701} (\bibinfo {year}
  {2002})%
  \bibAnnoteFile{NoStop}{li2002}%
\bibitem{erler2013}%
  \BibitemOpen
  \bibfield{author}{%
  \bibinfo {author} {\bibfnamefont{J.}~\bibnamefont{Erler}}, \bibinfo {author}
  {\bibfnamefont{C.~J.}\ \bibnamefont{Horowitz}}, \bibinfo {author}
  {\bibfnamefont{W.}~\bibnamefont{Nazarewicz}}, \bibinfo {author}
  {\bibfnamefont{M.}~\bibnamefont{Rafalski}},\ and\ \bibinfo {author}
  {\bibfnamefont{P.-G.}\ \bibnamefont{Reinhard}},\ }%
  \bibfield{journal}{%
  \bibinfo {journal} {Phys. Rev. C}\ }%
  \textbf{\bibinfo {volume} {87}},\ \bibinfo {pages} {044320} (\bibinfo {year}
  {2013})%
  \bibAnnoteFile{NoStop}{erler2013}%
\bibitem{engvik1997}%
  \BibitemOpen
  \bibfield{author}{%
  \bibinfo {author} {\bibfnamefont{L.}~\bibnamefont{Engvik}}, \bibinfo {author}
  {\bibfnamefont{M.}~\bibnamefont{Hjorth-Jensen}}, \bibinfo {author}
  {\bibfnamefont{R.}~\bibnamefont{Machleidt}}, \bibinfo {author}
  {\bibfnamefont{H.}~\bibnamefont{M{\"u}ther}},\ and\ \bibinfo {author}
  {\bibfnamefont{A.}~\bibnamefont{Polls}},\ }%
  \bibfield{journal}{%
  \bibinfo {journal} {Nucl. Phys. A}\ }%
  \textbf{\bibinfo {volume} {627}},\ \bibinfo {pages} {85} (\bibinfo {year}
  {1997})%
  \bibAnnoteFile{NoStop}{engvik1997}%
\bibitem{heiselberg}%
  \BibitemOpen
  \bibfield{author}{%
  \bibinfo {author} {\bibfnamefont{H.}~\bibnamefont{Heiselberg}}\ and\ \bibinfo
  {author} {\bibfnamefont{M.}~\bibnamefont{Hjorth-Jensen}},\ }%
  \bibfield{journal}{%
  \bibinfo {journal} {Phys. Rep.}\ }%
  \textbf{\bibinfo {volume} {328}},\ \bibinfo {pages} {237} (\bibinfo {year}
  {2000})%
  \bibAnnoteFile{NoStop}{heiselberg}%
\bibitem{myers1990}%
  \BibitemOpen
  \bibfield{author}{%
  \bibinfo {author} {\bibfnamefont{W.~D.}\ \bibnamefont{Myers}}\ and\ \bibinfo
  {author} {\bibfnamefont{W.~J.}\ \bibnamefont{Swiatecki}},\ }%
  \bibfield{journal}{%
  \bibinfo {journal} {Ann. of Phys.}\ }%
  \textbf{\bibinfo {volume} {204}},\ \bibinfo {pages} {401} (\bibinfo {year}
  {1990})%
  \bibAnnoteFile{NoStop}{myers1990}%
\end{thebibliography}

\bibliographystyle{apsrev}

\appendix

\section{Technical details}
\subsection{Relative momentum basis} \label{sec:relmom}

  Infinite nuclear matter is defined in the thermodynamic limit,
  that is, the limit where the volume $\Omega $ and the number of
  particles $A$ approach infinity, while the density of particles
  $\rho \equiv A/\Omega $ is kept constant. At the limit when $\Omega
  $ approaches infinity, the sums over momenta can be replaced by
  integrals:
  \begin{equation} \label{eq:sum2int}
    \sum_{\mathbf{k}} \rightarrow \frac{\Omega }{(2\pi )^{3}}\int
    d\mathbf{k}.
  \end{equation}
  We will in the following replace all sums over momenta by integrals
  according to Eq.~(\ref{eq:sum2int}).

  Taking the limit $\Omega
  \rightarrow \infty $, the ladder amplitude equations 
  (\ref{eq:pphhlad}) may be written in laboratory momentum 
  coordinates as
  \begin{align} \label{eq:pphh_lab}
    0 &=
    \braket{\mathbf{k}_{a}\mathbf{k}_{b}}{v|\mathbf{k}_{i}\mathbf{k}_{j}}
    \nonumber \\ & + \left( \vare(\mathbf{k}_{a}) +
    \vare(\mathbf{k}_{b}) - \vare(\mathbf{k}_{i}) -
    \vare(\mathbf{k}_{j}) \right)
    \braket{\mathbf{k}_{a}\mathbf{k}_{b}}{t|\mathbf{k}_{i}\mathbf{k}_{j}}
    \nonumber \\ &+ \frac{1}{2}\left( \frac{\Omega }{(2\pi
      )^{3}}\right)^{2}\int d\mathbf{k}_{k}\int d\mathbf{k}_{l}
    \braket{\mathbf{k}_{a}\mathbf{k}_{b}}{t|\mathbf{k}_{k}\mathbf{k}_{l}}\braket{\mathbf{k}_{k}\mathbf{k}_{l}}{v|\mathbf{k}_{i}\mathbf{k}_{j}}
    \nonumber \\ &\times
    \theta(k_{F}-|\mathbf{k}_{k}|)\theta(k_{F}-|\mathbf{k}_{l}|)
    \nonumber \\ &+ \frac{1}{2}\left( \frac{\Omega }{(2\pi
      )^{3}}\right)^{2}\int d\mathbf{k}_{c}\int
    \mathbf{k}_{d}\braket{\mathbf{k}_{a}\mathbf{k}_{b}}{v|\mathbf{k}_{c}\mathbf{k}_{d}}\braket{\mathbf{k}_{c}\mathbf{k}_{d}}{t|\mathbf{k}_{i}\mathbf{k}_{j}}
    \nonumber \\ &\times
    \theta(|\mathbf{k}_{c}|-k_{F})\theta(|\mathbf{k}_{d}|-k_{F}),
  \end{align}
  where $\theta(x) $ is the Heaviside step function and we have used
  the definition $\vare(\mathbf{k}) \equiv
  \braket{\mathbf{k}}{f|\mathbf{k}}$. Later, we will refer to $\vare(\mathbf{k})$ 
  as the single-particle energy.

We define the relative and center-of-mass (RCM) momentum coordinates
as
  \begin{align} \label{eq:rcm_lab}
    \mathbf{k} &= (\mathbf{k}_{i}-\mathbf{k}_{j})/2, \quad &\mathbf{K}
    &= \mathbf{k}_{i}+\mathbf{k}_{j} \nonumber \\ \mathbf{k}' &=
    (\mathbf{k}_{a}-\mathbf{k}_{b})/2, \quad &\mathbf{K}' &=
    \mathbf{k}_{a}+\mathbf{k}_{b} \nonumber \\ \mathbf{h} &=
    (\mathbf{k}_{k}-\mathbf{k}_{l})/2, \quad &\mathbf{H} &=
    \mathbf{k}_{k}+\mathbf{k}_{l} \nonumber \\ \mathbf{p} &=
    (\mathbf{k}_{c}-\mathbf{k}_{d})/2, \quad &\mathbf{P} &=
    \mathbf{k}_{c}+\mathbf{k}_{d},
  \end{align}
  where $i, j, k, l$ denote single-particle states occupied and 
  $a, b, c, d$ states unoccupied in the uncorrelated Fermi vacuum state. 
  Transforming to RCM coordinates, the PPHH-LAD equations become
  \begin{align} \label{eq:ladrcm}
    0 &= \braket{\mathbf{k}'}{v|\mathbf{k}} \nonumber \\ & + \left(
    \vare(|\mathbf{k}'+\mathbf{K}/2|)+\vare(|-\ma{k}'+\ma{K}/2|)
    \right. \nonumber \\ &
    \left. -\vare(|\ma{k}-\ma{K}/2|)-\vare(|-\ma{k}+\ma{K}/2|)\right)
    \braket{\ma{k}'}{t|\ma{k}} \nonumber \\ &+ \frac{1}{2}\int
    d\mathbf{h}
    \braket{\mathbf{k}'}{t(\mathbf{K})|\mathbf{h}}\braket{\mathbf{h}}{v|\mathbf{k}}
    \nonumber \\ &\times
    \theta(k_{F}-|\mathbf{h}+\mathbf{K}/2|)\theta(k_{F}-|-\mathbf{h}+\mathbf{K}/2|)
    \nonumber \\ &+ \frac{1}{2}\int d\mathbf{p}
    \braket{\mathbf{k}'}{v|\mathbf{p}}\braket{\mathbf{p}}{t(\mathbf{K})|\mathbf{k}}
    \nonumber \\ &\times
    \theta(|\mathbf{p}+\mathbf{K}/2|-k_{F})\theta(|-\mathbf{p}+\mathbf{K}/2|-k_{F}),
  \end{align}
  where the relation
  \begin{equation}
    \braket{\mathbf{k}_{p}\mathbf{k}_{q}}{v|\mathbf{k}_{r}\mathbf{k}_{s}}
    = \frac{\left(2\pi \right)^{3}}{\Omega
    }\braket{\mathbf{k}}{v|\mathbf{k}'}\delta_{\mathbf{K}\mathbf{K}'}
  \end{equation}
  has been used. The expressions (\ref{eq:eneref}) and (\ref{eq:eneccd}) 
  for the reference and correlation energies, respectively, 
  can be transformed to RCM coordinates in a similar way as 
  is shown here for the ladder amplitude equations.

Due to the isotropy of nuclear
  matter, we assume that the single-particle energy
  $\vare(\mathbf{k}_{p})$ depends only on the absolute value of the
  argument $\mathbf{k}_{p}$ \cite{ramos_phd,ramos_polls}. In laboratory frame momentum coordinates, the
  single-particle energy is then
  \begin{equation}
    \vare(|\mathbf{k}_{p}|) = \frac{\hbar^{2}k_{p}^{2}}{2m} +
    U(|\mathbf{k}_{p}|)
  \end{equation}
  where
  \begin{equation}
    U(|\mathbf{k}_{p}|) = \frac{\Omega }{(2\pi
      )^{3}}\sum_{m_{s}m_{t}}\int
    d\mathbf{k}_{q}\braket{\mathbf{k}_{p}\mathbf{k}_{q}}{v|\mathbf{k}_{p}\mathbf{k}_{q}}\theta(k_{F}-|\mathbf{k}_{q}|).
  \end{equation}
  Because of the isotropy, we choose the direction such that
  $\mathbf{k}_{p} = (0,0,k_{p})$. The single-particle potential energy
  can also be written as
  \begin{align}
    U(|\mathbf{k}_{p}|) =& \sum_{m_{s}m_{t}}\int d\mathbf{k}_{q}
    \left[
      \braket{\mathbf{p}}{v|\mathbf{p}}-\braket{\mathbf{p}}{v|-\mathbf{p}}\right]
    \nonumber \\ & \times \theta(k_{F}-|-\mathbf{p}+\mathbf{P}/2|),
  \end{align}
  where $\mathbf{p}$ and $\mathbf{P}$ are relative and center-of-mass
  momentum coordinates defined by $\mathbf{k}_{p}$ and $\mathbf{k}_{q}$, as in
  Eq.~(\ref{eq:rcm_lab}). 

\subsection{Momentum and angular momentum basis} \label{sec:angmom}

In this work, we assume that every interaction matrix element 
\begin{align}  
  \braket{k'(l'S)\mathcal{J}m_{\mathcal{J}}M_{T}}{v|k(lS)\mathcal{J}m_{\mathcal{J}}M_{T}} \nonumber
\end{align}
and every $t$-amplitude matrix element is multiplied by a factor
\begin{align} 
  \mathcal{A}^{l'lSM_{T}} = \left\{ \begin{array}{ll}
    1+(-1)^{l+l'}, & \text{ if } M_{T} = 0, \\
    \frac{1}{2}(1-(-1)^{l+S+1}) &\\
    \times (1-(-1)^{l'+S+1}), & \text{ if } |M_{T}| = 1,
  \end{array} \right.
  \label{eq:antisymm_app}
\end{align}
which ensures antisymmetry and conservation of parity. 

Using exact Pauli exclusion operators, the ladder 
$t$-amplitude equations (\ref{eq:ladrcm}) can be rewritten 
in the coupled angular momentum basis as
  \begin{align} 
    &\Delta \vare(\mathbf{k},\mathbf{k}',\mathbf{K})\langle
    k'(l'S)\mathcal{J}'m_{\mathcal{J}'}M_{T}|t(\mathbf{K},\hat{\mathbf{k}},\hat{\mathbf{k}}')|k(lS)\mathcal{J}m_{\mathcal{J}}M_{T}\rangle
    \nonumber
    \\ &=\braket{k'(l'S)\mathcal{J}'m_{\mathcal{J}'}M_{T}}{v|k(lS)\mathcal{J}m_{\mathcal{J}}M_{T}}\delta_{\mathcal{J}\mathcal{J}'}\delta_{m_{\mathcal{J}}m_{\mathcal{J}'}}
    \nonumber \\ &+
    \frac{1}{2}\sum_{\mathcal{J}''m_{\mathcal{J}''}}\sum_{l''l'''}\int_{0}^{k_{F}}h^{2}dh
    \nonumber \\ &\times \langle
    k'(l'S)\mathcal{J}'m_{\mathcal{J}'}M_{T}|t(\mathbf{K},\hat{\mathbf{k}},\hat{\mathbf{k}}')|h(l''S)\mathcal{J}''m_{\mathcal{J}''}M_{T}\rangle
    \nonumber \\ &\times
    \braket{h(l'''S)\mathcal{J}m_{\mathcal{J}}M_{T}}{v|k(lS)\mathcal{J}m_{\mathcal{J}}M_{T}}
    \nonumber \\ &\times
    Q_{hh}(l''\mathcal{J}''m_{\mathcal{J}''},l'''\mathcal{J}m_{\mathcal{J}};SM_{T}hK\theta_{K}\phi_{K})
    \nonumber \\ &+
    \frac{1}{2}\sum_{\mathcal{J}''m_{\mathcal{J}''}}\sum_{l''l'''}\int_{0}^{\infty
    }p^{2}dp \nonumber \\ &\times
    \braket{k'(l'S)\mathcal{J}'m_{\mathcal{J}'}M_{T}}{v|p(l''S)\mathcal{J}'m_{\mathcal{J}'}M_{T}}
    \nonumber \\ &\times \langle
    p(l'''S)\mathcal{J}''m_{\mathcal{J}''}M_{T}|t(\mathbf{K},\hat{\mathbf{k}},\hat{\mathbf{k}}')|k(lS)\mathcal{J}m_{\mathcal{J}}M_{T}\rangle
    \nonumber \\ &\times
    Q_{pp}(l''\mathcal{J}'m_{\mathcal{J}'},l'''\mathcal{J}''m_{\mathcal{J}''};SM_{T}pK\theta_{K}\phi_{K}),
    \label{eq:pphhlad_exact2}
  \end{align}
  where $\ket{(l S)\mathcal{J} m_{\mathcal{J}}}$ denotes a vector
  where $l$ and $S$ are coupled to $\mathcal{J}$. In 
  Eq.~(\ref{eq:pphhlad_exact2}) we have introduced the 
  shorthand notation
\begin{align} \label{eq:enedenom}
  \Delta \vare(\mathbf{k},\mathbf{k}',\mathbf{K}) &\equiv
  \vare(|\mathbf{k}+\mathbf{K}/2|) + \vare(|-\mathbf{k}+\mathbf{K}/2|)
  \nonumber \\ & - \vare(|\mathbf{k}'+\mathbf{K}/2|) -
  \vare(|-\mathbf{k}'+\mathbf{K}/2|)
\end{align}
for the energy denominator.
  
The sum of the single-particle energies corresponding to two hole
states can be expressed in terms of RCM coordinates as
\begin{align} \label{eq:spene2}
  \vare(|\mathbf{k}_{i}|) + \vare(|\mathbf{k}_{j}|) = &
  \vare(|\mathbf{k}+\mathbf{K}/2|) + \vare(|-\mathbf{k}+\mathbf{K}/2|)
  \nonumber \\ = & \frac{\hbar^{2}k^{2}}{m} +
  \frac{\hbar^{2}K^{2}}{4m} \nonumber \\ & +
  U^{M_{T},+}(|\mathbf{k}+\mathbf{K}/2|) \nonumber \\ & +
  U^{M_{T},-}(|-\mathbf{k}+\mathbf{K}/2|),
\end{align}
where
\begin{align}
  U^{M_{T},\pm }(|\mathbf{k}_{p}|) = &
  \frac{1}{4}\sum_{\mathcal{J}Sl}(2\mathcal{J}+1)\int_{0}^{k_{F}}dk_{q}
  k_{q}^{2}\int_{-1}^{1}d\cos \theta_{\mathbf{k}_{q}} \nonumber \\ &
  \times \mathcal{B}^{M_{T},\pm
  }\braket{p\mathcal{J}lS}{v|p\mathcal{J}lS},
\end{align}
and the variable $p = |\mathbf{k}_{p}-\mathbf{k}_{q}|/2$. 
The antisymmetrization operator $\mathcal{B}^{M_{T},\pm }$ 
is defined through the relation
\begin{align}
  & \mathcal{B}^{M_{T},\pm }\braket{p\mathcal{J}lS}{v|p\mathcal{J}l'S}
  = \braket{p\mathcal{J}lS}{v(M_{T}'=0)|p\mathcal{J}l'S} \nonumber
  \\ & + (1 - (-1)^{l'+S'+1}) \nonumber \\ & \times
  \braket{p\mathcal{J}lS}{v(M_{T}'=M_{T}\pm
    \delta_{M_{T},0})|p\mathcal{J}l'S}
\end{align}
for symmetric nuclear matter and the relation
\begin{align}
  & \mathcal{B}^{M_{T},\pm }\braket{p\mathcal{J}lS}{v|p\mathcal{J}l'S}
  = (1 - (-1)^{l'+S'+1}) \nonumber \\ & \times
  \braket{p\mathcal{J}lS}{v(M_{T}'=1)|p\mathcal{J}l'S}
\end{align}
for pure neutron matter.
 The expressions of $\mathbf{k}$ and
$\mathbf{K}$ are given in Eq.~(\ref{eq:rcm_lab}). The sum of
single-particle energies corresponding to two particle states,
$\vare(\mathbf{k}_{a}) + \vare({\mathbf{k}_{b}})$, is calculated in
the same way. We define the angular-averaged energy denominator as
\begin{align} \label{eq:angavdenom}
  \Delta \tilde{\varepsilon }(k,k',K) &\equiv \varepsilon(\overline{k_{i}}) + \varepsilon(\overline{k_{j}}) - \varepsilon(\overline{k_{a}}) - \varepsilon(\overline{k_{b}}),
\end{align}
where $\overline{k_{p}}$ for $p = i,j,a,b$ are angular-averaged 
input momenta defined in Eq.~(\ref{eq:k_lab_ave}). Observe that the energy denominator is assumed to be a function
of the two-particle isospin projection $M_{T}$.

The correlation energy can be written in the partial wave
expansion as
\begin{align} \label{eq:enecc_exact1}
  & \Delta E_{CCD}/A = \frac{3C}{64\pi
    k_{F}^{3}}\sum_{\mathcal{J}m_{\mathcal{J}}}\sum_{\mathcal{J}''m_{\mathcal{J}''}}\sum_{\mathcal{J}'''m_{\mathcal{J}'''}}\sum_{SM_{T}}\sum_{ll'l''l'''}
  \nonumber \\ & \times \int_{0}^{\sqrt{k_{F}^{2}-K^{2}/4}}k^{2}dk
  \int_{\sqrt{k_{F}^{2}-K^{2}/4}}^{\infty }k'^{2}dk'
  \int_{0}^{2k_{F}}K^{2}dK \nonumber \\ & \times \int_{-1}^{1}d\cos
  \theta_{K} \int_{0}^{2\pi }d\phi_{K}
  \braket{k(lS)\mathcal{J}M_{T}}{v|k'(l'S)\mathcal{J}M_{T}} \nonumber
  \\ & \times
  \hat{Q}_{hh}(l'''\mathcal{J}'''m_{\mathcal{J}'''},l\mathcal{J}m_{\mathcal{J}};SM_{T}kK\theta_{K}\phi_{K})
  \nonumber \\ & \times
  \hat{Q}_{pp}(l'\mathcal{J}m_{\mathcal{J}},l''\mathcal{J}''m_{\mathcal{J}''};SM_{T}k'K\theta_{K}\phi_{K})
  \nonumber \\ & \times \langle
  k'(l''S)\mathcal{J}''m_{\mathcal{J}''}M_{T}|t(\mathbf{K},\hat{\mathbf{k}},\hat{\mathbf{k}}')|k(l'''S)\mathcal{J}'''m_{\mathcal{J}'''}M_{T}\rangle
\end{align}
where the Pauli operators $\hat{Q}_{hh}$ and $\hat{Q}_{pp}$ are
defined in Eqs. (\ref{eq:qhh}) and (\ref{eq:qpp}). Here we use the
notations $\hat{Q}_{hh}$ and $\hat{Q}_{pp}$ instead of $Q_{hh}$ and
$Q_{pp}$ to emphasize that these are integral operators that operate
on the $t$-amplitude matrix. The $t$-amplitude depends on
$\hat{\mathbf{k}} \equiv (\theta_{\mathbf{k}},\phi_{\mathbf{k}})$ and
$\hat{\mathbf{k}}' \equiv (\theta_{\mathbf{k}'},\phi_{\mathbf{k}'})$
through the energy denominator $\Delta
\vare(\mathbf{k},\mathbf{k}',\mathbf{K})$, and the closed-form
expression (\ref{eq:qpp_simple}) can therefore generally not be used
in the energy equation (\ref{eq:enecc_exact1}).



When the ladder equations are written in the coupled partial wave
basis, it is possible to calculate the $t$-amplitude matrix for only
one angular direction of the CM momentum $\mathbf{K}$, and then obtain
the other matrix elements by performing a rotation
\cite{suzuki}. Using the same technique as Suzuki \emph{et al.}, an
amplitude matrix with a general CM momentum vector can be written
   \begin{align} \label{eq:rotation}
    &\braket{k'(l''S)\mathcal{J}''m_{\mathcal{J}''}}{t(\mathbf{K})|k(l'''S)\mathcal{J}'''m_{\mathcal{J}'''}}
     \nonumber \\ &=
     \sum_{m_{\mathcal{J}}m_{\mathcal{J}'}}D_{m_{\mathcal{J}''}m_{\mathcal{J}}}^{\mathcal{J}''}(\phi_{K},\theta_{K},0)D_{m_{\mathcal{J}'''}m_{\mathcal{J}'}}^{\mathcal{J}'''
       *}(\phi_{K},\theta_{K},0) \nonumber \\ &\times
     \braket{k'(l''S)\mathcal{J}''m_{\mathcal{J}}}{t(K)|k(l'''S)\mathcal{J}'''m_{\mathcal{J}'}},
  \end{align}
where $D_{m_{\mathcal{J}}m_{\mathcal{J}'}}^{\mathcal{J}}(\alpha ,\beta
,\gamma )$ is the Wigner $D$-function and $\alpha $, $\beta $, and
$\gamma $ are Euler angles, defined in for example Ref.~\cite{varshalovich}. Eq.~(\ref{eq:rotation}) can be used to obtain the correlation energy expression (\ref{eq:ene_exact}) from Eq.~(\ref{eq:enecc_exact1}).

\end{document}